\documentclass[a4paper,11pt]{article}

\pdfoutput=1

\usepackage{jheppub}
\usepackage{amsmath}
\usepackage{xspace}
\usepackage{hyperref}
\usepackage{mdwlist}
\usepackage[dvipsnames]{xcolor}
\usepackage[normalem]{ulem}

\definecolor{darkgreen}{rgb}{0.09, 0.45, 0.27}

\newcommand{\eq}[1]{eq.~\eqref{eq:#1}}
\newcommand{\eqs}[2]{eqs.~\eqref{eq:#1} and \eqref{eq:#2}}
\renewcommand{\sec}[1]{sec.~\ref{sec:#1}}
\newcommand{\secs}[2]{secs.~\ref{sec:#1} and \ref{sec:#2}}

\newcommand{\fig}[1]{fig.~\ref{fig:#1}}

\newcommand{\nnu}{\nonumber\\}
\newcommand{\nn}{\nonumber}
\newcommand{\bef}{\begin{figure}[t]\centering}
\newcommand{\eef}{\end{figure}}
\def\bea#1\eea{\begin{align}#1\end{align}}
\def \be  {\begin{equation}}
\def \ee  {\end{equation}}
\def \ba  {\begin{eqnarray}}
\def \ea  {\end{eqnarray}}

\newcommand{\f}{\frac}
\newcommand{\ord}[1]{\mathcal{O}(#1)}

\newcommand{\df}{\mathrm{d}}

\newcommand{\Li}{\textrm{Li}}

\newcommand{\sdt}{\!\cdot\!}

\newcommand{\al}{\alpha}
\newcommand{\bt}{\beta}
\newcommand{\ga}{\gamma}

\newcommand{\de}{\delta}

\newcommand{\si}{\sigma}
\newcommand{\zc}{z_{\rm cut}}
\newcommand{\fc}{f_{\rm cut}}
\newcommand{\tg}{\theta_g}
\newcommand{\tet}{\theta_t}

\newcommand{\cG}{{\mathcal G}}

\newcommand{\CS}{{\tilde{\mathcal S}}}

\newcommand{\bn}{\bar{n}}

\newcommand\as{\alpha_s}
\newcommand{\lqcd}{\Lambda_\mathrm{QCD}}

\newcommand{\ed}{\Delta_E}
\newcommand{\rs}{R_{\rm sub}}

\newcommand{\Pythia}{\textsc{Pythia}\xspace}

\allowdisplaybreaks[2]

\title{Jet energy drop}

\author[a,b]{Pedro Cal,}
\author[c,d]{Kyle Lee,}
\author[e,f]{Felix Ringer,}
\author[a,b]{Wouter J. Waalewijn}

\affiliation[a]{Institute for Theoretical Physics Amsterdam and Delta Institute for Theoretical Physics, University of Amsterdam, Science Park 904, 1098 XH Amsterdam, The Netherlands}
\affiliation[b]{Nikhef, Theory Group, Science Park 105, 1098 XG, Amsterdam, The Netherlands}
\affiliation[c]{C.N. Yang Institute for Theoretical Physics, Stony Brook University, Stony Brook, NY 11794,USA}
\affiliation[d]{Department of Physics and Astronomy, Stony Brook University, Stony Brook, NY 11794, USA}
\affiliation[e]{Nuclear Science Division, Lawrence Berkeley National Laboratory, Berkeley, CA 94720, USA}
\affiliation[f]{Physics Department, University of California, Berkeley, CA 94720, USA}
                                         
\emailAdd{p.cal@nikhef.nl}
\emailAdd{kunsu.lee@stonybrook.edu}
\emailAdd{fmringer@lbl.gov}
\emailAdd{w.j.waalewijn@uva.nl}

\abstract{
We study the jet energy drop, which is the relative difference between the groomed and ungroomed jet energy or  transverse momentum. It is one of the fundamental quantities that characterizes the impact of grooming on jets produced in high energy collisions. We consider three different grooming algorithms i)  soft drop, ii) iterated soft drop, and iii) trimming. We carry out the resummation of large logarithms of the jet energy drop, the jet radius as well as relevant grooming parameters at next-to-leading logarithmic (NLL$'$) accuracy. In addition, we account for non-global and clustering logarithms, and determine the next-to-leading order corrections. For soft drop we perform a joint resummation of the jet energy drop and the groomed jet radius, which is necessary to achieve the correct all-order structure of the cross section, in particular for the Sudakov-safe case of soft drop with $\beta=0$. 
We present numerical results for LHC energies and compare to \Pythia simulations as well as CMS data. 
Our factorization framework predicts the onset of nonperturbative effects in the jet energy distribution, in line with what we find in \Pythia.
The jet energy drop observables stand out because they only probe soft radiation, making them ideal candidates for the tuning of parton shower Monte Carlo event generators and for probing medium effects in heavy-ion collisions.}

\preprint{\vbox{%
\hbox{NIKHEF 20-017}
\hbox{YITP-SB-20-16}
}}
\begin{document}
\maketitle

\newpage

\section{Introduction~\label{sec:intro}}

Jet substructure techniques have become an important part of measurements at high energy particle colliders over the last decade. These techniques are used in searches for physics beyond the Standard Model, e.g.~tagging hadronic decays of heavy resonances or discriminating quark/gluon jets, the measurement of fundamental parameters, such as the strong coupling constant, and probing the modification of jets in heavy-ion collisions.  See refs.~\cite{Larkoski:2017jix,Asquith:2018igt,Marzani:2019hun} for reviews from a theoretical and experimental point of view. 

\begin{figure}[t]
\centering
\includegraphics[width=0.4\textwidth]{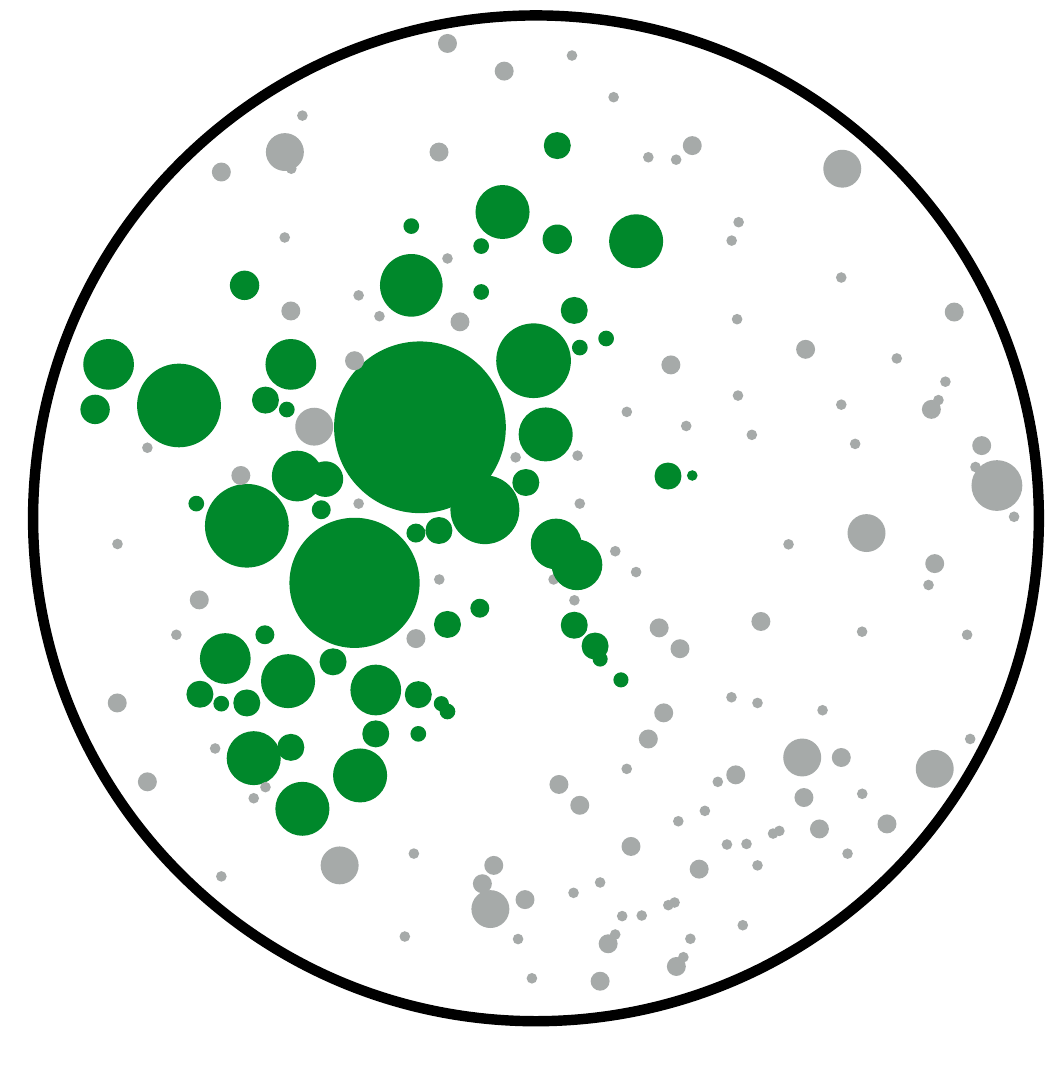} 
\caption{Schematic picture of a jet, with circles representing radiation whose size corresponds to its energy. Wide-angle soft radiation (grey) is groomed away, resulting in the groomed jet (green).~\label{fig:groomed_jet}}
\end{figure}

In the past couple of years, precision calculations of jet substructure observables have become available, that allow for direct comparisons of theoretical calculations and data. A crucial ingredient in making this possible, is the development of jet grooming techniques that are compatible with theoretical calculations. Grooming techniques address the highly-contaminated environment at hadron colliders, systematically removing soft wide-angle radiation from the observed jets, see \fig{groomed_jet},  thereby also reducing hadronization effects. Examples are trimming~\cite{Krohn:2009th}, pruning~\cite{Ellis:2009me}, soft killer~\cite{Cacciari:2014gra}, soft drop~\cite{Larkoski:2014wba}, iterated soft drop~\cite{Frye:2017yrw} and recursive soft drop~\cite{Dreyer:2018tjj}. 
Initially, grooming (desirable for experiment) and theoretical precision seemed mutually exclusive, but some of these grooming techniques are quite amenable to calculations in perturbative QCD. Specific examples include: the soft-drop groomed jet mass~\cite{Frye:2016aiz,Marzani:2017mva,Kang:2018jwa}, the groomed jet radius~\cite{Kang:2019prh} and the repositioning of the jet axes due to grooming~\cite{Cal:2019gxa}. Experimental results for soft-drop groomed jet observables can be found in refs.~\cite{Aaboud:2017qwh,Sirunyan:2017bsd,Kauder:2017cvz,Sirunyan:2018gct,Sirunyan:2018xdh,Acharya:2019djg,ATLAS:2019sol,Aad:2019vyi,Adam:2020kug,Aad:2020zcn}, and for related recent theoretical calculations of groomed jet substructure observables see refs.~\cite{Makris:2017hjk,Larkoski:2017cqq,Larkoski:2017iuy,Baron:2018nfz,Kang:2018vgn,Makris:2018npl,Kardos:2018kth,Chay:2018pvp,Napoletano:2018ohv,Lee:2019lge,Hoang:2019ceu,Gutierrez-Reyes:2019msa,Kardos:2019iwa,Marzani:2019evv,Mehtar-Tani:2019rrk,Kardos:2020ppl,Larkoski:2020wgx,Lifson:2020gua}.

In this work, we consider the jet energy drop, which is given by the relative transverse momentum (or energy) difference between the groomed and ungroomed (i.e.~original) jet,
\begin{equation}
\Delta_E=\frac{p_T-p_T^{\rm gr}}{p_T}=1-\frac{p_T^{\rm gr}}{p_T} \,.
\end{equation}
We consider the jet energy drop for three grooming procedures that have been used by experimental collaborations: i) trimming, ii) soft drop and iii) iterated soft drop. This observable is of great interest for characterizing the impact of grooming on the measured jets. In particular, the soft sensitivity of these observables makes them ideally suited for tuning parton shower event generators, see e.g.~refs.~\cite{ATLAS:2011gmi,Amoroso:2020lgh}. While we focus in this paper on the comparison to Pythia in the perturbative regime, studying the nonperturbative regime requires a field theoretic understanding of the effects on grooming, which have been discussed in refs.~\cite{Hoang:2019ceu,Cal:2019gxa}. For models describing the nonperturbative effect with grooming, see refs.~\cite{Dasgupta:2013ihk,Frye:2016aiz,Larkoski:2017iuy,Kang:2018vgn,Hoang:2017kmk,Lee:2019lge,Marzani:2017kqd}. Similarly, collinear drop~\cite{Chien:2019osu} also probes soft radiation, by ``taking the difference" of two soft drop grooming procedures with different parameters. However,  this removes the softest radiation, which is kept in our case.

For each grooming procedure, we develop the factorization formula for jet energy drop within Soft Collinear Effective Theory (SCET)~\cite{Bauer:2000ew, Bauer:2000yr, Bauer:2001ct, Bauer:2001yt,Beneke:2002ph}, which allows for the resummation of large logarithmic corrections to all orders at next-to-leading logarithmic (NLL$'$) order. The logarithms we resum are those of the jet energy drop, as well as the  jet radius and grooming parameters. We will treat all logarithms as independent, but in principle, one can refine predictions (e.g.~near the endpoint in $\ed$) by considering parametric relations between the jet energy drop and grooming parameters. To obtain our predictions, we included the one-loop expression for the ingredients of the factorization theorem, the one-loop anomalous dimensions, and two-loop cusp anomalous dimension, and the non-global~\cite{Dasgupta:2001sh} and Abelian logarithms including clustering effects~\cite{Appleby:2002ke}. While there has been significant progress in the study of NGLs~\cite{Banfi:2002hw,Weigert:2003mm,Hornig:2011tg,Hornig:2011iu,Dasgupta:2012hg,Hagiwara:2015bia,Caron-Huot:2015bja,Larkoski:2015zka,Becher:2016mmh,Larkoski:2016zzc,Balsiger:2020ogy}, including clustering effects~\cite{Delenda:2006nf,KhelifaKerfa:2011zu,Delenda:2012mm,Kelley:2012kj,Kelley:2012zs,Neill:2018yet}, we restrict to their contribution at order $\al_s^2$, since the higher order terms are numerically irrelevant for our phenomenological results. 

In order to resum the relevant logarithms for soft drop, we perform a joint resummation of logarithms of $\Delta_E$ and the soft drop groomed jet radius $R_g$~\cite{Larkoski:2014wba,Kang:2019prh}. 
For these two variables, we develop a two-dimensional scale-setting technique in order to perform numerical calculations. Depending on the relative scaling of $\Delta_E$ and $R_g$, a different factorization formula is obtained, which are matched before integrating over a range of $R_g$ or integrating it out completely. 
For the special case of soft-drop parameter $\beta=0$, the corresponding cross section is not infrared safe but Sudakov safe~\cite{Larkoski:2014wba}, and so resummation is essential to obtain a prediction. Other examples of Sudakov-safe observables include the soft-drop momentum sharing fraction $z_g$~\cite{Larkoski:2015lea}, ratios of two angularities~\cite{Larkoski:2015uqa,Procura:2018zpn}, and the jet-pull angle~\cite{Gallicchio:2010sw,Larkoski:2019urm,Larkoski:2019fsm}. In this work, we extend previous results of the jet energy drop for soft drop with $\beta=0$ beyond (modified) leading logarithmic accuracy.

The remainder of this paper is organized as follows. In the three subsequent sections~\ref{sec:iteratedsoftdrop}-\ref{sec:trimming} we discuss the jet energy drop for the three grooming algorithms: Iterated soft drop, soft drop, and trimming.
In each section, we first introduce the grooming procedure (though soft drop is already described in \sec{iteratedsoftdrop}), and present results at fixed order. We then factorize the cross section to resum large logarithmic corrections to all orders, and give expressions for all necessary perturbative ingredients, as well as non-global and clustering logarithms. In addition, we discuss profile scales and present numerical results for LHC kinematics, which we compare to \Pythia. Note that \sec{iteratedsoftdrop} contains many of the basic ingredients that are also needed in subsequent sections, such as the collinear factorization for inclusive jet production.
In~\sec{conclusions}, we draw conclusions and present an outlook.

\section{Iterated soft drop~\label{sec:iteratedsoftdrop}}

We start in~\sec{fac_jet} by reviewing the factorization of the inclusive jet cross section in terms of PDFs, hard functions, and jet functions. This initial step exploits the collimated nature of jets, but is independent of further details of the jet measurement. It is therefore the same for the jet energy drop calculation for all three grooming procedures discussed in this work. In~\sec{SDandISD}, we review the (iterated) soft drop algorithm, and we present results for the corresponding one-loop jet function in~\sec{fixedorder_iteratedsoftdrop}. In~\sec{refactorization_ISD}, we discuss the refactorization of this jet function and resummation of the logarithms of the jet energy drop $\Delta_E$ and grooming parameter $z_{\rm cut}$. In particular, we include results for all relevant functions at one-loop order, needed for our numerical results at NLL$'$. Non-global and clustering logarithms are discussed in~\sec{NGL_iteratedsoftdrop}, and \sec{ISDprof} describes our central scale choice, as well as the scale variations used to assess the perturbative uncertainty. Finally, in~\sec{numerics_iteratedsoftdrop} we present numerical studies for LHC kinematics.

\subsection{Jet production}
\label{sec:fac_jet}

We consider generic jet substructure measurements performed on an inclusive jet sample, as the discussion in this section applies to all jet energy drop observables in this paper. To achieve factorization, we assume that the jet is collimated, keeping only terms at leading power in the jet radius $R$. This allows us to factorize the cross section in terms of parton distribution functions (PDFs), hard-scattering functions, and jet functions, which capture the formation and evolution of the observed jet.\footnote{In many cases it has been observed that the neglected ${\cal O}(R^2)$ power corrections are numerically small, even for relatively large values of the jet radius parameter $R$, see e.g.~refs.~\cite{Mukherjee:2012uz,Dasgupta:2014yra,Scott:2019wlk}.} 

The cross section differential in the jet rapidity $\eta$ and transverse momentum $p_T$ and the energy drop $\Delta_E$ is given by
\begin{align} \label{eq:hard_fact}
  \frac{\df \si}{\df \eta\, \df p_T\, \df \Delta_E}
  &= \sum_{ijk} 
  \int\! \frac{\df x_i}{x_i}\, f_i(x_i,\mu) \int\! \frac{\df x_j}{x_j}\, f_j(x_j,\mu)
  \int\! \frac{\df z}{z}\, {\cal H}_{ijk}(x_i, x_j, \eta, p_T/z, \mu)
  \nn \\ & \quad\times
   \cG_k(z, \Delta_E, p_T R,\mu) \big[1+\mathcal{O}(R^2)\big]
\,.\end{align}
Here the PDFs are denoted by $f_{i,j}$ and we integrate over the momentum fractions $x_{i,j}$ of the colliding partons with flavor $i,j$. The hard function ${\cal H}_{ijk}$ captures the hard scattering of the incoming partons $ij\to kX$, where we are inclusive over additional hard partons ($X$). The hard function depends on the incoming momentum fractions $x_{i,j}$, the jet rapidity $\eta$, and the partonic transverse momentum $p_T/z$ of the final state parton $k$. It is independent of the jet algorithm, is the same as for inclusive hadron production, and known analytically at one loop~\cite{Aversa:1988vb,Jager:2002xm,Mukherjee:2012uz}. The produced parton $k$ subsequently fragments inclusively into the observed jets, which carry a momentum fraction $z$ and thus have transverse momentum $p_T = z \times (p_T/z)$. The corresponding dynamics of the formation of inclusive jets is captured by the jet function ${\cal G}_k$, which is convolved with the hard function. 
The jet function ${\cal G}_k$ also accounts for the jet energy drop $\Delta_E$, and thus depends on the grooming parameters of the algorithm under consideration. Since the discussion so far is independent of the specific grooming procedure we have omitted dependence on the grooming parameters here, but will include them when describing specific cases below. As we focus on this jet function and its refactorization in the remainder of this work, we find it convenient to change the parton flavor index $\cG_k$ to $\cG_i$ from this point on.

The factorization in \eq{hard_fact} is a generalization of the factorized cross section for inclusive jet production~\cite{Kaufmann:2015hma,Kang:2016mcy,Dai:2016hzf}. The characteristic scales of the various ingredients are the same as for inclusive jet production
\begin{align}
    \mu_f &\sim \lqcd
    \,, &
    \mu_{\cal H} &\sim p_T
    \,, &
    \mu_{\cal G} &\sim p_T R
\,.\end{align}
The resummation of logarithms of the jet radius $R = \mu_\cG/\mu_{\cal H}$ is achieved by evolving the
jet function $\cG_i$ from the jet scale $\mu_\cG$ to the hard scale $\mu_{\cal H}$, using the time-like DGLAP evolution equation~\cite{Gribov:1972ri,Altarelli:1977zs,Dokshitzer:1977sg} 
\begin{align} \label{eq:G_evo}
\mu\f{\df}{\df\mu}\,\cG_i(z, \Delta_E, p_T R , \mu)=\sum_j\int_z^1\f{\df z'}{z'}\,\f{\as}{\pi} P_{ji}(z/z') \,\cG_j(z', \Delta_E, p_T R , \mu)
\,.\end{align}
The relevant Altarelli-Parisi splitting functions $P_{ji}$ are collected in \eq{split}.

Integrating the jet function ${\cal G}_i$ in \eq{hard_fact} over the jet energy drop variable $\Delta_E$, the semi-inclusive jet function $J_i$ of ref.~\cite{Kang:2016mcy} is obtained
\begin{align}
     \int_0^1 \! \df  \Delta_E\, \cG_i(z, \Delta_E, p_T R, \mu) = J_i(z,p_T R,\mu)
\,.\end{align}
At next-to-leading order (NLO), it is convenient to rewrite the jet function ${\cal G}_i$ as
\begin{align}\label{eq:Delta_iG}
     \cG_i(z, \Delta_E, p_T R, \mu) =  J_i(z, p_T R, \mu) \,\de(\Delta_E) +  \de(1-z) \Delta \cG_i(\Delta_E, p_T R,\alpha_s( \mu))
\,.\end{align}
At this order, the initial parton splits into at most two other partons. The distribution in  $\Delta_E$ is encoded in the second term, which only receives a contribution when both partons are inside the jet, so $z=1$. In the following sections, we only report on $\Delta \cG_i$, which encodes the dependence on the grooming procedure and only depends on the scale $\mu$ through the strong coupling. Using \eq{Delta_iG}, we write
\begin{align} \label{eq:reshuffle}
     \cG_i(z, \Delta_E, p_T R, \mu) &=  \sum_j J_{ij}(z, p_T R, \mu) \Big[\de(\Delta_E) + \Delta \cG_j\bigl(\Delta_E, p_T R, \al_s(\mu)\bigr)\Big] + \ord{\al_s^2}
     \nn \\ & 
     \equiv  \sum_j J_{ij}(z, p_T R, \mu)\, \tilde \cG_j\bigl(\Delta_E, p_T R, \al_s(\mu)\bigr) \,,
\end{align}
conveniently separating the formation process of inclusive jets ($J_{ij}$) from the grooming and $\Delta_E$ measurement ($\tilde \cG_i$)~\cite{Kaufmann:2015hma,Cal:2019hjc}. Note that \eq{reshuffle} is not a separation of physics at different scales. Upon summation over the flavor index $j$, we recover the semi-inclusive jet function
\begin{equation}
J_i(z,p_T R,\mu)=\sum_j J_{ij}(z,p_T R,\mu)\,.
\end{equation}
The coefficients $J_{ij}$ are given by~\cite{Cal:2019hjc}  
\begin{align}
 J_{qq}(z,p_{T} R,\mu) 
  &= \de(1-z) + \f{\as}{2\pi} \bigg\{\ln\Big(\f{\mu^2}{p_T^2 R^2} \Big) P_{qq}(z) 
\\
 & \quad
+C_F\bigg[-2(1+z^2)\Big(\frac{\ln(1-z)}{1-z}\Big)_+ + \Big(\frac{13}{2}- \f{2\pi^2}{3} \Big)\delta(1-z) - 1+z  \bigg]\bigg\}
\,, \nnu
 J_{qg}(z,p_{T} R,\mu) 
 &=\f{\as}{2\pi}\bigg[\Big(\ln\Big(\f{\mu^2}{p_T^2 R^2} \Big) - 2 \ln(1-z) \Big) P_{gq}(z) - C_Fz \bigg]
\,, \nnu
J_{gq}(z,p_{T} R, \mu) 
 & =  \f{\as}{2\pi}\bigg[\Big(\ln\Big(\f{\mu^2}{p_T^2 R^2} \Big) - 2\ln(1-z) \Big)  P_{qg}(z) - T_F 2z(1-z) \bigg]
\,, \nnu
J_{gg}(z, p_{T} R, \mu) 
& = \de(1-z) + \f{\as}{2\pi}\bigg\{\ln\Big(\f{\mu^2}{p_T^2 R^2} \Big)  P_{gg}(z)
\nnu 
& \quad 
  - \frac{4C_A (1-z+z^2)^2}{z}\, \Big(\frac{\ln(1-z)}{1-z}\Big)_+
  + \Big[C_A \Big(\frac{5}{12} - \frac{2\pi^2}{3}\Big) + \frac{23}{12} \bt_0\Big]  \delta(1-z)  \bigg\}
\,.\nn\end{align}

\subsection{The soft drop grooming algorithm and its variants~\label{sec:SDandISD}}

We start by reviewing the original soft drop (SD) algorithm, introduced in ref.~\cite{Larkoski:2014wba}, which iteratively goes through the clustering history of a jet, eliminating soft branches until the so-called soft drop criterion is satisfied. 
First, an inclusive jet sample is identified with the anti-k$_T$ algorithm~\cite{Cacciari:2008gp}, which clusters particles pairwise according to their geometric distance in the $(\eta,\phi)$ plane and the inverse square of the transverse momenta (relative to the beam). The transverse momenta of these jets correspond to the ungroomed jet $p_T$. Second, each of the obtained jets is reclustered with the Cambridge/Aachen (C/A) algorithm~\cite{Dokshitzer:1997in,Wobisch:1998wt}. Different from anti-k$_T$, the C/A algorithm only depends on the pairwise geometric distance of particles. Therefore, particles that are closest in distance are clustered first, yielding an angular-ordered clustering tree. Third, the obtained reclustered C/A jet is declustered recursively to identify soft branches. At each step of the declustering procedure, the transverse momenta $p_{Ti}$, $i=1,2$ of the two branches and their relative distance $\Delta R_{12}=((\eta_1 - \eta_2)^2+ (\phi_1-\phi_2)^2)^{1/2}$ are considered. Whether or not the softer branch is removed from the jet, depends on the soft drop criterion
\begin{equation}\label{eq:criterion}
\frac{\min \left[p_{T 1}, p_{T 2}\right]}{p_{T 1}+p_{T 2}}>z_{\mathrm{cut}}\left(\frac{\Delta R_{12}}{R}\right)^{\beta} \,.
\end{equation}
Here the soft threshold $z_{\rm cut}$ and the angular exponent $\beta$ are free parameters that specify the grooming procedure.\footnote{The case $\beta=0$ corresponds to the modified mass drop tagger (mMDT)~\cite{Dasgupta:2013ihk}.} If the branches fail the criterion, i.e.~the splitting is too soft, the softer branch is removed and the declustering sequence continues following the more energetic branch. Once the soft drop criterion is satisfied, the grooming algorithm terminates and all remaining particles in the two branches constitute the groomed jet. If no branching satisfies the soft drop criterion, the last single particle is considered to be the groomed jet. 

The observable we consider in this work is the relative difference $\Delta_E$ of the jet energy or jet transverse momentum before and after grooming, which will be discussed for the original soft drop algorithm in section~\ref{sec:SD}. In the case of soft drop grooming, the factorization and resummation of $\Delta_E$ involve the groomed jet radius $R_g$, which is the geometric distance between the two branches that satisfy the soft drop criterion, $R_g \equiv \Delta R_{12}$. For convenience, we often use the normalized groomed jet radius $\theta_g \equiv R_g/R$.

\begin{figure}[t]
    \centering
    \includegraphics[width=0.48\textwidth]{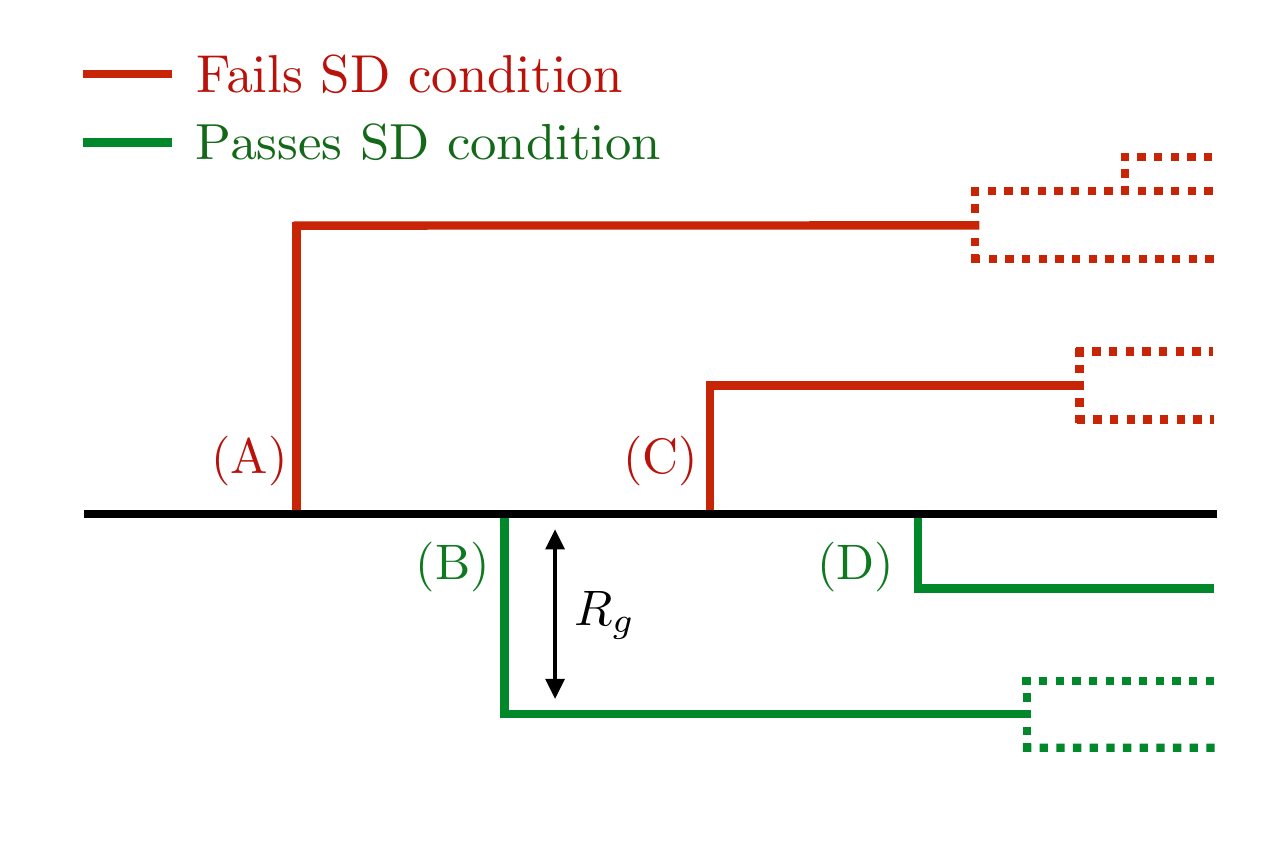}
    \caption{Illustration of the (iterated) soft drop grooming algorithm. Branches along the most energetic branch (black) are tested against the soft drop condition, starting from the left. The regular soft drop algorithm terminates when the first branch, here (B), passes the soft drop condition, which defines the soft drop groomed radius $R_g$. Instead, iterated soft drop continues testing all branches at smaller angular scales, here (C), (D), where (C) is also groomed away in this example. The dotted lines correspond to branchings that are not tested against the soft drop condition in either case.~\label{fig:Tree}}
\end{figure}

In this section we consider $\Delta_E$ for the iterated soft drop (ISD) algorithm~\cite{Frye:2017yrw}. This differs from the original soft drop by continuing with the grooming procedure, following the more energetic branch after the soft drop condition is satisfied\footnote{Alternatively, both branches can be followed, which is known as recursive soft drop~\cite{Dreyer:2018tjj}, and will not be considered in this paper.}. This continues until only one particle is left and, thus, the entire jet is declustered. The groomed jet is then given by all particles that are contained in branches that satisfy the soft drop condition along the way. In the remainder of this section, we present a calculation of the cross section differential in $\Delta_E$ for this grooming algorithm. See \fig{Tree} for an illustration of regular and iterated soft drop.

\subsection{Fixed-order results~\label{sec:fixedorder_iteratedsoftdrop}}

When the jet energy drop $\Delta_E$ and the grooming parameter $z_{\rm cut}$ are not parametrically small, i.e.~$\Delta_E, z_{\rm cut}$ are both order one, a fixed-order calculation of the relevant jet function $\Delta\cG_i^{\rm ISD}$ is sufficient, which we present here. In \sec{refactorization_ISD}, we will consider the case where they are parametrically small and lead to large logarithms in the jet function, requiring resummation.

To calculate the jet function $\Delta\cG_i^{\rm ISD}$, we can use the squared matrix element and the phase space in the collinear limit~\cite{Giele:1991vf}
\begin{align} \label{eq:coll}
\int\! \df \Phi_2\, \si_{2,q}^c &= 
\frac{\al_s}{\pi} \f{e^{\epsilon\gamma_E}}{\Gamma(1-\epsilon)}\left(\frac{\mu}{E}\right)^{2\epsilon}\int_0^1 \frac{\df x}{(x(1-x))^{2\epsilon}}\,
C_F\bigg[\frac{1+x^2}{1-x} - \epsilon\, (1-x) \bigg]
\int\f{\df \theta}{\theta^{1+2\epsilon}}
\,, \\
\int\! \df \Phi_2\, \si_{2,g}^c &= 
\frac{\al_s}{\pi} \f{e^{\epsilon\gamma_E}}{\Gamma(1-\epsilon)}\left(\frac{\mu}{E}\right)^{2\epsilon} \int_0^1 \frac{\df x}{(x(1-x))^{2\epsilon}} \,
\bigg\{C_A \Big[\frac{x}{1-x}+\frac{1-x}{x}+x(1-x)\Big] 
\nn \\ & \qquad
+ n_f T_F \Big[x^2+(1-x)^2-2\epsilon\, x (1-x)\Big]\bigg\}
\int\f{\df \theta}{\theta^{1+2\epsilon}}
\,,\end{align}
where $\theta$ is the angle between the two partons and $E$ is the energy of the parton initiating the jet.\footnote{For a jet at central rapidity $E=p_T$ and the distance in $(\eta,\phi)$ corresponds (approximately) to an angle. Boost invariance implies that our calculation is valid for general rapidity.}
The one-loop jet function for soft drop and iterated soft drop identical. However, differences appear at higher orders, leading to rather different factorization structures. The measurement function for the jet function for (iterated) soft drop is given by 
\begin{flalign}
&\Delta \cG^{\rm ISD}_i(\ed,p_T R,z_{\rm cut},\beta,\alpha_s(\mu) )
\nn \\
&\quad =
\int \df \Phi_2\, \si_{2,i}^c \Theta \left(\theta < R \right) \Big[ \Theta \bigl(x> \zc (\theta/R)^\beta \bigr)\Theta \bigl(1-x >\zc (\theta/R)^\beta \bigr)  \delta (\ed) \nn \\
&\qquad+ \Theta \bigl(x> \zc ( \theta/R )^\beta \bigr)\Theta \bigl(1-x <\zc (\theta/R )^\beta \bigr) \delta ( \ed -(1-x))\nn \\
&\qquad+ \Theta \bigl(x< \zc ( \theta/R )^\beta \bigr)\Theta \bigl(1-x >\zc (\theta/R )^\beta \bigr) \delta (\ed-x) - \delta(\ed) \Big].
\end{flalign}
The last term subtracts the semi-inclusive jet function, as required for $\Delta \cG^{\rm ISD}_i$, see \eq{Delta_iG}. Performing the integrals and expanding in distributions, we find for quark and gluon jets the following results 
\begin{align}\label{eq:ISDfixedorder}
 &\Delta\cG^{\rm ISD}_q( \Delta_E,p_TR,z_{\rm cut},\beta, \alpha_s(\mu)) 
 \nn \\
 &\quad = \frac{\alpha_s C_F}{\pi}\frac{1}{\beta}\bigg\{\Theta(\ed<z_{\rm cut})  \bigg[  -2 \left[\frac{\ln \ed}{\ed}\right]_+  + 2\ln z_{\rm cut}\frac{1}{[\ed]}_+ + \biggl(3-\frac{2}{1-\Delta_E}\biggr) \ln\Bigl(\frac{\ed}{z_{\rm cut}}\Bigr) \bigg]
\nn \\ & \qquad
 +\delta(\ed) \left[-\ln^2 z_{\rm cut}+3 z_{\rm cut} -2 \text{Li}_2(z_{\rm cut}) \right]\bigg\} \,,
\\
 &\Delta \cG^{\rm ISD}_g(\ed,p_TR,z_{\rm cut},\beta,\alpha_s(\mu)) 
 \nn \\
 &\quad = \frac{\alpha_s}{\pi}\f{1}{\beta}\bigg\{\Theta(\ed < \zc) \bigg( -2 C_A \left[\f{\ln \ed}{\ed} \right]_+ +2C_A \ln \zc\f{1}{[\ed]}_+ 
\nn\\&\qquad
 - \bigg[2C_A\left(\f{1}{1-\ed}-2 +\ed-\ed^2\right)+2 n_f T_F \left(\ed^2 + (1-\ed)^2\right) \bigg]\ln \Bigl(\f{\ed}{\zc} \Bigr)  \bigg)
\nn \\   &\qquad
+ \delta(\ed) \bigg[C_A \bigg(-\ln^2 \zc + 4\zc -\f{\zc^2}{2}+\f29 \zc^3 -2\Li_2 (\zc)  \bigg) 
\nn \\ & \qquad
+  n_f T_F \left(-\zc +\zc^2-\f49\zc^3 \right)  \bigg]\bigg\} \,.    
\end{align}
From \eq{ISDfixedorder} we read off that the jet energy drop is bounded by $\Delta_E<z_{\rm cut}$ at NLO. Note that by construction (see \eq{reshuffle}) we have
\begin{equation}
\int_0^1 \df\Delta_E\, \Delta\cG^{\rm ISD}_i(\ed,p_T R,z_{\rm cut},\beta,\alpha_s(\mu) )=0 \,.
\end{equation}

As a consistency check, we investigate several limits of the grooming parameters. First, we consider the limit $\beta\to\infty$. As can be seen from \eq{ISDfixedorder}, the entire jet function $\Delta \cG^{\rm ISD}_i$ is proportional to $1/\beta$ and vanishes in this limit. Indeed, for $\beta\to\infty$ the soft drop condition in \eq{criterion} is always trivially satisfied and no branches are removed from the jet. 
Second, since $\Delta\cG^{\rm ISD}_i$ is proportional to $1/\beta$ we cannot take the limit $\beta\to 0$ at fixed order. Indeed, for iterated soft drop, the jet energy drop with $\beta=0$ is not IRC safe. For regular soft drop the case $\beta=0$ is still Sudakov safe, as discussed in section~\ref{sec:SDbeta0}. Third, we consider the limit $z_{\rm cut}\to 0$, which (similar to $\beta\to \infty$) corresponds to the limit of no grooming. To see more clearly that $\Delta \cG^{\rm ISD}_i$ also vanishes in this limit, we rewrite it as follows: The plus distributions in \eq{ISDfixedorder} are defined such that they vanish when integrated over the interval $0<\Delta_E<1$. We can rewrite these distributions such that they instead vanish when integrating over the interval $0<\Delta_E< z_{\rm cut}$ of the theta function that multiplies the distributions, which we indicate by the subscript $\Theta+$. For example, for the quark case this yields
\begin{align}\label{eq:ISDfixedorder-Theta}
 \Delta\cG^{\rm ISD}_q &= \frac{\alpha_s C_F}{\pi}\frac{1}{\beta}\bigg\{\Theta(\ed<z_{\rm cut})  \bigg[  -2 \left[\frac{\ln \ed}{\ed}\right]_{\Theta +}  + 2\ln z_{\rm cut}\frac{1}{[\ed]}_{\Theta+}
\nn \\ & \quad
+ \left(3-\frac{2}{1-\Delta_E}\right) \ln\left(\frac{\ed}{z_{\rm cut}}\right) \bigg] +\delta(\ed) \left[3 z_{\rm cut} -2 \text{Li}_2(z_{\rm cut}) \right]\bigg\} \,,  
\end{align}
and similarly for the gluon, making it clear that $\Delta\cG_i^{\rm ISD}$ vanishes in the limit $z_{\rm cut}\to 0$.

\begin{figure}[t]
    \centering
    \includegraphics[width=0.48\textwidth]{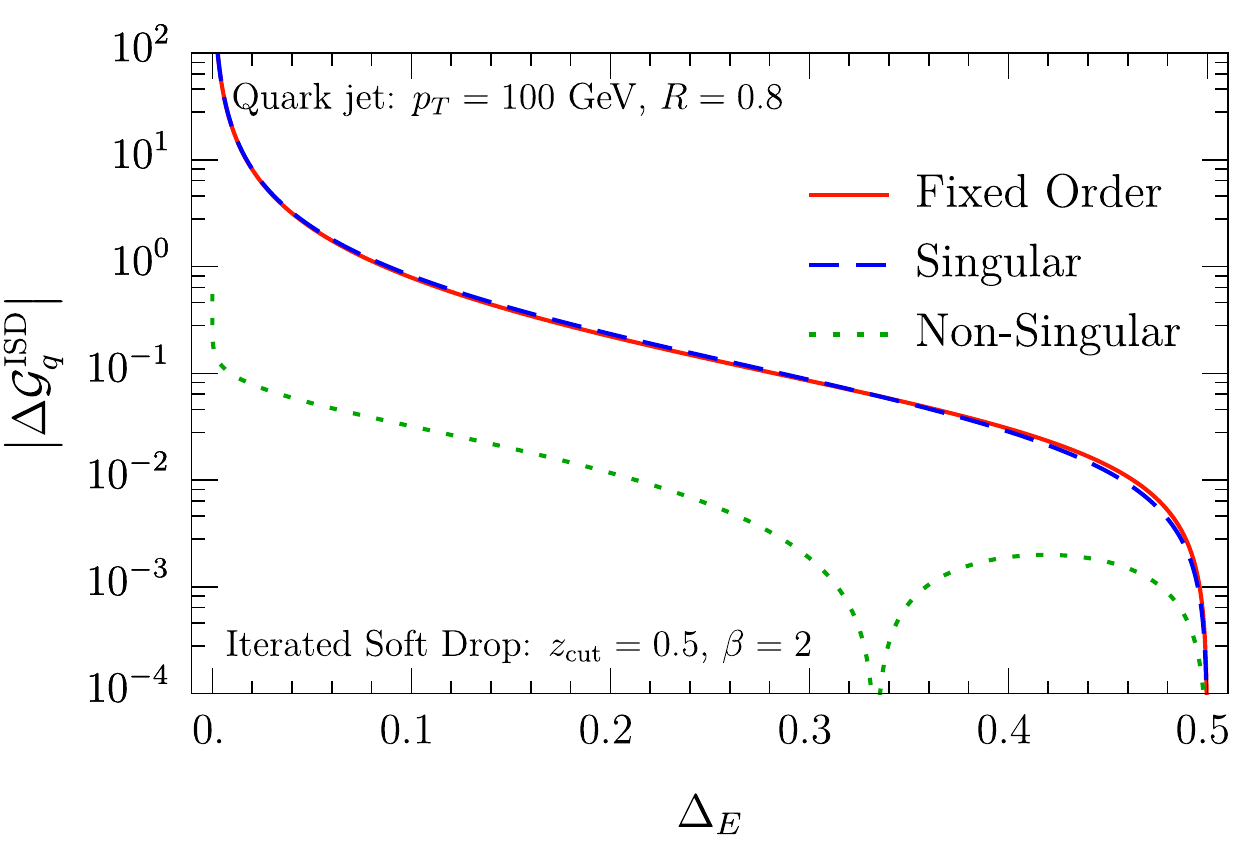} \hfill 
       \includegraphics[width=0.48\textwidth]{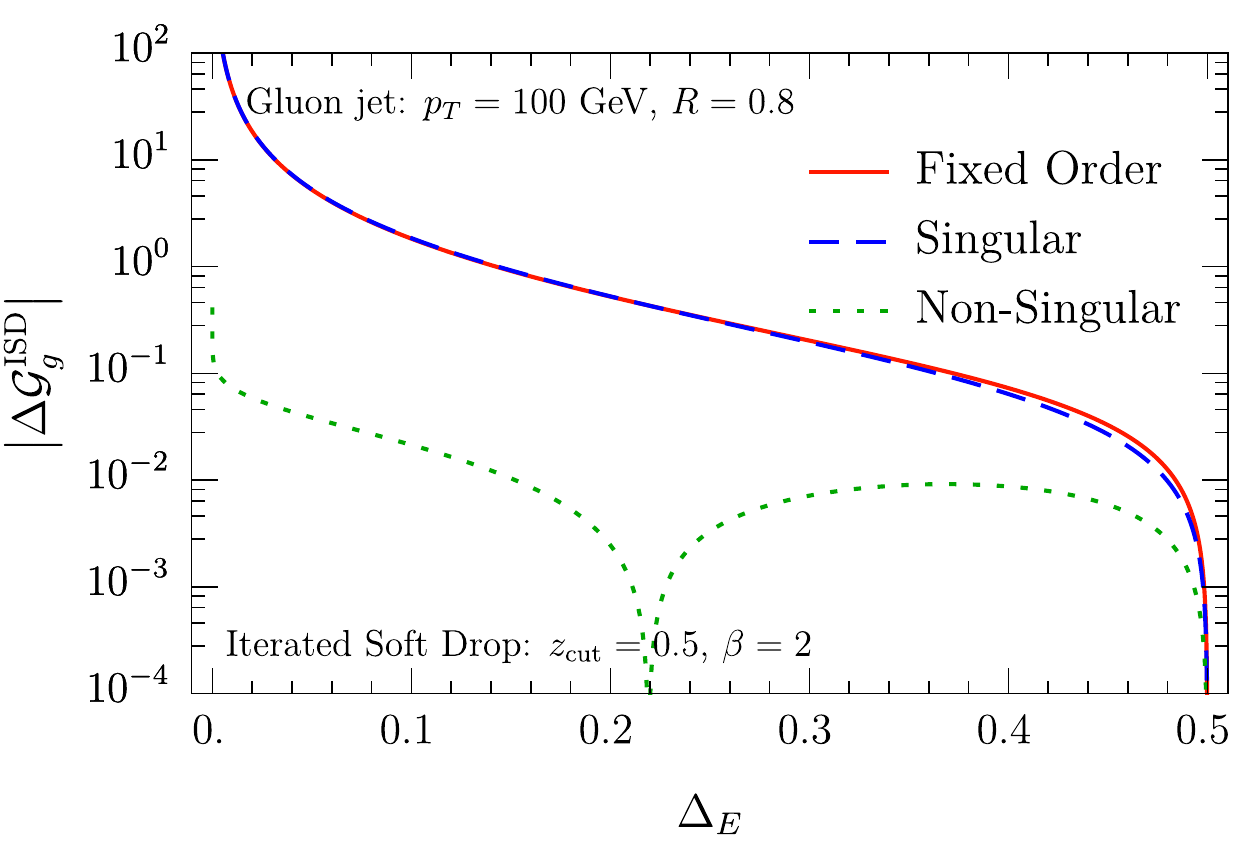} \hfill \phantom{.} \\
    \caption{Comparison of the full one-loop result for jet function for iterated soft drop $\Delta\cG^{\rm ISD}_i$ (solid red), its singular (dashed blue) and non-singular terms (dotted green), for jets initiated by quarks (left) and gluons (right).~\label{fig:refactorization_ISD}}
\end{figure}

We end this section by comparing the singular terms, obtained by expanding $\Delta\cG^{\rm ISD}_i$ in the limit $\Delta_E\ll\zc\ll 1$, to the full NLO expression of $|\Delta\cG^{\rm ISD}_i|$, shown in \fig{refactorization_ISD}. We chose representative values of the soft-drop parameters and jet kinematics, indicated in the figure. We observe at NLO the non-singular power corrections, which equals the difference between the singular terms and the fixed-order NLO, are very small compared to singular terms at NLO over the entire range of $\Delta_E$, suggesting the importance of all-order resummation, which is discussed in the next section. 
Because the non-singular is so small, we do not include it as a matching correction, since its impact on our results is negligible.

\subsection{Factorization and resummation~\label{sec:refactorization_ISD}}

\begin{figure}[t]
    \centering
    \includegraphics[width=0.48\textwidth]{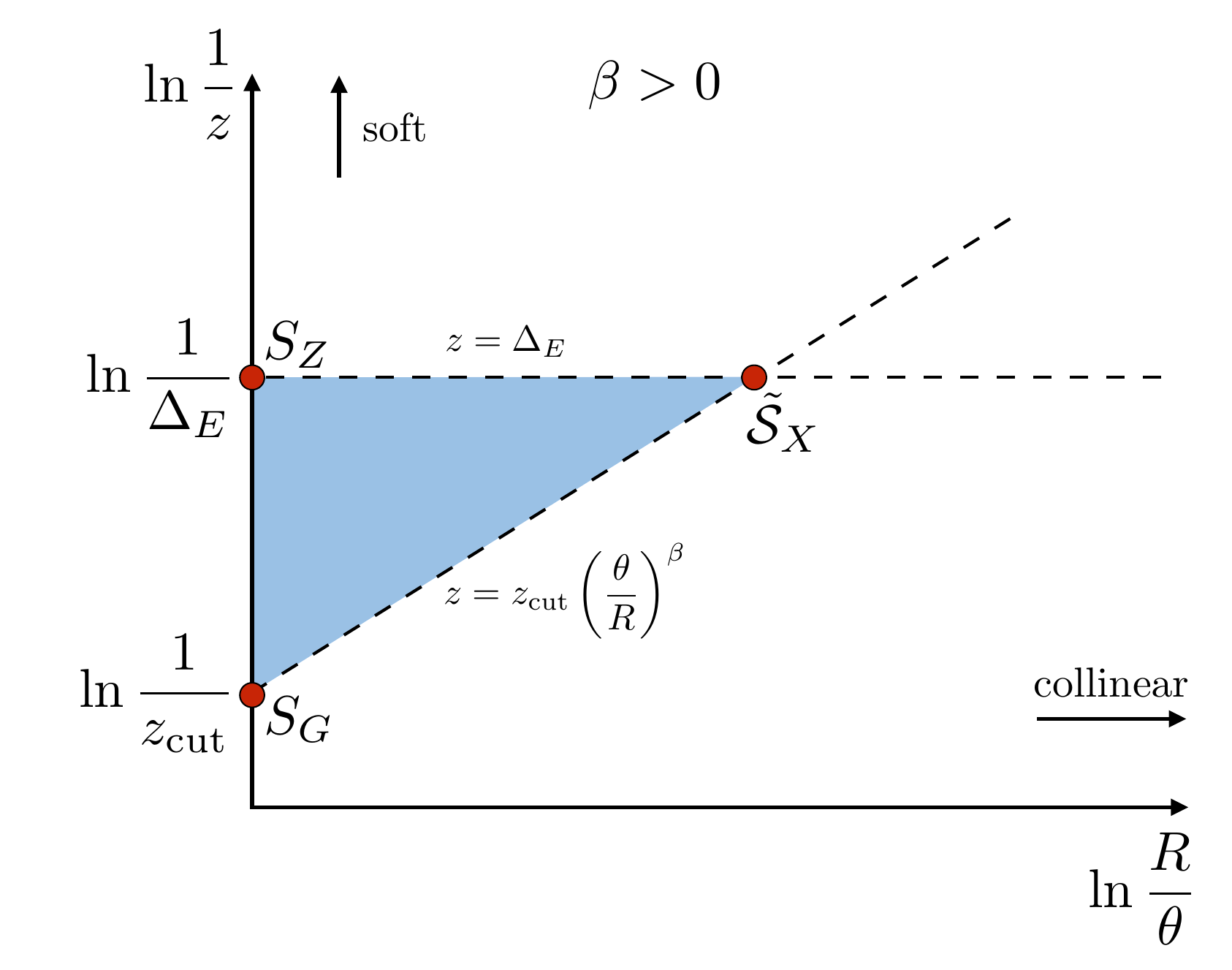} \\
    \caption{Lund diagram for the jet energy drop $\Delta_E$ for iterative soft drop grooming, with $\Delta_E\ll \zc\ll 1$.  At LL accuracy, emissions in the shaded triangle are vetoed. The relevant modes in SCET correspond to the red dots at the corners of the triangle, with their power counting summarized in table~\ref{tab:modes_ISD}.~\label{fig:LundISD}}
\end{figure}

In this section we discuss the refactorization of the jet function for iterated soft drop, which will enable the resummation of the logarithms of the jet energy drop $\Delta_E$ and grooming parameter $z_{\rm cut}$. Typically $z_{\rm cut}=0.1$ (we will consider larger values as well), so we assume the parametric scaling $\Delta_E\ll z_{\rm cut}\ll 1$. We start with a leading logarithmic (LL) analysis of the jet energy drop, by analyzing the Lund diagram~\cite{Andersson:1988gp} shown in \fig{LundISD}. 
By using the logarithm of the angle $\theta$ and momentum fraction $z$ on the horizontal and vertical axis, emissions have a uniform probability distribution in this plane at LL accuracy.
The grooming condition and the measurement are indicated by the two dashed lines. For the cross section with jet energy drop below some value $\Delta_E$, emissions inside the shaded triangular area in the Lund plane are not allowed. Such emissions are not groomed away and therefore lead to a value of the jet energy drop that is larger than $\Delta_E$. Here it is important to note that for iterated soft drop all branches along the leading branch are tested against the soft drop condition, whereas the original soft drop terminates once the criterion in \eq{criterion} is met. From the area of the vetoed region we can calculate the LL expression for the cross section cumulative in $\ed$, from which we obtain the differential result by taking the derivative:
\begin{align}
 \tilde \cG_{i}^{\rm ISD}\bigl(\ed , p_T R, \zc,\bt, \al_s(\mu)\bigr) &\stackrel{{\rm LL}}{=}\,\frac{\df}{\df \ed} \,\exp\biggl[- \frac{\al_s C_i}{\pi} \f{1}{\bt} \ln^2 \biggl(\f{\zc}{\ed} \biggr) \biggr]\,.
\end{align}
The color factors are $C_q=C_F$ ($C_g = C_A$) for jets initiated by a quark (gluon). 

We will now extend the resummation to NLL$'$ accuracy using SCET. The power counting of the relevant modes in the effective theory can be read off from the Lund diagram, and correspond to the red dots at the corners of the triangle. There are two soft modes $S_Z$ and $S_G$ located on the vertical axis, which are sensitive to the jet boundary as well as the energy drop and grooming condition, respectively. In addition, there is a collinear-soft~\cite{Bauer:2011uc,Procura:2014cba} mode $\CS_X$, located at the intersection (hence the subscript $X$) of the two dashed lines representing the measurement and the grooming condition, which it is therefore sensitive to. We note that for iterated soft drop there is no hard mode, which would correspond to the origin of the Lund diagram, in contrast to most jet substructure observables. Physically, this arises because energy drop with $\zc \ll 1$ only probes (collinear-)soft radiation. The relevant scaling of the three modes is summarized in table~\ref{tab:modes_ISD}, in terms of the light-cone components of their momenta,
\begin{align}
  p^\mu = \bn \sdt p\, \frac{n^\mu}{2} + n \sdt p\, \frac{\bn^\mu}{2} + p_\perp^\mu
\,,\end{align}
where $n^\mu = (1,0,0,1)$ is along the jet axis, $\bn^\mu = (1,0,0,-1)$, and $p_\perp^\mu$ denotes the transverse components. This leads to the following factorization formula for $\tilde\cG_i^{\rm ISD}$,
\begin{align}\label{eq:factorization_ISD}
 \tilde \cG_{i}^{\rm ISD}\bigl(\ed , p_T R, \zc,\bt, \al_s(\mu)\bigr) 
 & \stackrel{{\rm NLL}'}{=} S_{i, G}(\zc p_T R,\bt,\mu)  \, \int\! \df \ed'\, S_{i,Z}(\ed' , p_T R,\mu) 
 \\  & \qquad \times
\CS_{i,X}(\ed-\ed',\zc^{-1/\beta} p_T R, \beta, \mu) \, S_i^{\rm NG}\bigg(\f{\ed}{\zc} \bigg)
\,. \nn
\end{align}
To achieve NLL$'$ accuracy, we also include the contribution from non-global logarithms, which are accounted for by the non-global function $S_i^{\rm NG}$, discussed in \sec{NGL_iteratedsoftdrop}. Strictly speaking, the NGLs should also be included through a convolution in $\ed$, but the difference with the multiplicative treatment above is beyond the accuracy we are working at, see sec.~2.4 of ref.~\cite{Cal:2019hjc}.

\begin{table}
   \centering
   \begin{tabular}{l|l|l}
     \hline \hline
     Mode: & Function: & Scaling $(n \cdot p, \bar n\cdot p,p_\perp)$ \\ \hline
     soft  & $S_G$ & $z_{\rm cut}\,  p_T\, (R^2, 1, R) $ \\ 
     soft & $S_Z$ &  $\Delta_E\, p_T\,(R^2,1,R)$ \\     
     collinear-soft & $\CS_X$ & $ \ed\,  p_T \Bigl( \bigl(\f{\ed}{\zc}\bigr)^{2/\bt} R^2 ,1,\bigl(\f{\ed}{\zc}\bigr)^{1/\bt} R \Bigr)$ \\
     \hline \hline
   \end{tabular}
   \caption{The modes in SCET that enter in the refactorization of the jet function $\tilde \cG_i^{\rm ISD}$ for the jet energy drop with iterated soft drop, with $\Delta_E\ll z_{\rm cut}\ll 1$.~\label{tab:modes_ISD}}
\end{table}

The one-loop expressions for other three  functions in \eq{factorization_ISD} are given by
\begin{align}
S_{i,G}(\zc p_T R,\bt ,\mu)&=  1 + \f{\alpha_s C_i }{ \pi (1+\beta)}\left\{- \ln^2 \left( \f{\mu}{\zc p_T R} \right) +\f{\pi^2}{24} \right\}\,,~\label{eq:functions_ISD1}
\\[.5em]
S_{i,Z}(\ed, p_T R,\mu)& = \delta(\ed)+\f{\alpha_s C_i}{\pi}\bigg\{ 2 \left[\f{\ln \ed}{\ed} \right]_+  -\f{2}{[\ed]}_+ \ln \left(\f{\mu}{p_T R} \right)~\label{eq:functions_ISD2}
\nn \\& \quad
+ \delta (\ed) \left[\ln^2 \left( \f{\mu}{p_T R}\right)-\f{\pi^2}{24} \right] \bigg\} 
\,, \\[.5em]
\CS_{i,X}(\ed,\zc^{-1/\beta} p_T R, \beta, \mu)&= \delta (\ed) + \f{\as C_i}{\pi} \bigg\{-2 \frac{1+\beta}{\beta}\left[\f{\ln \ed}{\ed} \right]_+ +\f{2}{[\ed]}_+ \ln \biggl(\f{\mu}{z_{\rm cut}^{-1/\beta}p_T R} \biggr) 
\nn \\& \quad
 +\delta(\ed) \f{\beta}{1+\beta}\bigg[
- \ln^2 \biggl(\f{\mu}{\zc^{-1/\beta} p_T R} \biggr)  + \f{\pi^2}{24} \bigg] \bigg\} \,.~\label{eq:functions_ISD3}
\end{align}
We limited ourselves to reporting only the finite terms of the different functions, as the $1/\epsilon$ poles can be reconstructed from the $\ln\mu$ terms. We have verified that all $\ln\mu$ terms cancel in \eq{factorization_ISD}, and that the remainder agrees with the NLO result in \eq{ISDfixedorder} in the limit $\ed \ll \zc \ll 1$, providing a check on the refactorization. 

The natural scale of each mode is given by its virtuality. Reading off from table~\ref{tab:modes_ISD}, 
\begin{equation}\label{eq:scales_ISD}
\mu_{S_G}\sim z_{\rm cut}p_T R\,,\quad \mu_{S_Z}\sim \Delta_E\, p_T R\,,\quad \mu_{\CS_X}\sim \Delta_E^{(1+\beta)/\beta}z_{\rm cut}^{-1/\beta}p_T R \,.
\end{equation}
By evaluating each function in \eq{factorization_ISD} at its natural scale, and evolving them to a common scale $\mu$ through renormalization group (RG) equations, we achieve the joint resummation of logarithms of ${\Delta_E}$ and $z_{\rm cut}$. The RG equations are given by
\begin{align} \label{eq:RGE}
\mu\f{\df}{\df\mu}\,S_{i,G}(\zc p_T R,\bt ,\mu) =&\,\ga_i^{S_G}(\zc p_T R,\beta,\mu) \,S_{i,G}(\zc p_T R,\bt ,\mu)
\,,\\[.5em]
\mu\f{\df}{\df\mu}\,S_{i,Z}(\ed, p_T R,\mu) =&\,\int \df \ed'\,  \gamma^{S_Z}_i(\ed-\ed',p_TR,\mu)\,S_{i,Z}(\ed', p_T R,\mu)
\,, \\
\mu\f{\df}{\df\mu}\,{\CS }_{i,X}(\Delta_E,\zc^{-1/\beta} p_T R,\bt ,\mu) =&\,\int\! \df \ed' \, \ga^{\CS_X}_i(\Delta_E-\Delta_E',z_{\rm cut}^{-1/\beta} p_TR,\beta,\mu) \,
\nn\\ & \times
{\CS}_{i,X}(\Delta_E',\zc^{-1/\beta} p_T R,\bt ,\mu)
\,,
\end{align}
where the corresponding anomalous dimensions can be found in the appendix~\ref{app:anom}. 

Next we discuss in some detail how we solve the different evolution equations, as similar techniques will be employed for the other grooming techniques discussed in subsequent section. Evolving the function $S_{i,G}$ from initial scale $\mu_0$ to the scale $\mu$, using the multiplicative RG equation  in \eq{functions_ISD1}, 
\begin{align}
S_{i,G}(z_{\rm cut}p_T R,\beta,\mu)&=U_{i,S_G}(z_{\rm cut} p_T R,\beta,\mu,\mu_0)\,S_{i,G}(z_{\rm cut}p_T R,\beta,\mu_0) \,,
\nn \\
U_{i,S_G}(z_{\rm cut} p_T R,\beta,\mu,\mu_0) &= e^{-\frac{2}{1+\beta} K_{i}\left(\mu, \mu_{0}\right)}\biggl(\frac{\mu_{0}}{z_{\mathrm{cut}} p_{T} R}\biggr)^{-\frac{2}{1+\beta} \eta_i\left(\mu, \mu_{0}\right)} \,.
\end{align}
The two functions $K$ and $\eta$ are given by 
\begin{align}\label{eq:Keta1}
K_{i}(\mu,\mu_0)=&\,\int_{\alpha_{s}\left(\mu_{0}\right)}^{\alpha_{s}(\mu)} \frac{\df \alpha}{\beta(\alpha)} \Gamma_i(\alpha) \int_{\alpha_{s}\left(\mu_{0}\right)}^{\alpha} \frac{\df \alpha^{\prime}}{\beta(\alpha^{\prime})} 
\,,\\
\eta_i(\mu,\mu_0)= &\, \int_{\alpha_{s}\left(\mu_{0}\right)}^{\alpha_{s}(\mu)} \frac{\df \alpha}{\beta(\alpha)} \Gamma_i(\alpha)  \,,\label{eq:Keta2}
\end{align}
following the convention of refs.~\cite{Fleming:2007xt,Stewart:2010qs}. Here, $\beta(\alpha)$ is the QCD beta function and $\Gamma_i$ is the cusp anomalous dimension, which allow for a perturbative expansion
\begin{equation}
\beta(\alpha_{s})=-2 \alpha_{s} \sum_{n=0}^{\infty} \beta_{n}\left(\frac{\alpha_{s}}{4 \pi}\right)^{n+1}\,, \quad \Gamma_{\mathrm{cusp}}^{i}\left(\alpha_{s}\right)=\sum_{n=0}^{\infty} \Gamma_{n}^{i}\left(\frac{\alpha_{s}}{4 \pi}\right)^{n+1}\,.
\end{equation}
The relevant coefficients $\bt_i$ and $\Gamma_i$ are given in \eqs{bt}{ga}.
We evaluate the integrals in \eq{Keta1} and \eq{Keta2} up to NLL accuracy
\begin{align}
K(\mu_{0}, \mu)&=\, -\frac{\Gamma_{0}}{4 \beta_{0}^{2}}\bigg[\frac{4 \pi}{\alpha_{s}(\mu_{0})}\left(1-\frac{1}{r}-\ln r\right)+\left(\frac{\Gamma_{1}}{\Gamma_{0}}-\frac{\beta_{1}}{\beta_{0}}\right)(1-r+\ln r)+\frac{\beta_{1}}{2 \beta_{0}} \ln ^{2} r\bigg]
\,,\\
\eta(\mu_{0}, \mu)&=\, -\frac{\Gamma_{0}}{2 \beta_{0}}\bigg[\ln r+\frac{\alpha_{s}\left(\mu_{0}\right)}{4 \pi}\left(\frac{\Gamma_{1}}{\Gamma_{0}}-\frac{\beta_{1}}{\beta_{0}}\right)(r-1)\bigg]\,,
\end{align}
where $r=\alpha_s(\mu)/\alpha_s(\mu_0)$. Similarly, for the evolution equations of the functions $S_{i,Z}$ and $S_{i,X}$, we find
\begin{align}\label{eq:Sevol2}
S_{i,Z}(\ed, p_T R,\mu) =&\,\int \df \ed' \,  U_{i,Z}(\ed-\ed',p_TR,\mu,\mu_0)\,S_{i,Z}(\ed', p_T R,\mu_0)
\,, \\
{\CS}_{i,X}(\Delta_E,\zc^{-1/\beta} p_T R,\bt ,\mu) =&\,\int  \df \ed' \,U_{i,\CS_X}(\Delta_E-\Delta_E',z_{\rm cut}^{-1/\beta} p_TR,\beta,\mu,\mu_0) \,
\nn\\ & \times
{\CS}_{i,X}(\Delta_E',\zc^{-1/\beta} p_T R,\bt ,\mu_0)
\,,
\end{align}
where the corresponding evolution factors can be written as 
\begin{align}
U_{i,S_Z}(\ed,p_TR,\mu,\mu_0) = 
&\,
\frac{e^{2 K_{i}\left(\mu, \mu_{0}\right)}}{\Gamma\left[-2 \eta_i(\mu, \mu_{0})\right]}\biggl(\frac{\mu_{0} e^{\gamma_{E}}}{p_{T} R}\biggr)^{2 \eta_i\left(\mu, \mu_{0}\right)}\biggl[\frac{\Theta(\Delta_E)}{\Delta_E^{1+2 \eta(\mu,\mu_0)}}\biggr]_{+}
\,, \\
U_{i,\CS_X}(\Delta_E,z_{\rm cut}^{-1/\beta} p_TR,\beta,\mu,\mu_0) = 
&\,
\frac{e^{-\frac{2 \beta}{1+\beta} K_{i} \left(\mu, \mu_{0}\right)}}{\Gamma\left[2 \eta_i(\mu, \mu_{0})\right]}\biggl[e^{\gamma_{E}} \biggl(\frac{\mu_{0}}{z_{\rm cut}^{-1/\beta}p_{T} R}\biggr)^{\frac{\beta}{1+\beta}}\biggr]^{-2 \eta_i\left(\mu, \mu_{0}\right)}
\nn\\&\times\,
\biggl[\frac{\Theta(\Delta_E)}{\Delta_E^{1-2 \eta(\mu,\mu_0)}}\biggr]_{+}
\,.
\end{align}
\begin{table}[t]
  \centering
  \begin{tabular}{l l | c c c c c c}
  \hline \hline
  & & Fixed-order & $\beta$ & $\gamma$ & NGLs \\ \hline
  $\ln R$ & LL & tree & $1$-loop & $1$-loop  & - \\
  & NLL & $1$-loop & $2$-loop & $2$-loop  & - \\
  & NNLL & $2$-loop & $3$-loop & $3$-loop  & - \\ \hline
  $\ln \Delta_E,\ln z_{\rm cut}$ & LL & tree & $1$-loop & $1$-loop  & - \\
  & NLL & tree & $2$-loop & $2$-loop  & LL \\ 
  & NLL$'$ & $1$-loop & $2$-loop & $2$-loop  & LL \\   
  & NNLL & $1$-loop & $3$-loop & $3$-loop & NLL \\
  \hline\hline
  \end{tabular}
  \caption{The required perturbative ingredients needed at different orders (rows) for the resummation of logarithms of the jet radius $R$, jet energy drop $\ed$ and grooming parameter $z_{\rm cut}$ for iterated soft drop. The columns indicate the order of fixed-order ingredients in the factorization, the QCD beta function $\beta$, the anomalous dimension $\gamma$ and  NGLs.~\label{tab:orders}}
\end{table}
The convolutions of the above evolution factors and the soft functions at the initial scale $\mu_0$ in \eq{Sevol2} can be carried out following e.g.~refs.~\cite{Fleming:2007xt,Ligeti:2008ac}. 
For completeness, we summarize the required perturbative ingredients in table~\ref{tab:orders}. An analogous counting of the perturbative accuracy applies to the other grooming techniques discussed in subsequent sections.

\subsection{Non-global logarithms~\label{sec:NGL_iteratedsoftdrop}}

\begin{figure}[t]
\centering
\includegraphics[width=0.55\textwidth]{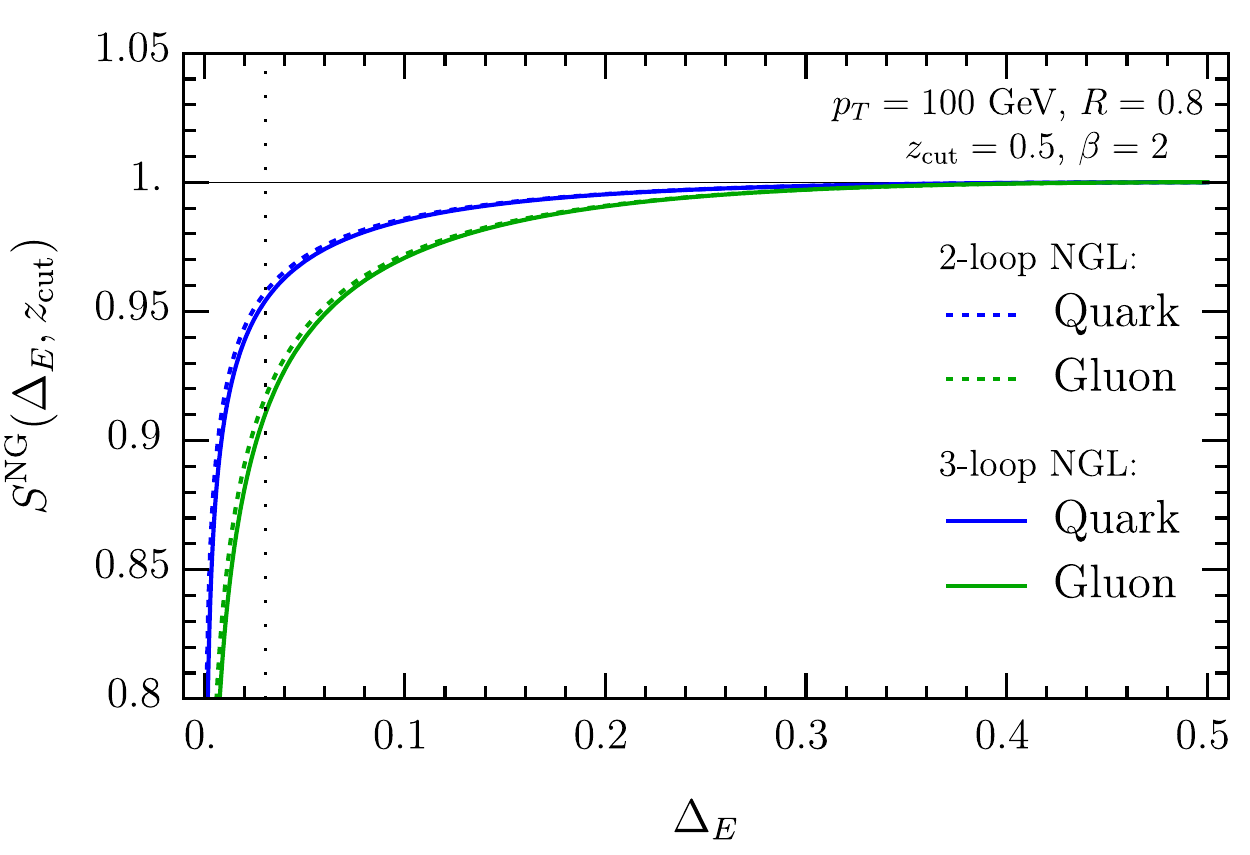} 
\caption{Non-global soft contribution $S^{\rm NG}_i$ for the jet energy drop for iterated soft drop, for $i=q$  (blue) and $i=g$ (green), including two-loop (dotted) and three-loop (solid) contributions. The dotted vertical line shows the onset of the nonperturbative region.~\label{fig:NGL_ISD}}
\end{figure}

Non-global logarithms (NGLs) start contributing to the jet energy drop at next-to-next-to-leading order (NNLO). For iterated soft drop, we will show that the NGLs are related to the hemisphere case, for which a fit to the leading logarithmic resummation~\cite{Dasgupta:2001sh} or a perturbative expansion is available~\cite{Schwartz:2014wha}, in the large-$N_c$ limit. We will show that for our phenomenological results, the effect of NGLs beyond their leading NNLO contribution is negligible, and therefore limit ourselves to this contribution for the other grooming procedures.

The NGLs for iterated soft drop originate from correlations between the two soft modes $S_G$ and $S_Z$, see \fig{LundISD}. As jets were identified using the anti-k$_T$ jet algorithm, which provides a hard boundary for soft radiation, we do not have to take into account clustering effects. Starting with the NGL at NNLO, we exploit the small $R$ limit to map the in- and out-of-jet region to two hemispheres~\cite{Banfi:2010pa}. In the strong energy-ordered limit of two soft gluon emissions~\cite{Dasgupta:2001sh} 
\begin{align}
S_i^{\rm NG}(\Delta_E,\zc)=&\,1+ 8C_i C_A \Bigl( \f{\as}{2\pi} \Bigr)^2\! \int \frac{\df x_{1}}{x_{1}} \frac{\df x_{2}}{x_{2}} \int \mathrm{d} c_{1} \frac{\mathrm{d} \phi_{1}}{2 \pi} \int \mathrm{d} c_{2} \frac{\mathrm{d} \phi_{2}}{2 \pi}  \Theta(x_1> x_2) \nn \\
&\times \frac{\cos \phi_{2}}{\left(1-c_{1} c_{2}-s_{1} s_{2} \cos \phi_{2}\right) s_{1} s_{2}}\, 
[ \Theta ( c_1 < 0)+\Theta ( c_1 >0) \Theta (x_1 > \zc \theta_1^\bt )] 
\nn \\ &
\times [ \Theta ( c_2 < 0)+\Theta (c_2>0) \Theta (x_2 < \ed ) ]
\,,
\end{align}
where $\theta_i$ are the polar angles of the emissions with $c_i=\cos\theta_i$ and $s_i=\sin\theta_i$, and $x_i=k_{Ti}/p_T$ their energy fractions. The constraints on the soft radiation that are specific to the measurement can be read off from the Lund plane in \fig{LundISD} and are encoded in the theta functions on the second and third lines. Specifically, the energy fraction of the most energetic emission has to pass the soft drop criterion if it is inside the jet, whereas the second one has to be less than $\Delta_E$ if it is inside. We can replace $x_1 > \zc \theta_1^\bt$ by $x_1 > \zc$, up to subleading NGLs, because $\theta_1 \sim 1$ in the frame where the in and out-jet region are different hemispheres.
Outside the jet both emissions are unconstrained. Out of the four resulting contributions, the term $\sim \Theta ( c_1<0) \Theta ( c_2<0)$ is scaleless, and the other three terms add up to give\footnote{Integrals over $x_i$ that include 0 in the integration domain are divergent. To calculate these, we note that the integral over $0 \leq x_i \leq 1$ does not yield a large logarithm, allowing us to rewrite the original integral as minus the integral over the complement, which is convergent. The infinities cancel between real and virtual contributions. A similar approach can be used for the angular integral for emissions in the same hemisphere.}
\begin{align} \label{eq:SNG}
S_i^{\rm NG}(\Delta_E,\zc)=&\,1-\f{\pi^2}{3}C_i C_A  \Bigl( \f{\as}{2\pi} \Bigr)^2 \ln^2\Bigl(
\f{\ed}{\zc}\Bigr) \, .
\end{align}
This is the usual result for the leading NGL in the hemisphere case, where the argument of the logarithm squared is now given by the ratio of the characteristic scales of the two functions $S_G$ and $S_Z$ in \eq{scales_ISD}. We emphasize, however, that this is NGL does not arise in the standard way, as both the high and low energy restrictions are imposed on the \emph{same} hemisphere.

We plot the numerical size of the NNLO non-global contribution to the jet energy drop distribution in \fig{NGL_ISD} for quarks and gluons as a function of $\ed$. The region to the left of the dotted vertical line is nonperturbative, as the softest scale in the factorization formula $\mu_{\CS_X}$ (see \eq{scales_ISD}) drops below 0.5~GeV. Outside the nonperturbative region, the effect of NGLs is less than 10\%. Although this NGL does not arise in the standard way, we still expect that the higher-order corrections are also the same as for the hemisphere case. Thus we explore the effect of higher order corrections using the solution~\cite{Schwartz:2014wha} of the BMS equation~\cite{Banfi:2002hw}. We find that  the effect of the three-loop contribution is (much) below the percent level, outside the nonperturbative region, as shown in \fig{NGL_ISD}. The two-loop NGL is thus sufficient for our numerical results in \sec{numerics_iteratedsoftdrop}, and we adopt the same practical approach for the grooming algorithms discussed in the subsequent sections.

\subsection{Profile functions and scale variations ~\label{sec:ISDprof}}

We will now describe our central scale choice, taking particular care to avoid the Landau pole in the nonperturbative region. The scale variations used to estimate the perturbative uncertainty will also be discussed. 

We observe from \eq{scales_ISD} that  the softest scale $\mu_{\mathcal{S}_X}$ determines the nonperturbative region of the $\Delta_E$ distribution,
\bea
\label{eq:ISDNPonset}
\Delta_E < \biggl(\frac{\Lambda_{\rm NP}z_{\rm cut}^{1/\beta}}{p_T R} \biggr)^{\beta/(1+\beta)}\,.
\eea
Here we take $\Lambda_{\rm NP} = 1.5$~GeV as the value where the scale starts becoming nonperturbative. For instance, we used this value of $\Lambda_{\rm NP}$  in \eq{ISDNPonset} to determine the position of the dotted vertical line in \fig{NGL_ISD}.

To prevent the strong coupling constant $\alpha_s$ from running into the Landau pole for small $\ed$, we use profile functions~\cite{Ligeti:2008ac} to freeze the scales at some value $\Lambda_{\rm freeze}$ above the Landau pole. The transition to the fixed-order region (where $\ed$ is large) does not require special care, because the non-singular contribution is so small, see \fig{refactorization_ISD}.
We make the following choice to smoothly transition
\bea
\label{eq:ISDfreeze}
f_{\text{pro}}(x;x_0)=&\left\{
    \begin{array}{ll}
      x \hspace{3.7cm} x>2x_0  \hspace{1cm} \text{region I}\,,\\ 
      x_0[1+(x/x_0)^2/4] \hspace{1cm}x\leq 2x_0 \hspace{1cm}\text{region II}\,.
    \end{array}
  \right.
\eea
Our central scale choice is given by
\bea
\mu_{\CS_X}^{\text{cent}} &= f_{\text{pro}}(\Delta_E^{(1+\beta)/\beta}z_{\rm cut}^{-1/\beta}p_T R;\Lambda_{\rm freeze})\,,\notag\\
\mu_{S_Z}^{\text{cent}} &= \bigl(\mu_{\CS_X}^{\text{cent}}(z_{\rm cut} p_T R)^{\frac{1}{\beta}}\bigr)^{\frac{\beta}{1+\beta}}\,,\notag\\
\mu_{S_G}^{\text{cent}} & = z_{\rm cut} p_T R\,,\notag\\
\mu_{\mathcal{G}}^{\text{cent}} &= p_T R\,,\notag\\
\mu_{\mathcal{H}}^{\text{cent}} &= p_T\,,
\label{eq:ISDcan}
\eea
where it is important to relate the two scales which depend on $\Delta_E$, such that $\mu_{S_Z}^{\text{cent}}$ also stops running when the softer scale $\mu_{\CS_X}^{\text{cent}}$ enters the nonperturbative region. Note that the latter two scales, the hard scale and jet scale, enter our calculation through the jet production described in \sec{fac_jet}. We make the choice 
\bea
\label{eq:x0ISD}
\Lambda_{\rm freeze} = 0.2\;\text{GeV}
\eea
throughout this paper, which ensures that we see the Sudakov peak.

QCD scale uncertainties are obtained by varying the scales of $\CS_X$, $S_Z$ individually up and down by a factor of 2 around their central value. We also vary the scales of $S_G, \mathcal{G}, \mathcal{H}$ simultaneously because there is not a large hierarchy between them, since $R=0.8$ and we generally take $\zc = 0.5$. Finally, we vary all scales simultaneously up and down, and take the envelope of these variations to obtain the uncertainty band.

\subsection{Numerical results~\label{sec:numerics_iteratedsoftdrop}}

\begin{figure}[t]
     \hfill \includegraphics[width=0.48\textwidth]{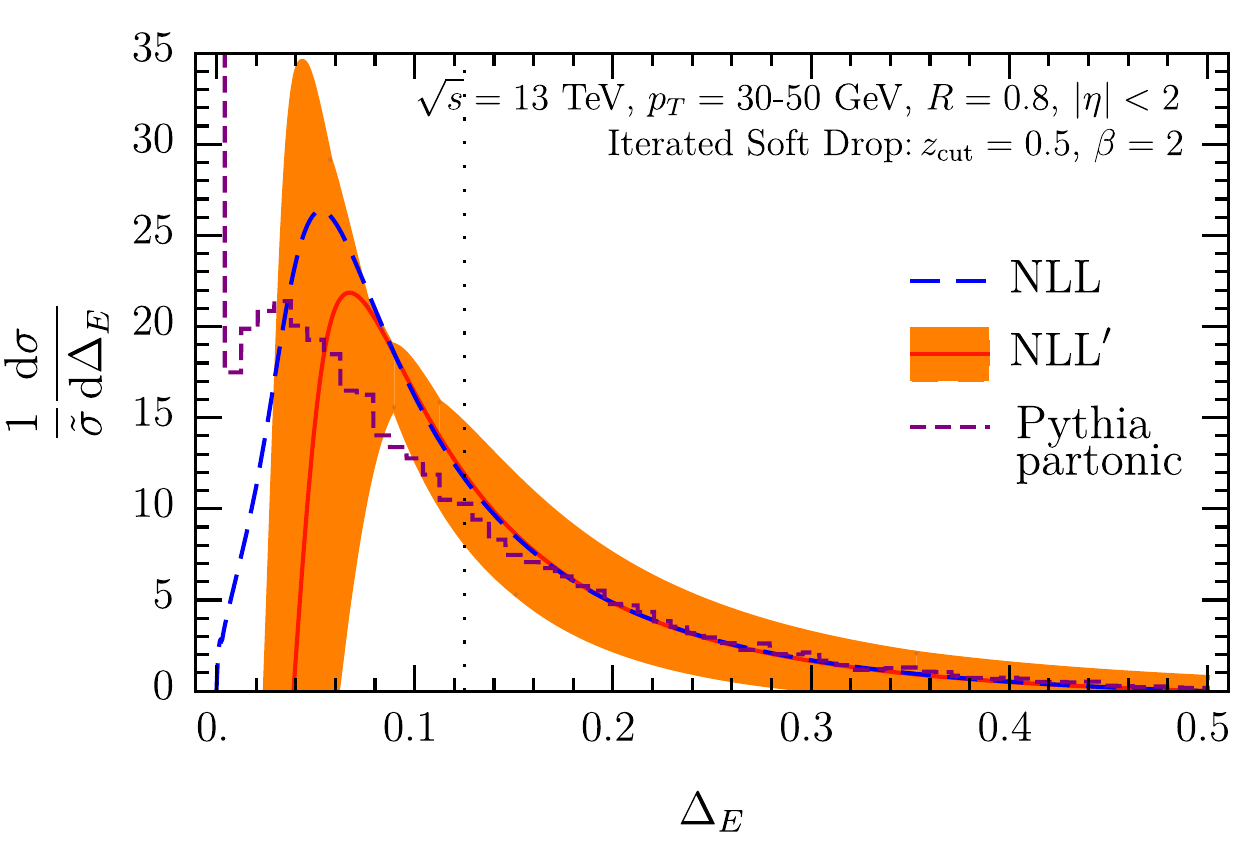} \hfill 
     \includegraphics[width=0.48\textwidth]{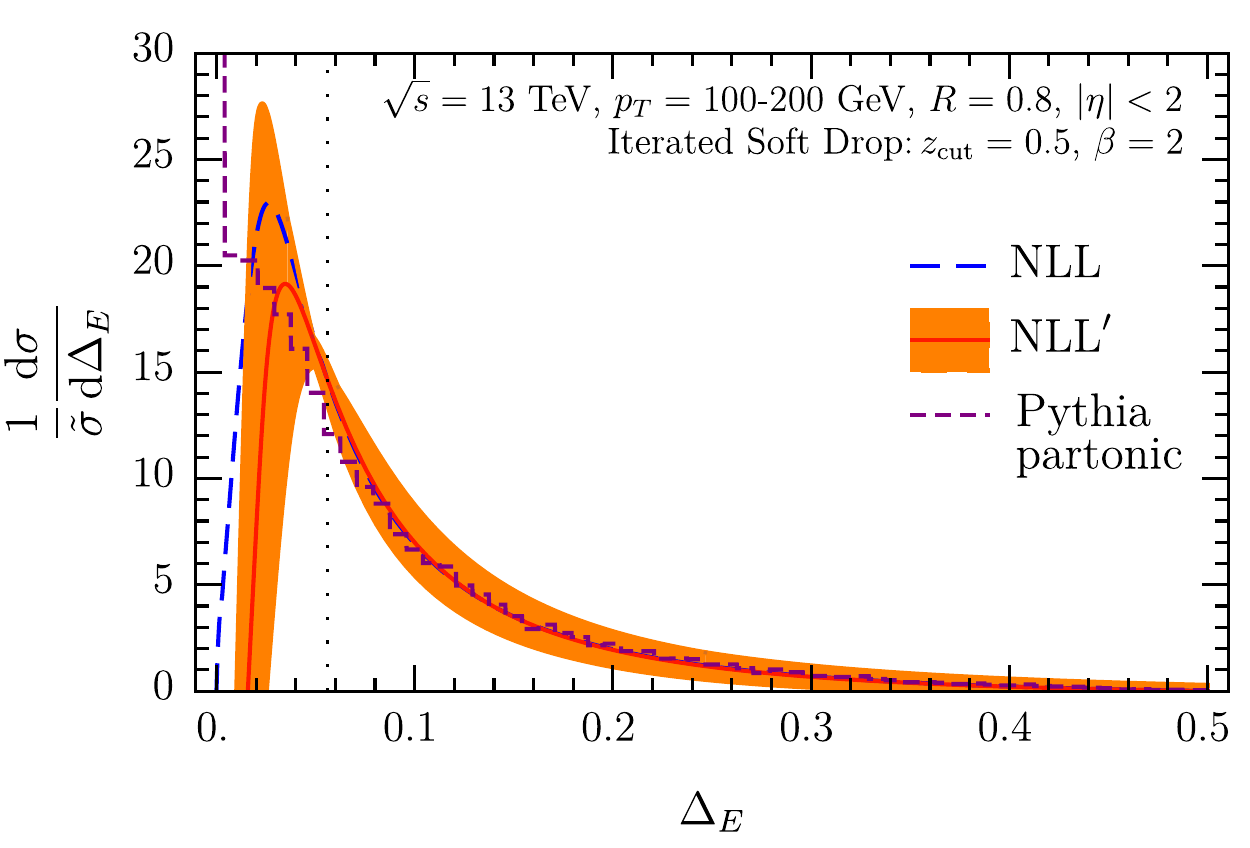} \hfill \phantom{.} \\
     \hfill \includegraphics[width=0.48\textwidth]{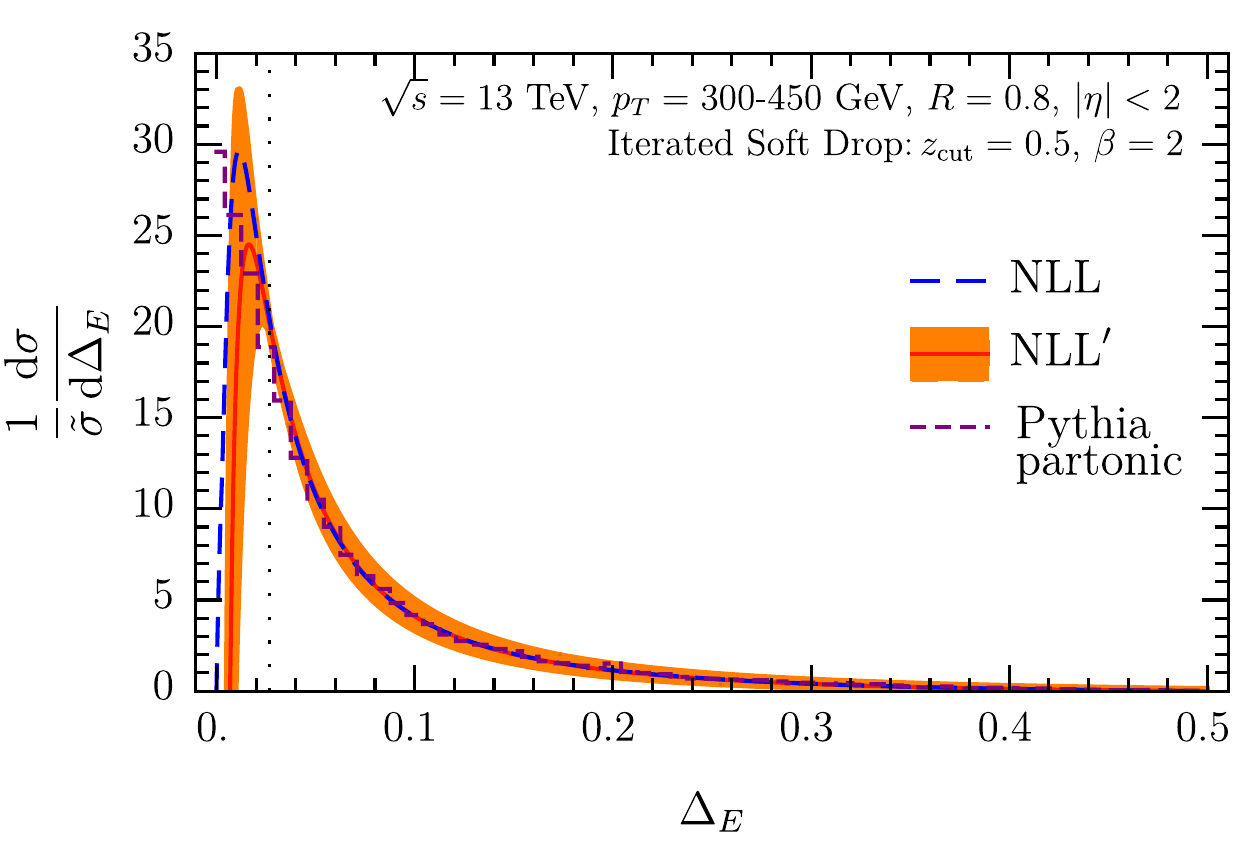} \hfill 
     \includegraphics[width=0.48\textwidth]{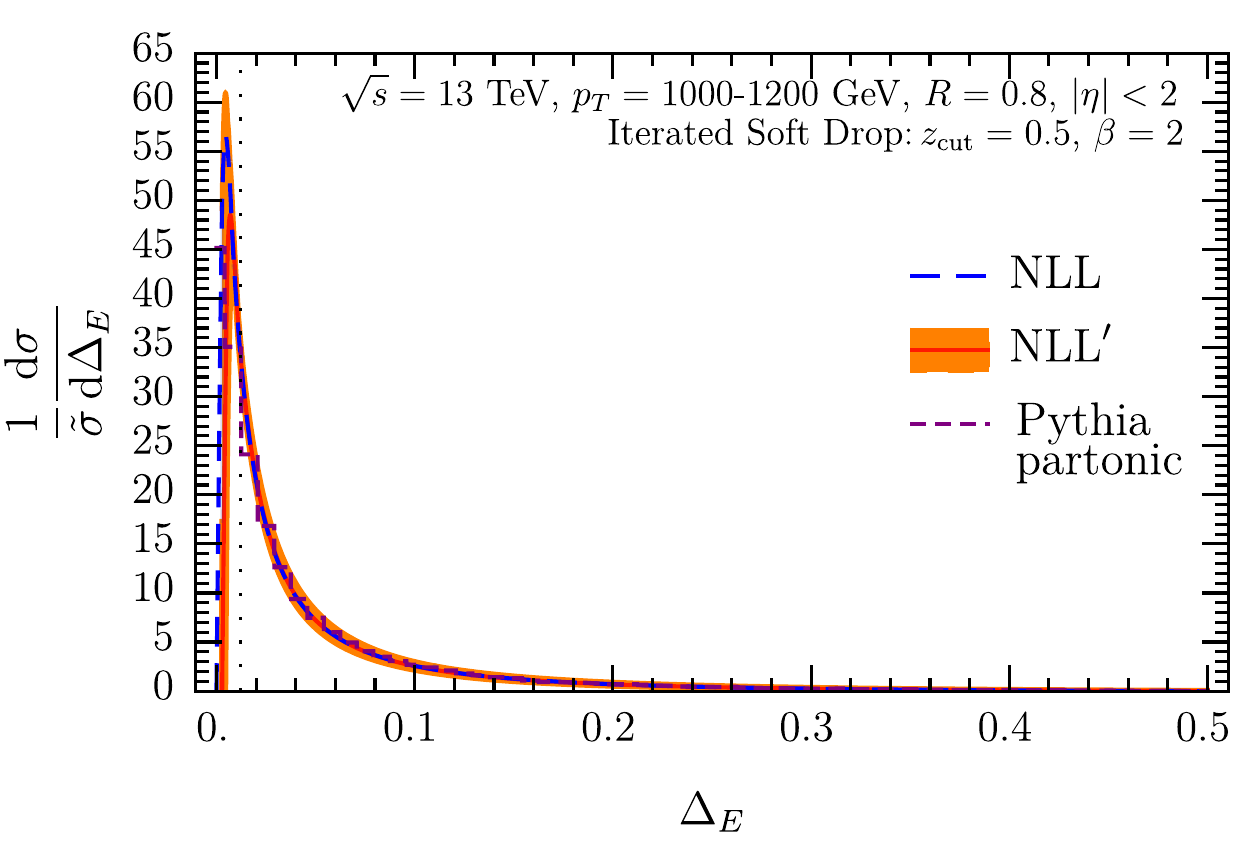} \hfill \phantom{.} 
    \caption{Jet energy drop distribution with $z_{\rm cut}=0.5$ and $\beta=2$ at NLL (dashed blue) and NLL$'$ accuracy (orange curve and band), compared to \Pythia (dashed purple). The different panels correspond to different jet transverse momenta.  The central curves are normalized to unity between the dotted vertical line and the endpoint $\Delta_E=z_{\rm cut}$.~\label{fig:numerics_ISD1}} 
\end{figure}

In this section we present our numerical results for the jet energy drop for the iterated soft drop algorithm, comparing to \Pythia~8.2 simulations~\cite{Sjostrand:2014zea}. We consider LHC kinematics at $\sqrt{s}=13$~TeV, reconstructing jets with the anti-k$_T$ algorithm and $R=0.8$ in the rapidity range of $|\eta|<2$. Throughout this work, we use the CT14 NLO PDF set~\cite{Dulat:2015mca}. 

\begin{figure}[t] 
    \centering
    \includegraphics[width=0.52\textwidth]{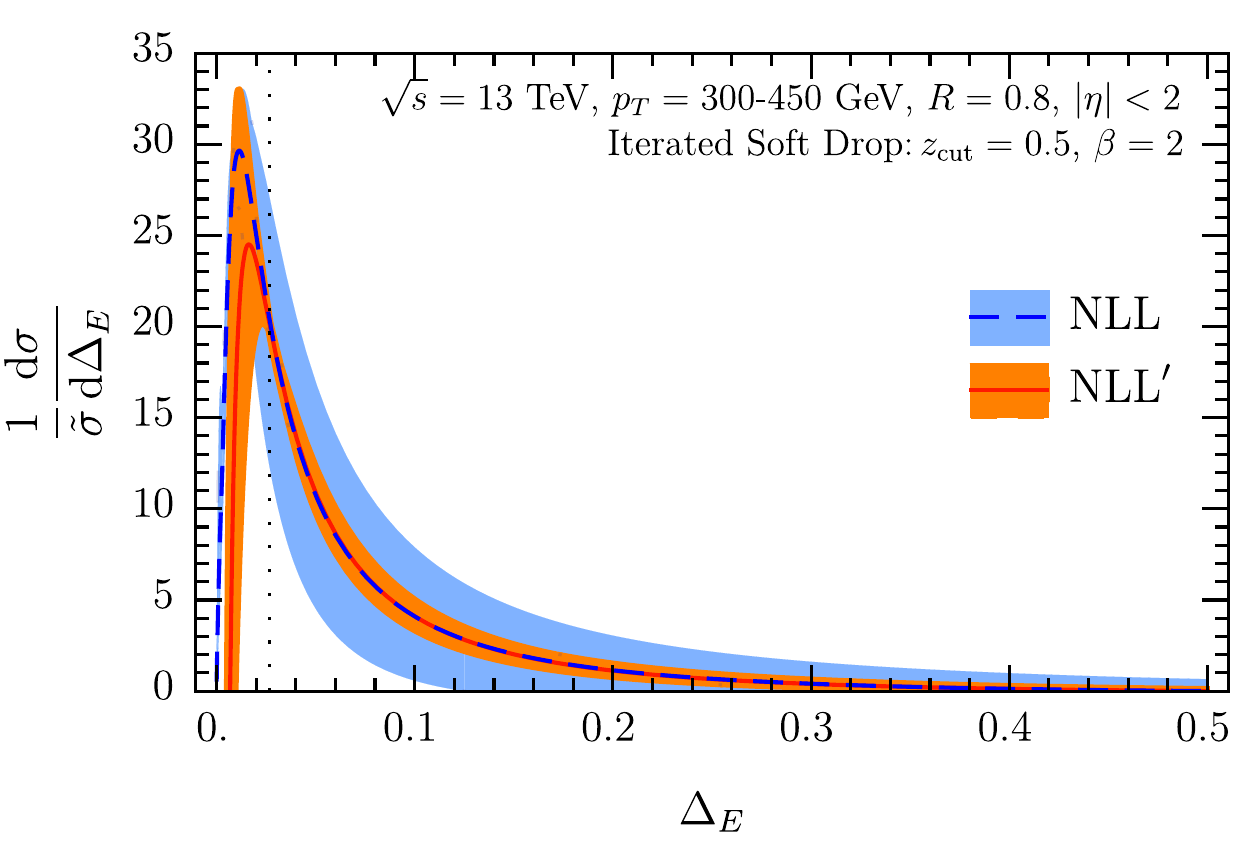}
    \caption{Comparison of the QCD scale uncertainties at NLL and NLL$'$, for the lower left panel of \fig{numerics_ISD1}.~\label{fig:ISDconvergence}} 
\end{figure}

In \fig{numerics_ISD1}, we show our results at NLL and NLL$'$ accuracy for the jet energy drop, and the corresponding results for \Pythia at parton level, including initial- and final-state radiation. The different panels correspond to different jet transverse momentum intervals ranging from $p_T=30$ to 1200 GeV, and we choose the grooming parameters $z_{\rm cut}=0.5$ and $\beta=2$. For the NLL$'$ curves, we include the perturbative uncertainty bands, following the procedure in \sec{ISDprof}. We indicate the onset of the nonperturbative region by a dotted vertical line, corresponding to  $\mu_{\CS_X}\sim 1.5$~GeV (see \eq{ISDNPonset}). We use a differential scale setting, which leads to a good prediction for the shape but only ensures the correct normalization up to higher-order corrections. We address this by simply normalizing our distribution, though there are more refined proposals (see e.g.~ref.~\cite{Bertolini:2017eui} for a discussion in the context of the thrust event shape). The NLL$'$ result becomes unreliable (negative) at small $\Delta_E$, because of large perturbative corrections from $\CS_X$, and would anyway need to be supplemented by a nonperturbative model. We therefore use the respective NLL curve (which is always positive) to obtain the normalization factor for the individual quark/gluon predictions and apply this to the NLL$'$ curves as well.  After combining these with the appropriate quark/gluon fractions we normalize the prediction by the cross section $\tilde \sigma$ on the interval between the vertical dotted line and the endpoint at $\Delta_E=z_{\rm cut}$, to limit the sensitivity to nonperturbative physics in the perturbative region. 
We note that the NLL results  lie within the uncertainty band of the NLL$'$, instilling confidence in the convergence of resummed perturbation theory. We also find generally good agreement with \Pythia, with the largest differences in the nonperturbative region, as expected. We observe that for lower jet $p_T$ the jet energy drop distribution peaks at larger values and is generally broader, which arises from the larger value of $\alpha_s$.

As an example, we show the QCD scale uncertainty at NLL and NLL$'$ accuracy in \fig{ISDconvergence}. We observe a dramatic reduction of the uncertainty band at NLL$'$. This illustrates the need to perform perturbative calculations at least at NLL$'$ accuracy, where scale variations in the RG evolution kernels are partially canceled by the NLO results of the different functions, and is the reason we omit the uncertainty band for NLL curves in subsequent plots.

\begin{figure}[t]
     \hfill \includegraphics[width=0.48\textwidth]{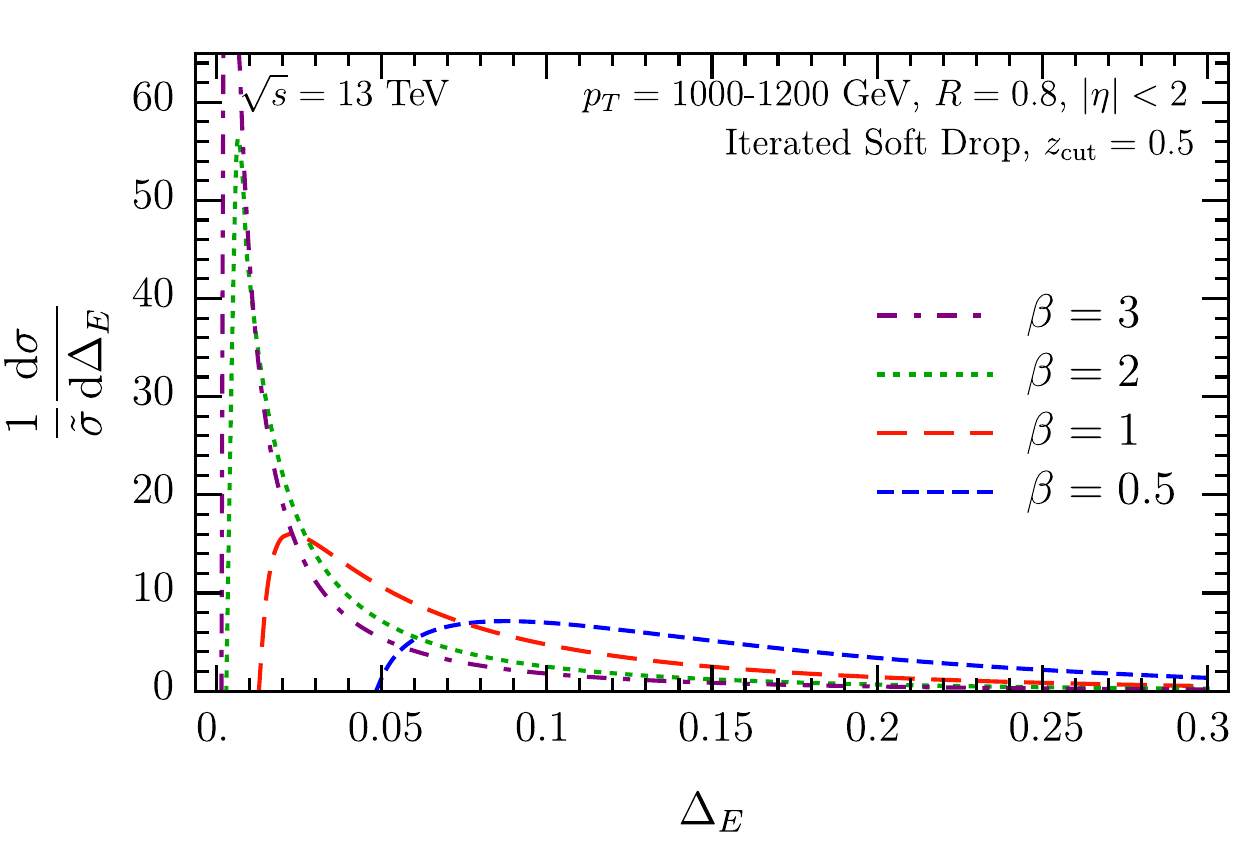} \hfill 
     \includegraphics[width=0.48\textwidth]{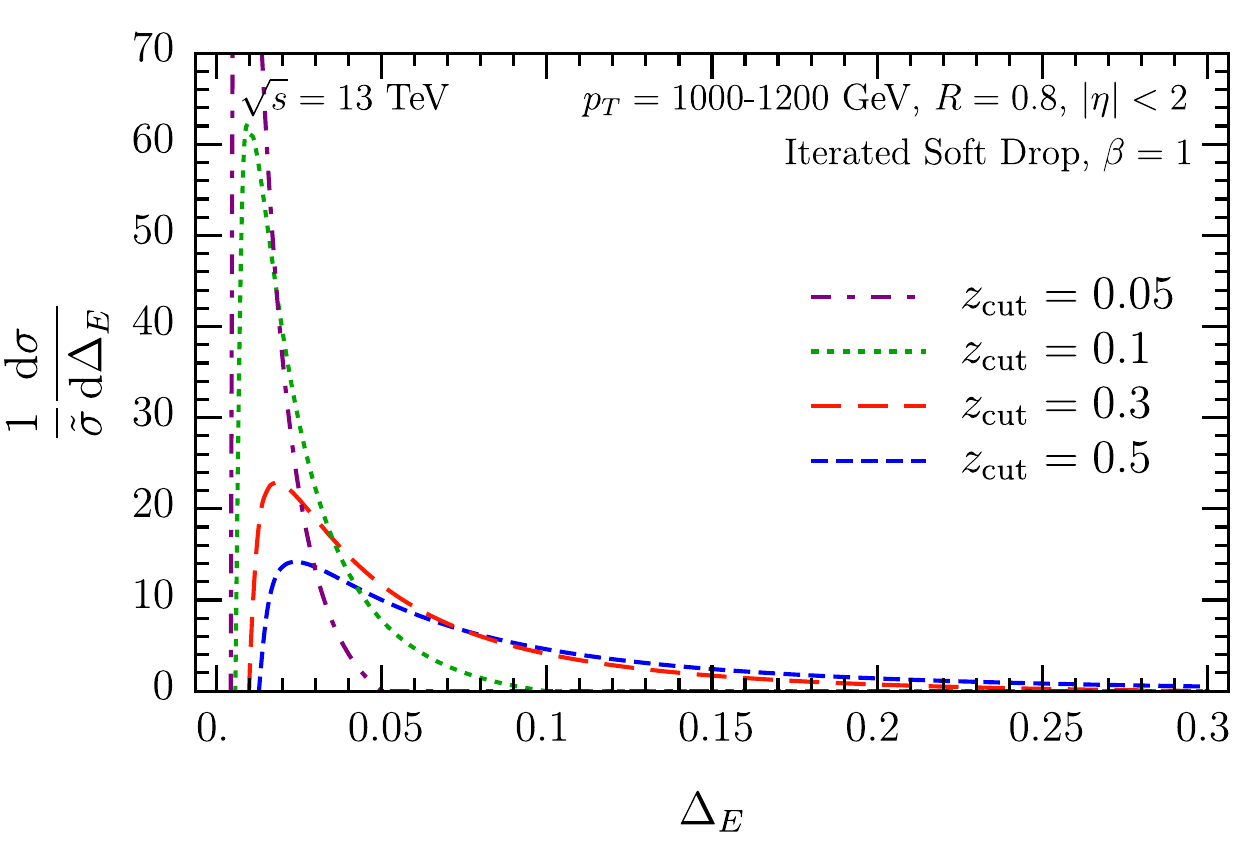} \hfill \phantom{.}
    \caption{The jet energy drop for iterated soft drop at NLL$'$ for several values of $\bt$ (left) and $\zc$ (right).~\label{fig:numerics_ISD2}} 
\end{figure}

\begin{figure}[t]
     \hfill \includegraphics[width=0.48\textwidth]{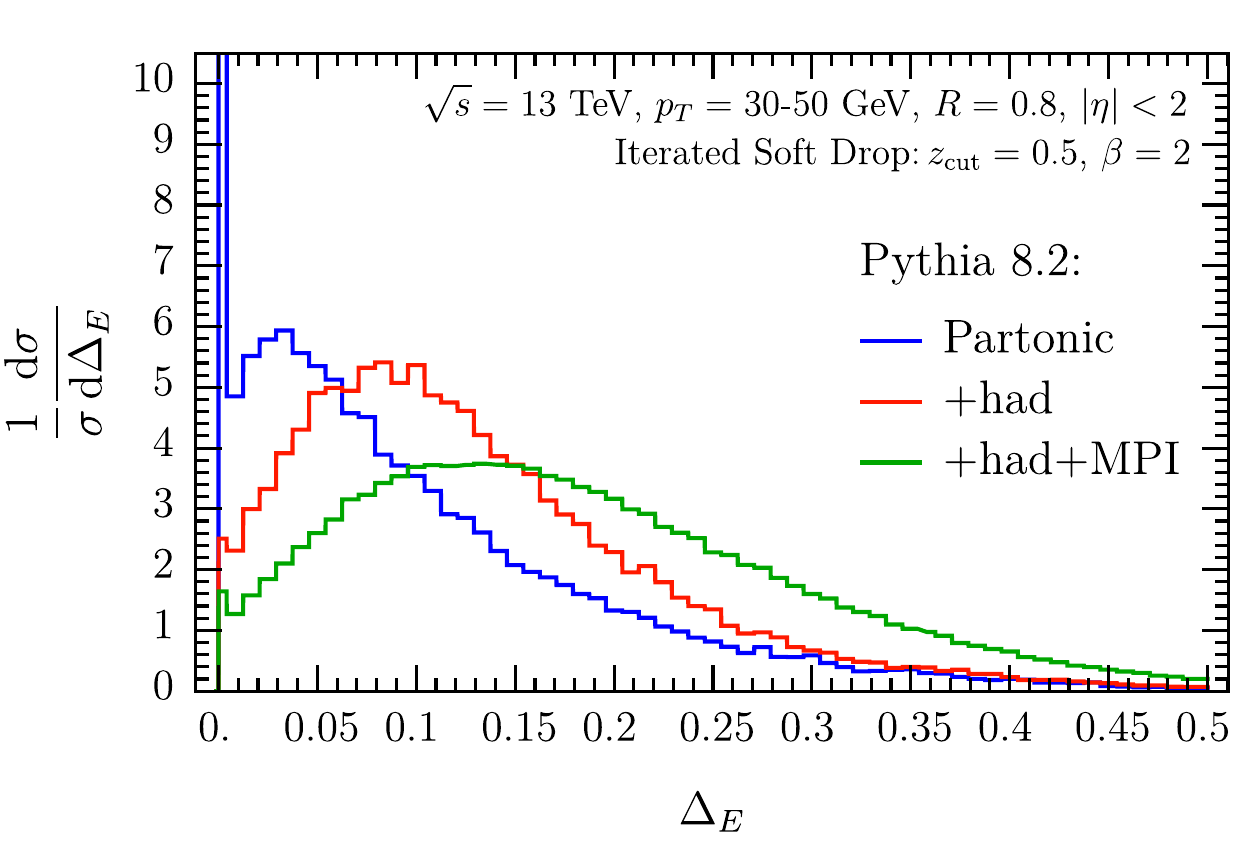} \hfill 
     \includegraphics[width=0.48\textwidth]{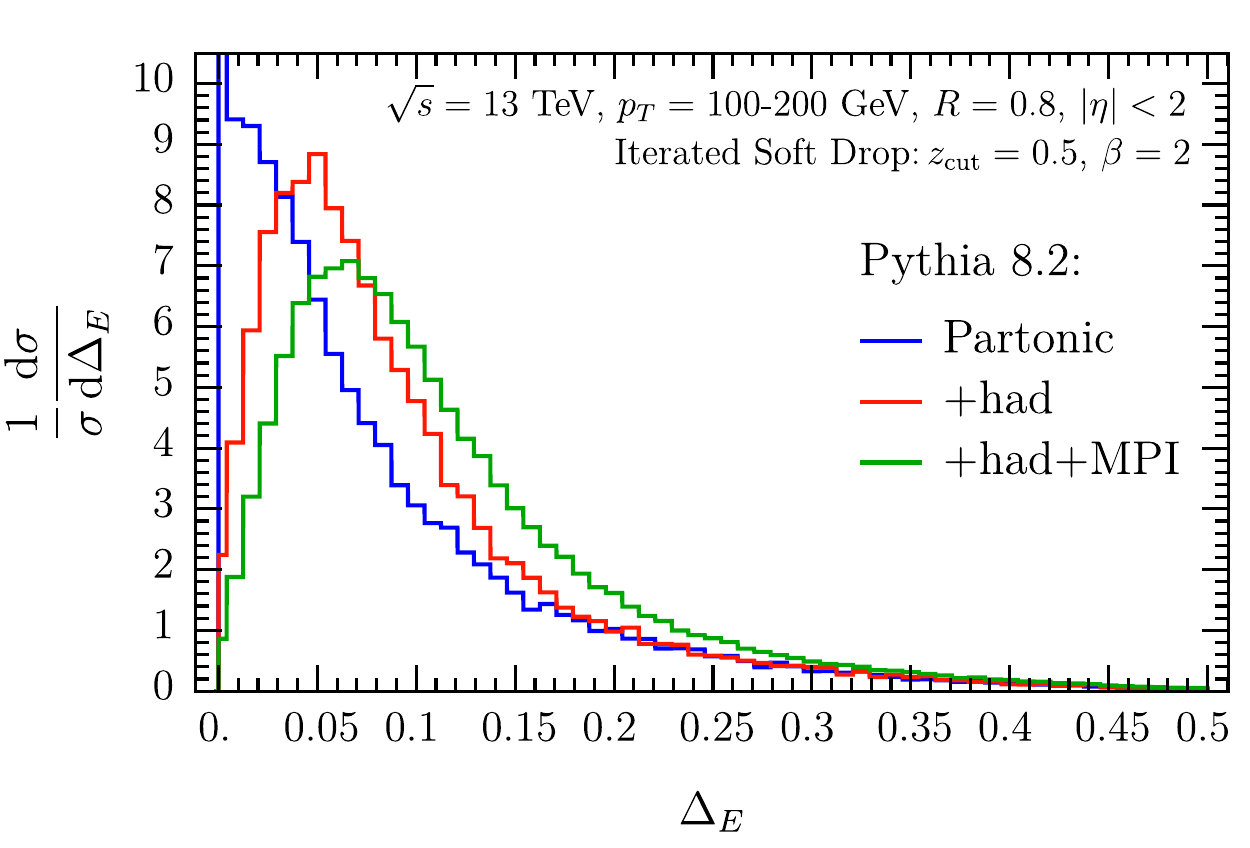} \hfill \phantom{.}
    \caption{\Pythia results for the jet energy drop with iterated soft drop at parton level (blue), including hadronization (red) and also MPI (green), for $p_T = 30 -50$ GeV (left) and $100- 200$ GeV (right). Note that these curves are normalized on the full $\ed$ interval.~\label{fig:numerics_ISD3}} 
\end{figure}

In \fig{numerics_ISD2}, we show the dependence of the jet energy drop on the grooming parameters $z_{\rm cut}$ and $\beta$. We consider jets with $p_T=1000-1200$~GeV to limit the effect of the nonperturbative region. In the left panel we fix $z_{\rm cut}=0.5$ and vary $\beta$, while in the right panel we choose $\beta=1$ and vary $z_{\rm cut}$. As expected from \eq{criterion}, the energy drop becomes smaller in the limit $\beta\to \infty$ and $z_{\rm cut}\to 0$. Indeed, in these limits, the jet energy drop distributions approach a delta function at $\Delta_E=0$ (apart from nonperturbative effects). 

In \fig{numerics_ISD3}, we study the effect of hadronization and multiple parton interactions (MPI) on the jet energy drop in \Pythia, for two different bins in the jet transverse momentum. The effect of hadronization is huge: In particular, for jet $p_T = 30-50$~GeV about 17\% of the jets at parton level are unaffected by grooming (i.e.~$\ed = 0$).  The effects of MPI are sizable and affect the whole distribution: the radiation due to MPI is uncorrelated to the primary scattering and therefore fairly uniformly distributed over the jet, such that grooming always removes a substantial part of them, independent of the value of $\ed$. This has of course the desired effect of removing them from the groomed jet, but makes our observable particularly sensitive to MPI. Hadronization mostly affects the peak region, shifting its location to the right. The effect of hadronization extends over a larger range of $\ed$ than one would expect from the onset of nonperturbative effects estimated in \eq{ISDNPonset}. The effect of both hadronization and MPI is reduced at higher jet energies.

\section{Soft drop\label{sec:SD}}

In this section we present the calculation of the jet energy drop using regular soft drop grooming. As discussed in~\sec{SDandISD} above, the soft drop algorithm terminates once a pair of branches satisfies the criterion in \eq{criterion}. Therefore, the soft drop condition is not applied to emissions that are at smaller angles than the opening angle between the two branches that satisfy the grooming condition. This leads to a different and more complicated factorization structure than for iterated soft drop, involving the angle between the two branches that satisfy the soft drop condition.

In our factorization analysis, we consider the cross section differential in both the energy drop $\ed$ and the opening angle of the two branches  $R_g= \theta_g  R$. We identify two separate regimes, depending on the relative size of $\Delta_E,\,\theta_g$, and $z_{\rm cut}$, as discussed below. After the resummation is performed, we can remove the dependence on $\theta_g$ by integrating over it, or, alternatively, calculate the cross section for jet energy drop with a cut on $\theta_g < \theta_g^{\rm cut}$.  A related factorization structure was found in ref.~\cite{Cal:2019gxa}, for the angle between the standard and the groomed jet axis. Both observables probe the radiation which is groomed away by soft drop, and are therefore very soft sensitive. Indeed, imposing a cut on $\theta_g$ reduces the soft sensitivity of the jet energy drop, as will be demonstrated in \sec{numerics_softdrop}. In addition, it can be advantageous for experimental measurements.  
The (modified) LL jet energy drop cross section was calculated in ref.~\cite{Larkoski:2014wba}, by means of a conditional probability. We will also explain the connection between this approach and our double differential factorization.

We start by presenting results for the jet function differential in both $\Delta_E$ and $\theta_g$ at NLO in~\sec{fixedorder_softdrop}. In~\sec{refactorization_softdrop}, we discuss in detail the refactorization of the jet function, separated into two factorization regimes, and the resummation of logarithms of $\Delta_E,\, \theta_g$ and $\zc$, including non-global logarithms. We show how the global logarithms can be reproduced by means of a conditional probability in \sec{Cond_Prob}, and discuss the Sudakov-safe case $\beta=0$ in \sec{SDbeta0}. In \sec{SDnonp} nonperturbative effects are discussed (in particular for the case where there is a cut on $\theta_g$), and our scale choices are presented in \sec{SDprofiles}. Finally, in~\sec{numerics_softdrop} we present numerical results for LHC kinematics and compare to \Pythia results. 

\subsection{Fixed-order results~\label{sec:fixedorder_softdrop}}

In our factorization analysis, we need to account for the groomed jet radius $\theta_g$, since its value modifies the structure of the large logarithms in the jet energy drop $\ed$. In particular, we will jointly resum large logarithms involving $\ed$ and $\theta_g$, to all orders in $\al_s$. We therefore calculate the double-differential jet function at NLO, which will provide a check on our factorization. However, only the jet function differential in $\ed$ enters in the final result (unless a cut on $\theta_g$ is imposed). 

At NLO, $\Delta\cG_i^{\rm SD}$ is calculated from 
\begin{flalign}\label{eq:grJF1}
&\Delta \cG^{\rm SD}_i(\ed,\tg,p_TR, z_{\rm cut},\beta,\alpha_s(\mu)) 
\nn \\
&\quad = \int\! \df \Phi_2\, \si_{2,i}^c \, \Theta(\theta < R) 
\Bigl[ \Theta \bigl(x> \zc (\theta/R)^\beta \bigr)\Theta \bigl(1-x >\zc (\theta/R)^\beta \bigr)  \delta (\ed)  \delta \bigl( \tg - \theta/R \bigr)\nn \\
&\qquad+ \Theta \bigl(x> \zc \bigl( \theta/R \bigr)^\beta \bigr)\Theta \bigl(1-x <\zc (\theta/R)^\beta \bigr) \delta ( \ed -(1-x)) \delta (\tg)\nn \\
&\qquad+ \Theta \bigl(x< \zc \bigl( \theta/R \bigr)^\beta \bigr)\Theta \bigl(1-x >\zc (\theta/R)^\beta \bigr) \delta (\ed-x) \delta (\tg) - \delta(\ed) \delta (\tg) \Bigr]\,.
\end{flalign}
At this order terms are either $\propto\delta(\theta_g)$, when one parton fails the soft drop criterion, or $\propto\delta(\Delta_E)$, when both pass. The final term subtracts off the contribution already contained in the semi-inclusive jet function, see \eq{Delta_iG}. For quarks and gluons we find to NLO 
\begin{align}\label{eq:grJF2}
&\Delta \cG_q^{\rm SD}(\ed,\tg,p_TR,z_{\rm cut},\beta,\alpha_s(\mu)) 
\nn \\
&\quad = \frac{\alpha_s C_F}{\pi}\bigg\{\delta(\ed)\Theta(\tg<1)\biggl[ (-2+3 \zc)\bt \bigg[\f{\ln \tg }{\tg} \biggr]_+
+ \Bigl(-2 \ln \zc-\f32+3\zc \Bigr)\f{1}{[\tg]}_+ 
\nn \\ &\qquad
+\f{2}{\tg} \ln (1-\tg^\bt \zc)\bigg] 
+\frac{1}{\beta}\Theta(\ed <z_{\rm cut})\delta(\tg) \bigg[-2 \biggl[\frac{\ln \ed}{\ed}\biggr]_+ + 2\ln \zc \frac{1}{[\ed]}_+
  \nn \\&\qquad
+\Bigl(3-\frac{2}{1-\ed}\Bigr) \ln\Bigl(\frac{\ed}{z_{\rm cut}}\Bigr) \bigg]+\frac{1}{\beta}\delta(\ed) \delta(\tg) [-\ln^2 z_{\rm cut}+3 z_{\rm cut}]\bigg\} 
\\[.5em]
&
\Delta \cG_g^{\rm SD}( \ed, \tg,p_TR, \zc, \bt ,\alpha_s(\mu)) 
 \\ & \quad
 =\frac{\al_s}{\pi} \bigg(\delta(\ed)\Theta(\tg<1)\bigg\{ C_A \bigg[ \Big(-2 -\f32\zc +9\zc^2 -9\zc^3 \Big)\bt   \biggl[\f{\ln \tg}{\tg}\biggr]_+ 
\nn \\  &\qquad
 + \Big(-2 \ln \zc -\f32 \zc   +\f92 \zc^2 -3\zc^3 \Big)\f{1}{[\tg]}_+ + \f{2}{\tg} \ln \Big(1-\zc \tg^\bt \Big)\bigg]
 \nn\\  &\qquad
 +\f{\bt_0}{2} \bigg[(3\zc-6\zc^2+6\zc^3)\bt \biggl[\f{\ln \tg}{\tg}\biggr]_+ +(-1 +3\zc -3\zc^2 +2\zc^3)\f{1}{[\tg]}_+ \bigg]
\nn \\  &\qquad
+\f{1}{\bt}\delta(\tg) \Theta(\ed < \zc) \Bigg[ C_A \bigg[ -2 \biggl[\f{\ln \ed}{\ed} \biggr]_+ +2\ln \zc\f{1}{[\ed]}_+\bigg] 
\nn \\  &\qquad
- \bigg[C_A\Bigl( \f{2}{1-\ed}-4 +2\ed-2\ed^2\Bigr) +2n_f T_F \Bigl(\Delta_E^2+(1-\Delta_E)^2 \Bigr) \bigg] \ln \biggl(\f{\ed}{\zc} \biggr)\Bigg]
\nn \\  &\qquad
  +\f{1}{\bt}\delta(\tg) \delta(\ed) \bigg[C_A \Bigl( \!-\!\ln^2 \zc \!+\! 4\zc \!-\!\f{\zc^2}{2}\!+\!\f29 \zc^3\Bigr) 
  + n_f T_F \Bigl(\!-\!\zc \!+\!\zc^2 \!-\! \f49\zc^3 \Bigr)  \bigg]  \bigg\}.  \nn
\end{align}
As is clear from the $1/\bt$ poles in the above expressions, the jet energy drop is not IRC safe for soft drop with $\beta=0$. However, unlike for iterated soft drop, $\bt=0$ is Sudakov safe, as will be discussed in \sec{SDbeta0}. Alternatively, it is also IRC safe if a cut on $\theta_g$ is imposed, which removes the singularity at $\theta_g=0$.

\begin{figure}[t]
    \centering
\includegraphics[width=0.48\textwidth]{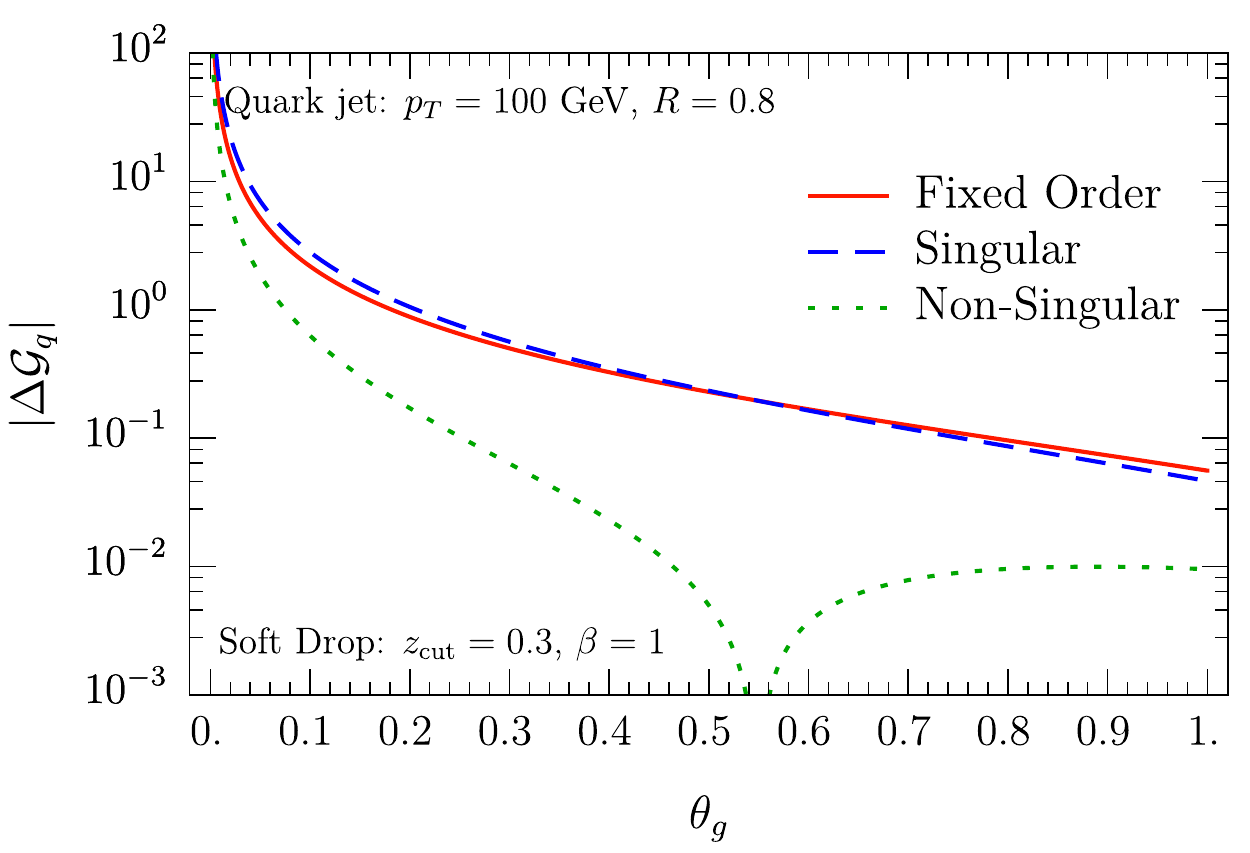} \hfill 
\includegraphics[width=0.48\textwidth]{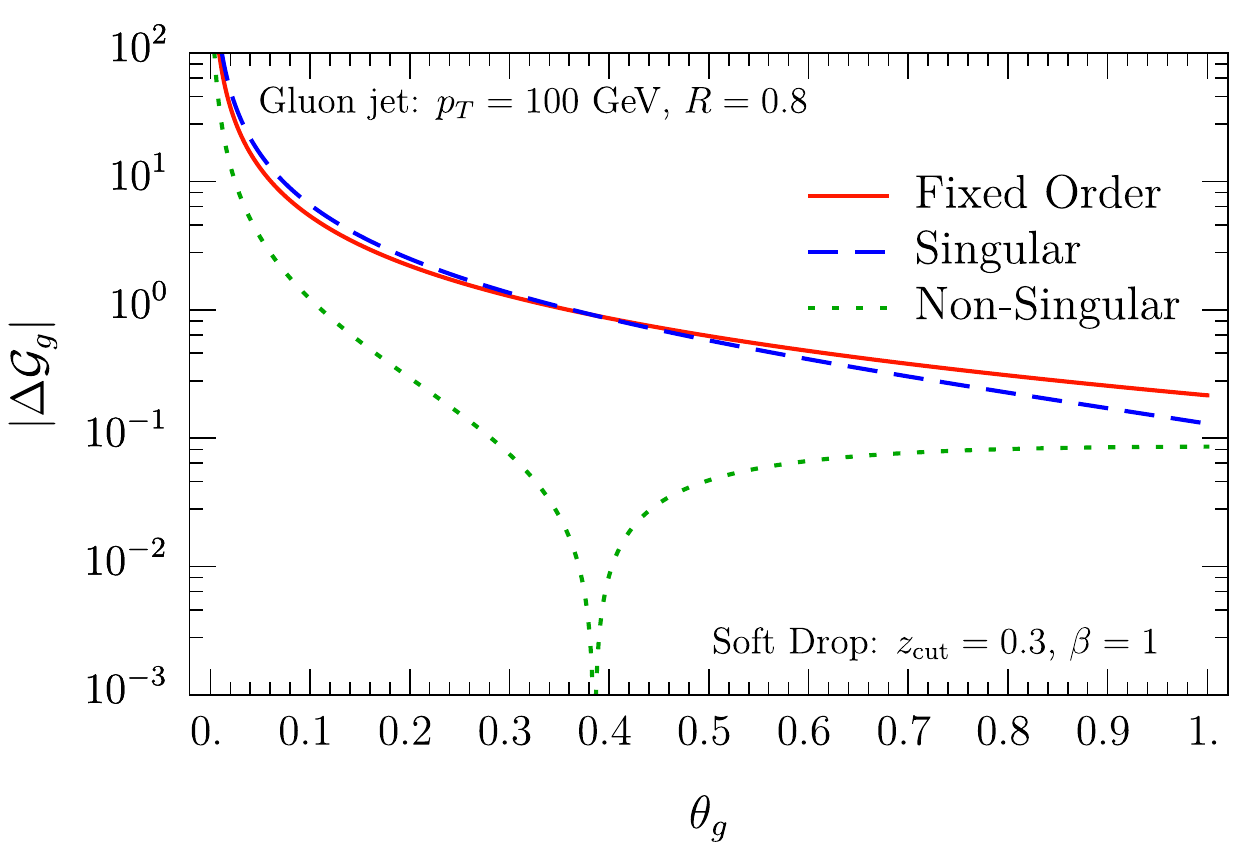} \hfill \phantom{.} \\
    \caption{Numerical results for the singular, non-singular and fixed-order result of $\Delta\cG_i^{\rm SD}$ for soft drop as a function of the groomed jet radius $\theta_g$,
    for the same jet kinematics as in \fig{refactorization_ISD} and with grooming parameters $\beta=1$ and $z_{\rm cut}=0.3$.~\label{fig:fixedorderthetag_SD}}
\end{figure}

Upon integration over the groomed radius $\theta_g$, 
\begin{equation}
\int_0^1\df\theta_g\, \Delta \cG^{\rm SD}_i(\ed,\tg,p_TR,z_{\rm cut},\beta,\alpha_s(\mu)) \stackrel{{\rm NLO}}{=} \Delta \cG^{\rm ISD}_i(\ed,p_TR,z_{\rm cut},\beta,\alpha_s(\mu)) \,.
\end{equation}
we obtain the jet function for iterated soft drop in \eq{ISDfixedorder}. Consequently, the size of the logarithmically enhanced terms in the jet function is the same as for iterated soft drop, shown in \fig{refactorization_ISD}. For completeness, we also plot the NLO jet function $\Delta\cG_i^{\rm SD}$ as a function of $\theta_g$ to further assess the numerical size of the power corrections to the singular terms in the limit $\Delta_E\ll z_{\rm cut}\ll 1$. The results in \fig{fixedorderthetag_SD} show that the power corrections are small as long as $\theta_g$ is not too small. We impose a sufficiently large $\theta_g^{\rm cut}$ in our phenomenological studies below, and thus do not need to include a matching correction.

\subsection{Factorization and resummation~\label{sec:refactorization_softdrop}}

We will consider the kinematic regime where $\Delta_E\ll\zc\ll 1$ and $\theta_g\ll 1$. We obtain two different factorization formulae, depending on whether $\theta_g$ is (parametrically) larger or smaller than $(\Delta_E/z_{\rm cut})^{1/\beta}$, discussed in \secs{SD_regimeA}{SD_regimeB}, respectively.

\subsubsection{Regime A\label{sec:SD_regimeA}}

\begin{figure}[t]
    \centering
    \includegraphics[width=0.48\textwidth]{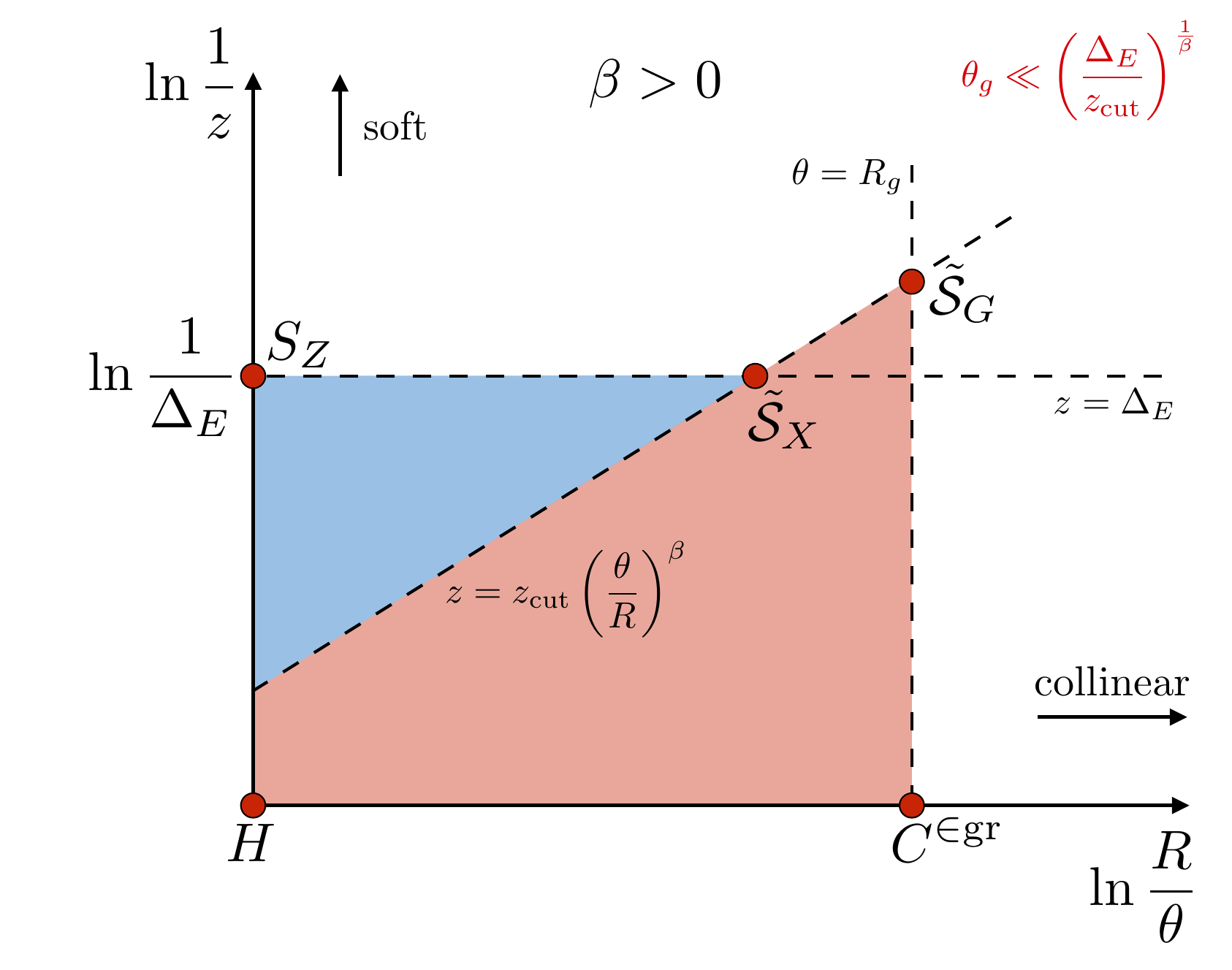} \hfill 
       \includegraphics[width=0.48\textwidth]{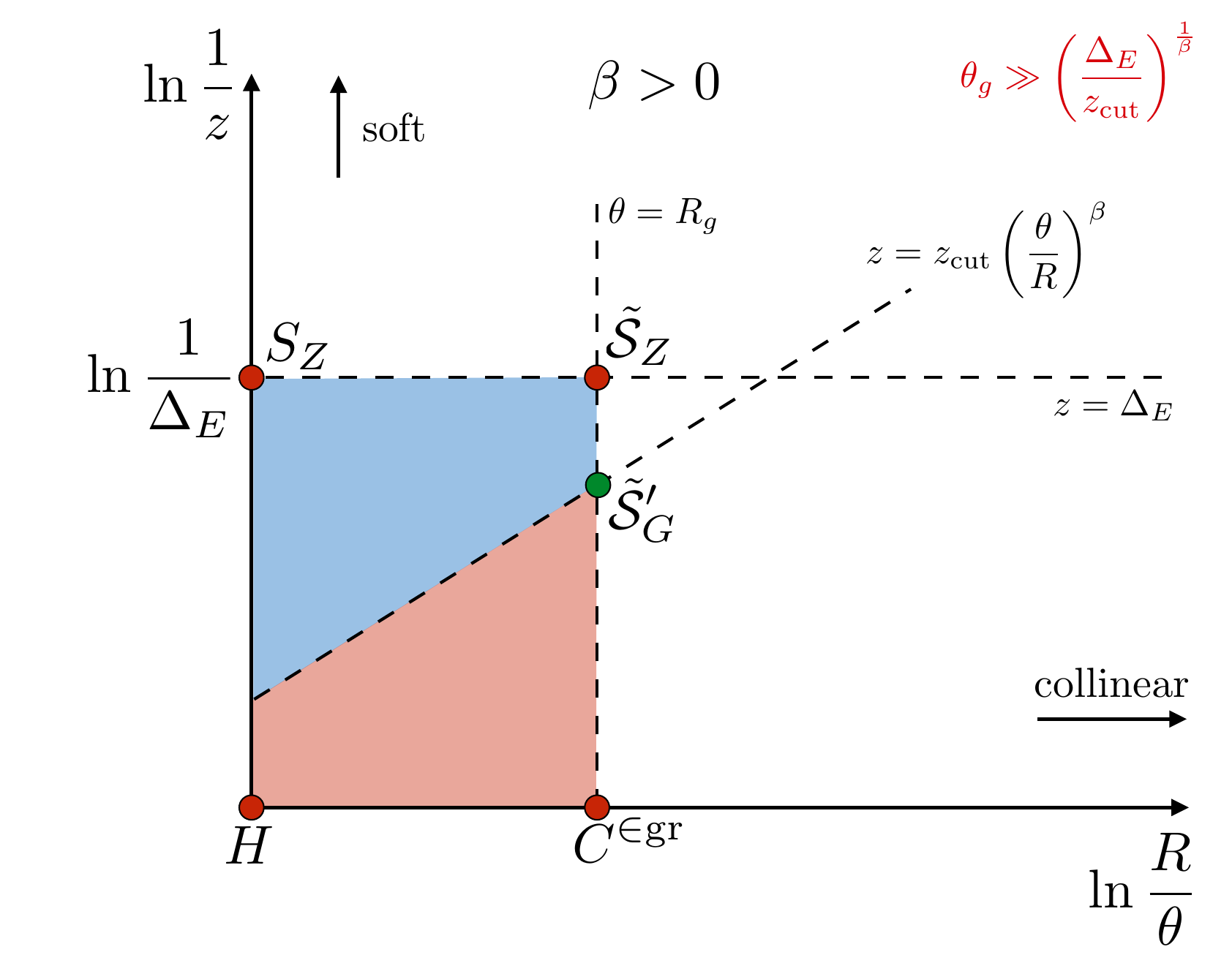} \hfill \phantom{.} \\
    \caption{Lund diagrams for the energy drop of a soft-drop groomed jet in the region $\ed \ll \zc\ll 1$ and $\theta_g \ll1$, for regime $A$ (left) and $B$ (right). The relevant SCET modes are indicated by red and green dots, and their power counting can be read off, see table~\ref{tab:modes_SD}.~\label{fig:LundSD}}
\end{figure}

The Lund diagram for regime $A$ is shown in the left panel of \fig{LundSD}. The dashed lines show the measurements of $\Delta_E$ and $\theta_g$ as well as the grooming condition, as indicated in the figure. The cumulative measurement of the groomed radius $\theta_g<\theta_g^c$ vetoes emissions in the red region with momentum fraction $z>z_{\rm cut}(\theta/R)^\beta$ and angles $\theta/R>\theta_g^c$, see also refs.~\cite{Larkoski:2014wba,Kang:2019prh}. In addition, we now measure the jet energy drop $\Delta_E<\Delta_E^c$. Emissions with $z<\zc(\theta/R)^\beta$ and $\theta/R>\theta_g$ are groomed away, and contribute to the measured value of $\Delta_E$. Therefore, emissions are vetoed in the blue region. For $\theta_g<(\Delta_E/z_{\rm cut})^{1/\beta}$, we thus obtain the Lund diagram as shown in \fig{LundSD}. Note that to simplify the notation, we omit the superscript $c$ for cumulative variables in \fig{LundSD}.

We start with the resummed result at LL accuracy, which can be calculated from the vetoed red and blue shaded areas of the Lund diagram. This gives the cumulant from which we obtain the double-differential result by taking derivatives with respect to both $\ed$ and $\theta_g$, 
\begin{align}
 &\tilde \cG_{i,A}^{\rm SD}\bigl(\ed , \theta_g,  p_T R, \zc,\bt, \al_s(\mu)\bigr) \nn \\
 & \quad \stackrel{{\rm LL}}{=} \frac{\df}{\df \ed}\,\frac{\df}{\df \theta_g}\,\exp\bigg\{- \frac{\al_s C_i}{\pi}\bigg[\f{1}{\bt} \ln^2 \Bigl(\f{\zc}{\ed} \Bigr) +2 \ln \zc \ln \tg+\bt \ln^2  \tg  \bigg] \bigg\} \,.\end{align}

\begin{table}
   \centering
   \begin{tabular}{l|l|ll}
     \hline \hline
     Mode: & Function: & Regime $A$ & Regime $B$ \\ \hline
     hard & $H$ & \multicolumn{2}{c}{$p_T(R^2,1,R)$} \\
     soft & $S_Z$ & \multicolumn{2}{c}{$\ed\, p_T (R^2, 1, R) $} \\ 
     collinear & $C^{\in {\rm gr}}$ & \multicolumn{2}{c}{$p_T (R_g^2,1, R_g)$} \\
     collinear-soft & $\tilde {\cal S}_G$ & \multicolumn{2}{c}{$\zc \theta_g^\bt p_T (R_g^2,1, R_g)$} \\
     collinear-soft & $\tilde {\cal S}_X$ & $ \ed\, p_T \Bigl( \bigl(\f{\ed}{\zc}\bigr)^{2/\bt} R^2 ,1,\bigl(\f{\ed}{\zc}\bigr)^{1/\bt} R \Bigr)$ \quad \phantom{a} & \\
     collinear-soft & $\tilde {\cal S}_Z$ &  & $ \ed\, p_T (R_g^2,1, R_g)$ \\
     \hline \hline
   \end{tabular}
   \caption{The scaling of the modes that enter the factorization formulae of the jet energy drop cross section in the kinematic region where $\ed \ll \zc \ll 1 $ and $\theta_g=R_g/R \ll 1$. Regime $A$ ($B$) correspond to $\theta_g$ being smaller (larger) than $(\ed/\zc)^{1/\bt}$.~\label{tab:modes_SD}}
\end{table}

We extend this beyond LL using a factorization formula within SCET, for which the modes correspond to red points at the intersections of the dashed lines in the left panel of \fig{LundSD}. The parametric scaling of the momenta of the hard, collinear, soft, and collinear-soft modes are summarized in table~\ref{tab:modes_SD}. (Because this is a refactorization of a collinear function, these modes are strictly speaking hard-collinear, collinear, etc.) The resulting factorization, differential in $\ed$ and $\theta_g$, is given by 
\begin{align} \label{eq:refact_A}
&\tilde \cG_{i,A}^{\rm SD}\bigl(\ed, \theta_g, p_T R, \zc,\bt, \al_s(\mu)\bigr)
\\ &\quad
  \stackrel{{\rm NLL'}}{=} \frac{\df}{\df \theta_g} \bigg[\tilde H_i(p_{T} R,\mu)\, 
 C_i^{\in \rm gr}(\theta_g p_T R,\mu) \,  \CS_{i,G}\bigl(\zc \theta_g^{1+\bt} p_T R, \bt, \mu\bigr)
 \CS_i^{\rm NG+AC}(\zc \theta_g^\bt)  
 \nn \\ & \qquad\times\,
 \int\! \df \ed'\, \CS_{i,X}( \ed' , p_T R, \zc, \bt, \mu)  
 S_{i,Z}(\ed - \ed' , p_T R, \mu) \, S_i^{\rm NG}(\ed) \bigg]
\,.\nn
\end{align}
The hard function $\tilde H_i$ is only sensitive to the jet scale and does not depend on $\Delta_E$ and $\theta_g$. Note that for iterated soft drop this hard function was absent. The collinear function $C^{\in{\rm gr}}_i$ does not depend on $\ed$, since collinear radiation is never groomed away. It can set the measurement of $\theta_g$ (when the derivative acts on the collinear function), or account for collinear emissions at smaller angles (when the derivative does not act on it). Next, the collinear-soft function $\CS_{i,G}$ is sensitive to the soft drop grooming condition and can also set the groomed radius of the jet $\theta_g$. The three functions discussed so far also appear in the NLL$'$ factorization of the soft drop groomed radius~\cite{Kang:2019prh}. The collinear-soft function $\CS_{i,X}$ is sensitive to both the $\Delta_E$ measurement and the grooming condition, as the corresponding emissions  contribute to the jet energy drop if they fail the soft drop criterion. Finally, the soft function $S_{i,Z}$ accounts for soft wide-angle radiation which is always groomed away. The same functions $\CS_{i,X}$ and $S_{i,Z}$ enter in the factorization for iterated soft drop in \eq{factorization_ISD}. Interestingly, in regime A the dependence on $\ed$ and $\theta_g$ appears in separate parts of the factorization formula.

There are two types of non-global logarithms in \eq{refact_A} associated with ungroomed and groomed jet boundary, $R$ and $R_g$, respectively. These can be treated independently as long as they are sufficiently separated, i.e.~$R_g \ll R$. The NGLs at the ungroomed jet boundary arise due to correlations of the out-of-jet region, where the radiation is unconstrained (and thus has energies of order $p_T$), and the in-jet region, where wide-angle radiation must have energies below $\ed p_T$. This is taken into account by the same non-global soft function as the hemisphere case, $S_i^{\rm NG}(\ed)$. The $ \CS_i^{\rm NG+AC}(\zc \theta_g^\bt)$ arises at the boundary of the groomed jet. Unlike the hard boundary of the initial ungroomed anti-k$_T$ jet, it is sensitive clustering effects from C/A. This same contribution entered in the resummation of the groomed jet radius~\cite{Kang:2019prh}, and is given by 
\begin{align}
\label{eq:NGAC}
\CS_i^{\rm NG+AC}(\zc \theta_g^\bt)=&\,1-\f{4}{9}\f{\pi^2}{3}C_i C_A  \Bigl( \f{\as}{2\pi} \Bigr)^2 \ln^2\bigl(\zc \theta_g^\bt\bigr) \, .
\end{align}
The factor 4/9 compared to \eq{SNG} is due to clustering effects.

Here we present the one-loop expressions for the functions in \eq{refact_A} that did not appear for iterated soft drop. The hard function $\tilde H_i$~\cite{Kang:2017mda,Cal:2019hjc}, the collinear function $C^{\in {\rm gr}}_i$ and the collinear-soft function $\CS_{i,G}$~\cite{Kang:2019prh} are given by 
\begin{align}
\tilde H_q(p_{T} R,\mu) =&\, 1 + \frac{\al_s C_F}{\pi} \biggl[- \ln^2\bigg(\f{\mu}{p_T R} \bigg)  - \frac32 \ln\bigg(\f{\mu}{p_T R} \bigg)  - \frac{13}{4} + \frac{3\pi^2}{8} \biggr]
 \,, \\[.5em]
\tilde H_g(p_{T} R,\mu) =&\, 1 + \frac{\al_s}{\pi} \biggl[ C_A\bigg( -  \ln^2\bigg(\f{\mu}{p_T R} \bigg)  - \frac{5}{24} + \frac{3\pi^2}{8}\bigg) 
\nn \\ & \quad 
+ \frac{\beta_0}{2} \bigg(- \ln\bigg(\f{\mu}{p_T R} \bigg)  - \frac{23}{12}\bigg) \biggr]
\,,\\[.5em]
C^{\in \rm{gr}}_q(\tg p_T R\, \mu)=&\,1+\f{\as C_F}{\pi} \bigg[ \ln^2\left(\frac{\mu}{\theta_g p_T R}\right) 
+ \frac32 \ln\left(\frac{\mu}{\theta_g p_T R}\right) + \frac{13}{4}-\frac{3\pi^2}{8} \bigg]
\,, \\
C^{\in \rm{gr}}_g( \tg p_T R, \mu)=&\,1+\f{\as}{\pi} \bigg[ C_A  \ln^2\left(\frac{\mu }{\theta_g p_T R }\right)   + \frac{\beta_0}{2} \ln\left(\frac{\mu}{\theta_g p_T R}\right) 
 \nn\\&
+C_A \left(\frac{67}{18}-\frac{3\pi^2}{8}\right)-T_F n_f\frac{23}{18} \bigg]
\,, \\[.5em]
\CS_{i,G}(\zc\tg^{1+\beta} p_T R, \beta, \mu )=& \,1+  \f{\alpha_s C_i }{ \pi (1+\beta)}\bigg[
-\ln ^2 \bigg( \f{\mu}{\zc\tg^{1+\beta} p_T R} \bigg) +\f{\pi^2}{24} \bigg]
\,.
\end{align}
The functions $\CS_{i,X}$ and $S_{i,Z}$ are given in \eq{functions_ISD1} above. We have verified that combining these ingredients agrees with the fixed-order result for $\Delta \cG_i$ in \sec{fixedorder_softdrop}, in the limit where the factorization holds.

To resum the logarithms of $\ed$, $\theta_g$ and $\zc$, we evaluate each of the ingredients in the factorization formula in \eq{refact_A} at their natural scale,
\begin{align}\label{eq:scales_SD}
\mu_H\sim &\, p_TR\,,\quad\mu_{C^{\in {\rm gr}}}\sim\theta_g p_T R\,,\quad\mu_{\CS_G}\sim z_{\rm cut}\theta_g^{1+\beta}p_T R\,,\quad \mu_{S_Z}\sim \Delta_E\, p_T R\,,
\nn\\[.5em]
\mu_{\CS_X}\sim &\, \Delta_E^{(1+\beta)/\beta}z_{\rm cut}^{-1/\beta}p_T R \,,
\end{align}
and evolve them to a common scale $\mu$. The RG equations for the new ingredients are
\begin{align}
\mu \frac{\mathrm{d}}{\mathrm{d} \mu}\, \tilde{H}_{i}\left(p_{T} R, \mu\right)=&\,\gamma_{i}^{\tilde{H}}\left(p_{T} R, \mu\right) \tilde{H}_{i}\left(p_{T} R, \mu\right) 
\,\\[.5em]
\mu \frac{\mathrm{d}}{\mathrm{d} \mu}\, C_{i}^{\in \mathrm{gr}}\left(\theta_{g} p_{T} R, \mu\right)=&\,\gamma_{i}^{C^{\in \mathrm{gr}}}\left(\theta_{g} p_{T} R, \mu\right) C_{i}^{\in \mathrm{gr}}\left(\theta_{g} p_{T} R, \mu\right) 
\,,\\[.5em]
\mu \frac{\mathrm{d}}{\mathrm{d} \mu}\, \CS_{i,G}(z_{\mathrm{cut}}\theta_{g}^{1+\beta} p_{T} R, \beta, \mu)=&\,\gamma_{i}^{\CS_{i,G}}(z_{\mathrm{cut}}\theta_{g}^{1+\beta} p_{T} R, \beta, \mu)\, \CS_{i,G}(z_{\mathrm{cut}}\theta_{g}^{1+\beta} p_{T} R, \beta, \mu) \,,
\end{align}
and the one-loop expressions for these anomalous dimensions are given in appendix~\ref{app:anom}.

\subsubsection{Regime B~\label{sec:SD_regimeB}}

The Lund diagram for regime $B$ is shown on the right side of fig.~\ref{fig:LundSD}. Since in this case $\theta_g>(\Delta_E/z_{\rm cut})^{1/\beta}$, there is a white triangle between the dashed lines representing the measurement of $\ed$ and the soft drop criterion which is not vetoed. This triangle corresponds to emissions that fail the soft drop criterion and would give a value of $\ed$ that is too large, except that the soft drop procedure has already terminated. Different than in regime $A$, the measurements of $\Delta_E$ and $\theta_g$ are not independent here. For regime $B$ we work differentially in the groomed radius $\theta_g$. One emission sets the value of $\theta_g$ and other emissions must be outside the shaded region with boundary $\theta/R=\theta_g$ and $z = \ed$. At LL accuracy, the resummed result is  
\begin{align}
\label{eq:LLregimeB}
  \tilde \cG_{i,B}^{\rm SD}\bigl(\ed, p_T R, \zc,\bt, \theta_g, \al_s(\mu)\bigr) &\stackrel{{\rm LL}}{=} -\frac{\alpha_s C_i}{\pi} \frac{2}{\theta_g} \ln (\zc \theta_g^\bt)  \frac{\df}{\df \ed} \exp\bigg\{- \frac{2\al_s C_i}{\pi}  \ln \ed \ln \tg \bigg\}
\,.\nn\\
\end{align}

The LL result can again be extended to NLL$'$ using SCET. We identify a total of five modes  that contribute. Four of them correspond to the corners of the shaded region in \fig{LundSD},  indicated by red points. In addition, there is a mode indicated by the green point which is located at the intersection of the grooming condition and $\theta/R=\theta_g$. Since only the emission that sets $\theta_g$ is sensitive to the soft drop condition, the collinear-soft $\CS_{i,G}'$ mode only contributes if it sets $\theta_g$ and has a single emission (see the discussion in ref.~\cite{Cal:2019gxa} and \sec{Cond_Prob}). We find that the extension to NLL$'$, including non-global logarithms, can be written as 
\begin{align} \label{eq:refact_B}
&\tilde \cG_{i,B}^{\rm SD}\bigl(\ed , p_T R, \zc,\bt, \theta_g, \al_s(\mu)\bigr) 
\nn \\ & \quad \stackrel{{\rm NLL'}}{=}
 \tilde H_i(p_{T} R,\mu) \bigg[\frac{\df}{\df \theta_g}\, C_i^{\in \rm gr}(\theta_g p_T R,\mu)
 +C_i^{\in \rm gr}(\theta_g p_T R,\mu)\,\CS_{i,G}'\bigl(\theta_g, \zc \theta_g^\bt p_T, \bt, \mu\bigr)\bigg]
  \nn \\ & \qquad
\times \int\! \df \ed'\,  \CS_{i,Z}(\ed', \tg p_T R, \mu)\, S_{i,Z}(\ed - \ed',p_T R,\mu) 
 \CS_i^{\rm NG+AC}(\ed) S_i^{\rm NG}(\ed) 
\,.\end{align}
The function $\CS_{i,G}'$ at one-loop order is given by derivative of $\CS_{i,G}$ (in regime $A$) with respect to $\theta_g$. However, emissions in $\CS_{i,G}'$ which do not set $\theta_g$ are scaleless and hence, the associated RG equation is given by
\begin{equation}
\mu \frac{\mathrm{d}}{\mathrm{d} \mu} \CS_{i,G}^{\prime}(\theta_{g}, z_{\mathrm{cut}} \theta_{g}^{\beta} p_{T}, \beta, \mu)=-\frac{\mathrm{d}}{\mathrm{d} \theta_{g}} \gamma_{i}^{C^{\in \mathrm{gr}}}\left(\theta_{g} p_{T} R, \mu\right) \,,
\end{equation}
as required for consistency of the factorization formula in \eq{refact_B}. The new collinear-soft function $\CS_{i,Z}$ in \eq{refact_B} is at one-loop order given by
\begin{align} \label{eq:B_CSZ}
\CS_{i,Z}(\Delta_E, \tg p_T R , \mu)=&\,\delta(\Delta_E)+ \f{\as C_i}{\pi} \bigg\{ -2\left[ \f{\ln \ed}{\ed} \right]_++ \f{2}{[\ed]}_+ \ln \left( \f{\mu}{\tg p_T R}\right)  \nn \\
&+\delta(\ed) \bigg[- \ln^2 \left( \f{\mu}{\tg p_T R}\right) +\f{\pi^2}{24}\bigg] \bigg\} \,.  
\end{align}
It satisfies the RG equation
\begin{equation}
\mu \frac{\mathrm{d}}{\mathrm{d} \mu} \CS_{i, Z}\left(\Delta_{E}, \theta_g p_{T} R, \mu\right)=\int \mathrm{d} \Delta_{E}^{\prime}\, \gamma_{i}^{\CS_{Z}}\left(\Delta_{E}-\Delta_{E}^{\prime}, \theta_g p_{T} R, \mu\right) \CS_{i, Z}\left(\Delta_{E}^{\prime}, \theta_g p_{T} R, \mu\right)\,,
\end{equation}
with the anomalous dimension again given in appendix~\ref{app:anom}. The characteristic scales of the ingredients in the factorization formulae for regime $B$ are given by
\begin{align}\label{eq:scales_SD-B}
\mu_H\sim &\, p_TR\,,\quad\mu_{C^{\in {\rm gr}}}\sim\theta_g p_T R\,,\quad\mu_{\CS_G'}\sim z_{\rm cut}\theta_g^{1+\beta}p_T R\,,\quad \mu_{S_Z}\sim \Delta_E\, p_T R\,,
\nn\\[.5em]
\mu_{\CS_Z}\sim&\, \Delta_E\,\theta_g\, p_T R \,.
\end{align}

The structure of the non-global logarithms is very similar to regime A, except that the argument of $\CS_i^{\rm NG+AC}$, describing the NGLs at the groomed boundary, is now $\ed$ instead of $\zc \theta_g^\bt$. To understand this change, remember that these NGLs arise from correlated emissions inside and outside the groomed jet radius. Emissions inside are unconstrained (i.e.~have energy of order $p_T$), while the energy of emissions outside is constrained to be below $\ed p_T$, which for regime B is more restrictive than the grooming condition. 

In principle one can also consider the intermediate regime $\theta_g \sim (\Delta_E/z_{\rm cut})^{1/\beta}$. 
The effective theory corresponding to this intermediate case can be obtained from regime A, for which $\tilde{\mathcal{S}}_X$ and $\tilde{\mathcal{S}}_G$ merge, or regime B, for which $\tilde{\mathcal{S}}_Z$ and $\tilde{\mathcal{S}}_G'$ merge into one function. The anomalous dimensions are smooth in this merging, but the function describing these merged modes in the factorization theorem can have different fixed-order expressions (due to terms that become power suppressed in regime A or B). The matching between A and B is thus automatically valid at NLL accuracy, and this intermediate case only needs to be considered if one wants to ensure NLL$'$ accuracy throughout the intermediate matching regime.

\subsection{Conditional probability~\label{sec:Cond_Prob}}
\begin{figure}[t]
    \centering
    \includegraphics[width=0.3\textwidth]{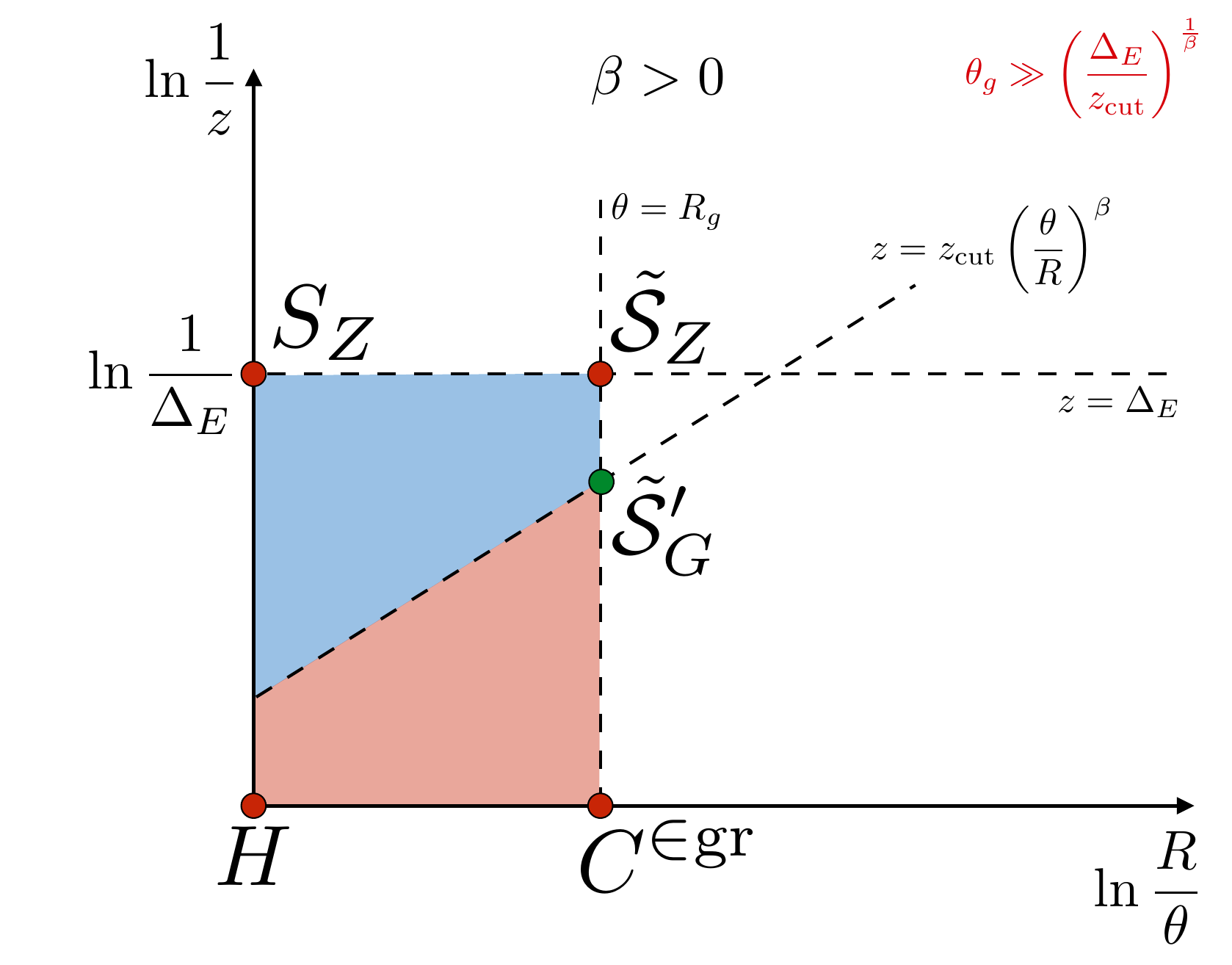} \hfill 
      \includegraphics[width=0.3\textwidth]{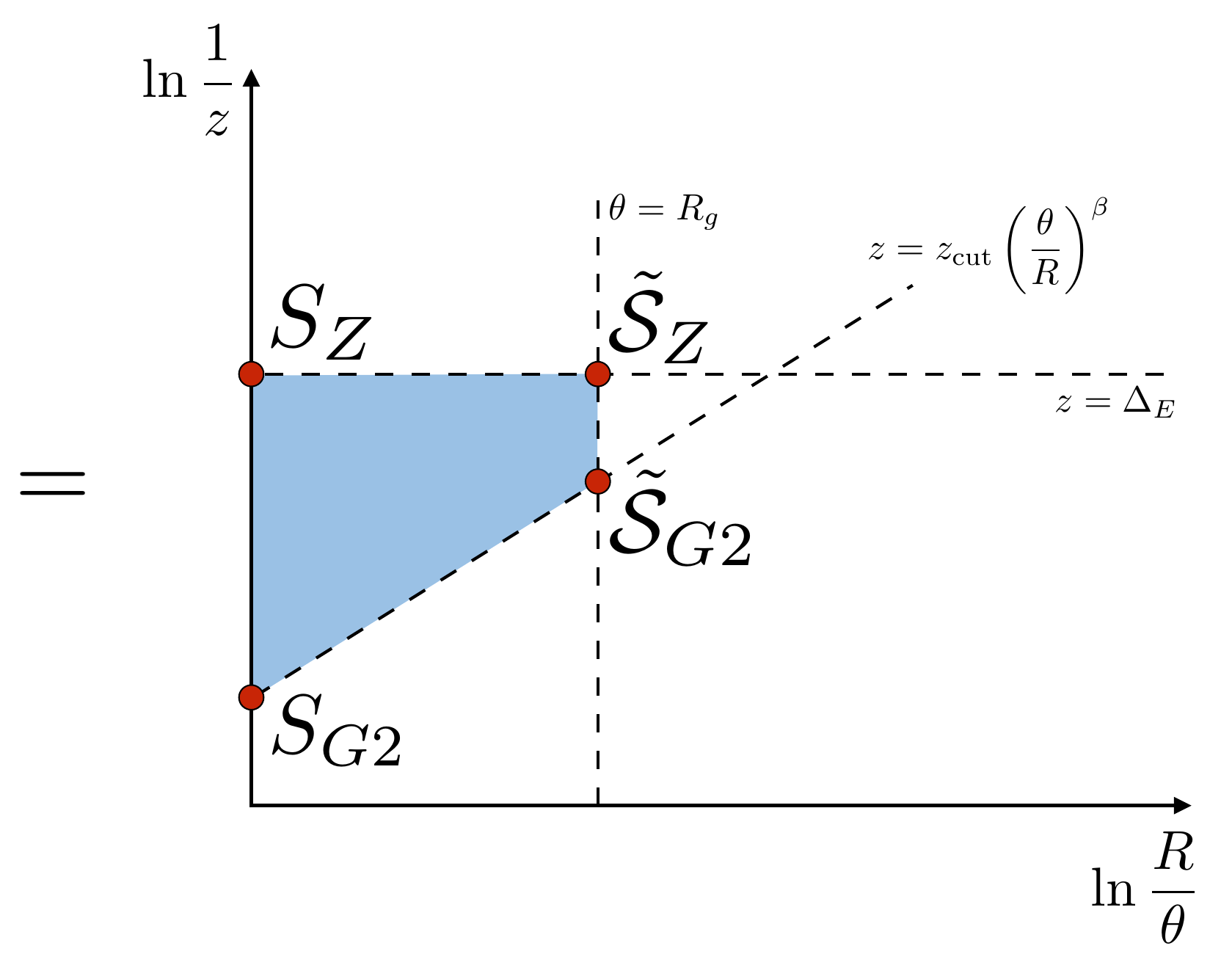} \hfill
     \includegraphics[width=0.3\textwidth]{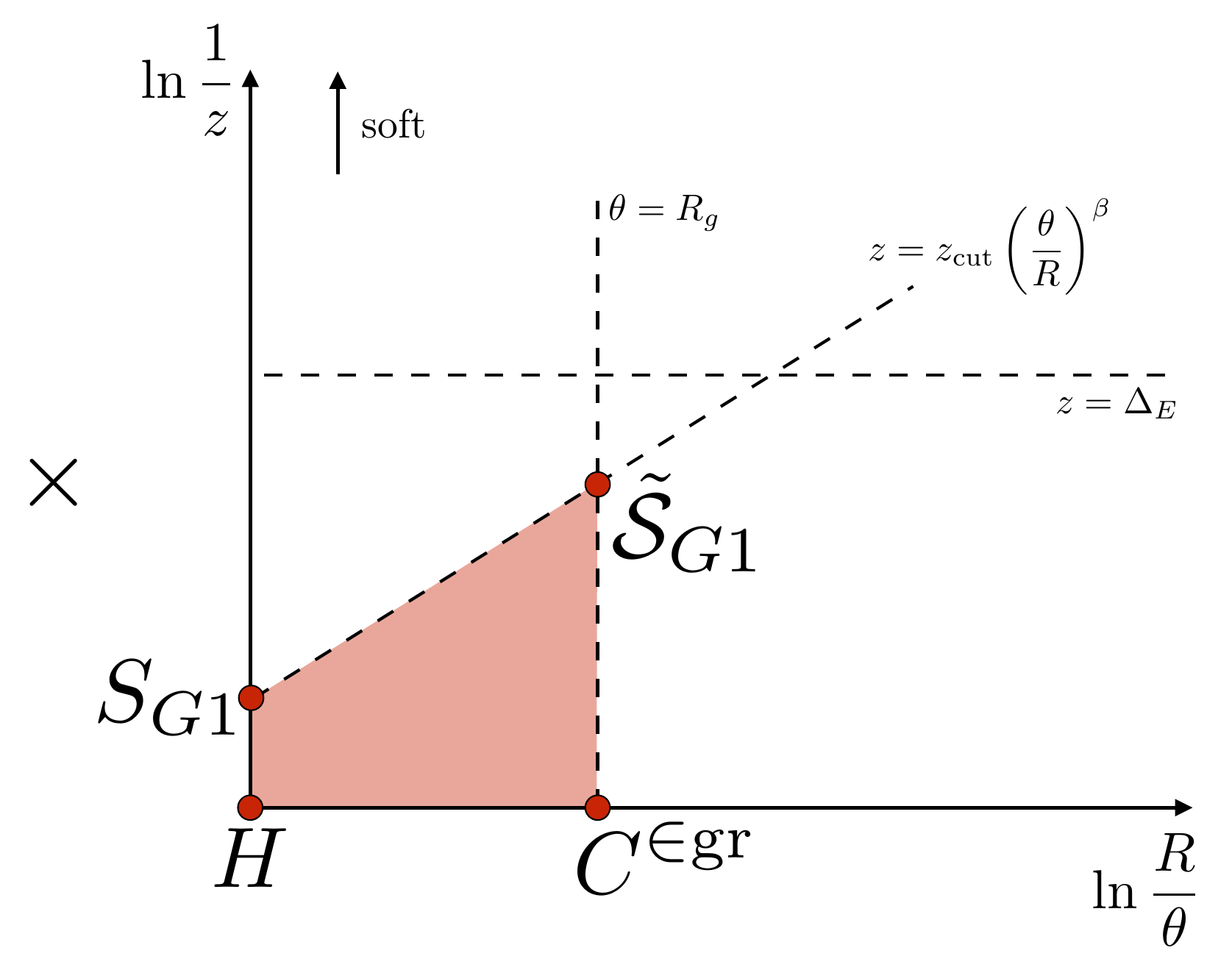} \hfill \phantom{.} \\
    \caption{Lund diagram analysis in which regime B is written as a conditional probability for $\ed$ given a value of $\theta_g$, and the $\theta_g$ cross section.  \label{fig:CondProb}}
\end{figure}

An alternative way of obtaining the global structure of the factorization for regime B is by means of a conditional probability, which is how the jet energy drop was calculated in ref.~\cite{Larkoski:2014wba} at (modified) LL accuracy. We describe how this arises in our SCET framework, providing additional insight on the origin of $\CS_{i,G}'$.

The starting point is to write the double differential $\cG_{i,B}^{\rm SD}\bigl(\ed, \theta_g \bigr)$ as 
\begin{align}
 \tilde \cG_{i,B}^{\rm SD}\bigl(\ed, \theta_g \bigr)=  \tilde {\cG} _{i}^{\rm SD}(\ed | \, \tg) \times \tilde{\cG}_{i}^{\rm SD} (\tg),
 \end{align} 
where for brevity we suppress arguments other than $\ed$ and $\theta_g$. $ \tilde {\cG} _{i}^{\rm SD}(\ed | \, \tg)$ denotes the conditional probability to obtain a certain energy drop $\ed$ for a given $\tg$, i.e.~it treats $\tg$ as a fixed parameter just like the jet radius $R$. $\tilde{\cG}_{i}^{\rm SD} (\tg)$ represents the probability distribution of having a specific value of $\tg$, calculated to NLL$'$ in ref.~\cite{Kang:2019prh}. 

This decomposition can also be understood in terms of the Lund plane, as seen in \fig{CondProb}. Here the conditional probability is depicted by the Lund plane with a blue vetoed area, while the one with a red vetoed area represents the $\tg$ distribution. In contrast to the double differential case, for the conditional probability emissions outside the groomed radius that pass the SD condition are now allowed since $R_g$ is treated as a fixed parameter. The relevant modes are again indicated as red dots on the corners, leading to the following factorization for each of these pieces: 
\begin{align} \label{eq:CP_refact}
 \tilde {\cG} _{i}^{\rm SD}(\ed | \, \tg )& =  \CS_{i,G2}\bigl( \zc \theta_g^{1+\bt} p_T, \bt, \mu\bigr) S_{i,G2}\bigl( \zc  p_T R, \bt, \mu\bigr)  \\ & \quad
\times \int\! \df \ed'\,  \CS_{i,Z}(\ed', \tg p_T R, \mu)\, S_{i,Z}(\ed - \ed',p_T R,\mu)\,,   \nn \\ 
 \tilde{\cG}_{i}^{\rm SD} (\tg) &= \f{\df}{\df \tg} \bigl[ \tilde H_i(p_{T} R,\mu) C_i^{\in \rm gr}(\theta_g p_T R,\mu)\,\CS_{i,G1}\bigl( \zc \theta_g^{1+\bt} p_T, \bt, \mu\bigr) S_{i,G1}\bigl( \zc  p_T R, \bt, \mu\bigr) \bigr].    \nn 
\nn\end{align}

We now note that the modes $S_{G1}$ and $S_{G2}$ are defined in a similar way: the former vetoes emissions that pass the grooming condition, whereas the latter vetoes emissions that fail said condition. This translates to a minus sign difference in the $\as$ term of these two functions, i.e.~$S_{G1} \times S_{G2}= 1 +\mathcal{O}(\as^2)$, removing their contribution. Of course this holds also for multiple independent emissions, exponentiating the one-loop soft function, which is why there is no such mode in the factorization for regime B in \eq{refact_B}. Similarly, for $\CS_{G1}$ and $\CS_{G2}$ we have $\CS_{G1} \times \CS_{G2}= 1 +\mathcal{O}(\as^2)$. However, in this case the derivative with respect to $\tg$ in \eq{CP_refact} between $\CS_{i,G1}$ and $\CS_{i,G2}$ prevents them from cancelling, leaving exactly $\CS_{i,G}'$ as remainder. Thus we reproduce the factorization theorem for regime B in \eqref{eq:refact_B}, apart from non-global logarithms.

\subsection{$\beta=0$ and Sudakov safety~\label{sec:SDbeta0}}

\begin{figure}[t]
 \includegraphics[width=0.48\textwidth]{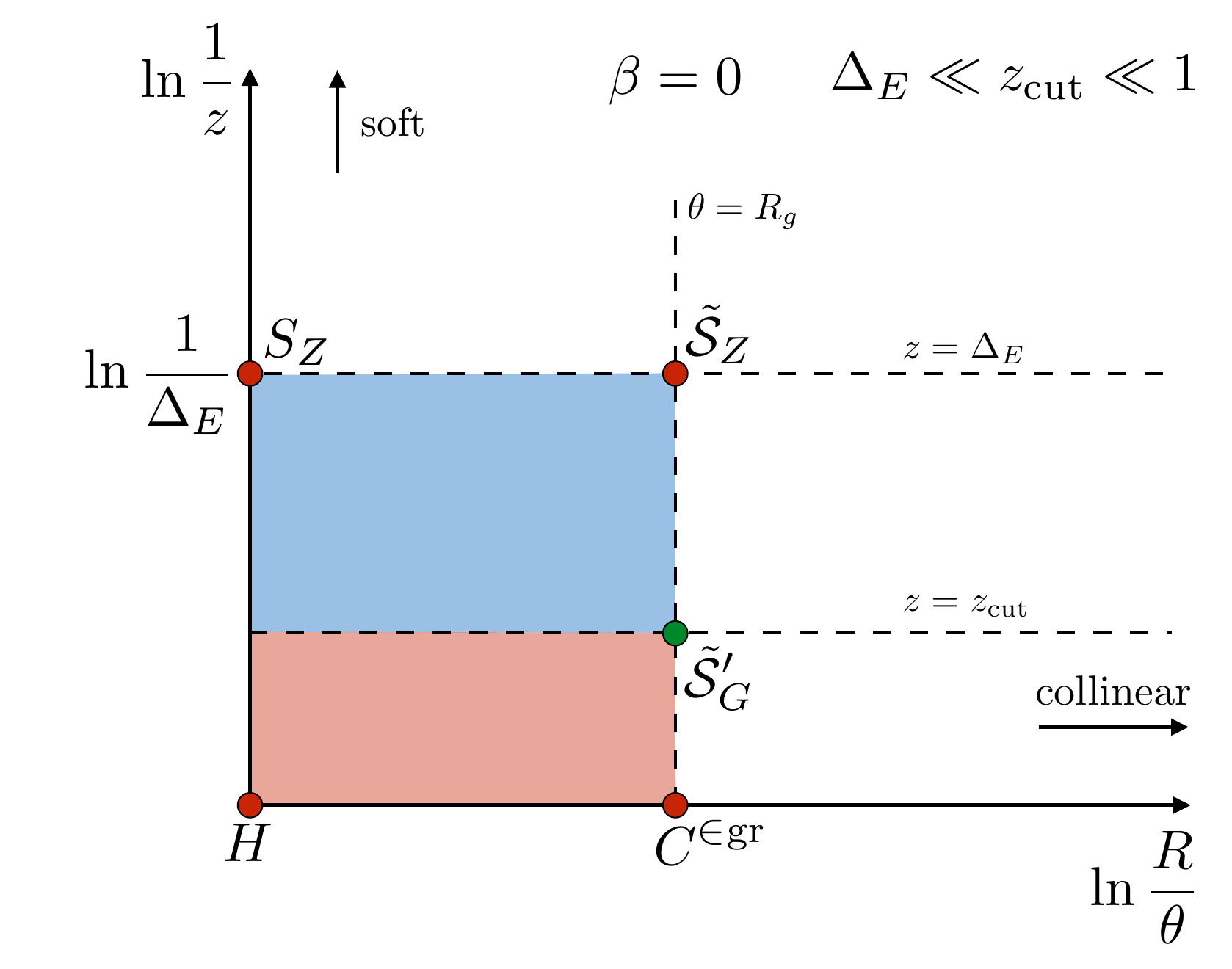}  
 \centering
    \caption{The Lund diagram for the jet energy drop with soft drop grooming and $\beta=0$.~\label{fig:LUNDb0}}
\end{figure}

Soft drop with $\beta=0$ corresponds to the modified mass drop tagger of ref.~\cite{Dasgupta:2013ihk}. It is a special case, because the jet energy drop is not IRC safe, as is clear when taking $\beta \to 0$ in the fixed-order result for $\cG^{\rm SD}_i$ in \eq{grJF2}. With a cut $\theta_g > \theta_g^c > 0$, it is IRC safe. As was found in ref.~\cite{Larkoski:2014wba}, the jet energy drop is Sudakov safe, allowing us to safely take the $\beta \to 0$ limit of the cross section in which the logarithms of $\Delta_E$ and $\theta_g$ are jointly resummed. This works because the Sudakov factor arising from the resummation regularizes the divergence. (See ref.~\cite{Larkoski:2015lea} for a discussion of Sudakov safety in the context of the momentum sharing fraction $z_g$.) In the absence of a cut on $\theta_g$, the jet energy drop is particularly sensitive to nonperturbative effects, as discussed in \sec{SDnonp}.

For $\bt=0$, only regime B contributes, which was shown in the right panel of \fig{LundSD} for $\beta>0$. In this case, the line corresponding to the soft drop criterion now has slope $\bt=0$, as shown in \fig{LUNDb0}. It is clear from the figure that the $\theta_g$ measurement is necessary to regulate the collinear divergence for $\Delta_E$ measurements at $\beta = 0$. In fact, we can immediately take $\bt\to 0$ in the ingredients in this regime, and there is no difficulty in obtaining resummed predictions. The IR divergence is in regime A, whose range of applicability is shrunk to the point $\theta_g = 0$ for $\bt \to 0$. In the resummed cross section, it is Sudakov suppressed. With the collinear divergence regulated by the $\theta_g$ measurement, we can obtain a $\Delta_E$ distribution by integrating over a desired range of $\theta_g$. It is worth noting that LL formulation given in \eq{LLregimeB} yields an $\alpha_s$ independent result when integrated over the entire range of $\theta_g$,
\bea
\int_0^1 d\theta_g \, \tilde \cG_{i,B}^{\rm SD}\bigl(\ed, p_T R, \zc,\bt=0, \theta_g, \al_s(\mu)\bigr) &\stackrel{{\rm LL}}{=}\frac{\ln z_{\rm cut}}{\ln \Delta_E}\,.
\eea
This surprising feature was already pointed out in ref.~\cite{Larkoski:2014wba}, where they also kept the subleading terms in the splitting functions, compared to the LL expression in \eq{LLregimeB}. Our full NLL$'$ result includes many more contributions but its analytic expression is not particularly tractable. 

\subsection{$\theta_g$ and nonperturbative effects \label{sec:SDnonp}}

We will now discuss the size of nonperturbative effects, considering the case where we completely integrate over $\theta_g$, as well as imposing a minimum cut on $\theta_g$. As we will see below, introducing a cutoff reduces the sensitivity to nonperturbative effects. It is also advantageous from an experimental point of view, if the tracking efficiency is limited at small $\theta_g$~\cite{Sirunyan:2017bsd}. 
\begin{figure}[t]
     \hfill \includegraphics[width=0.495\textwidth]{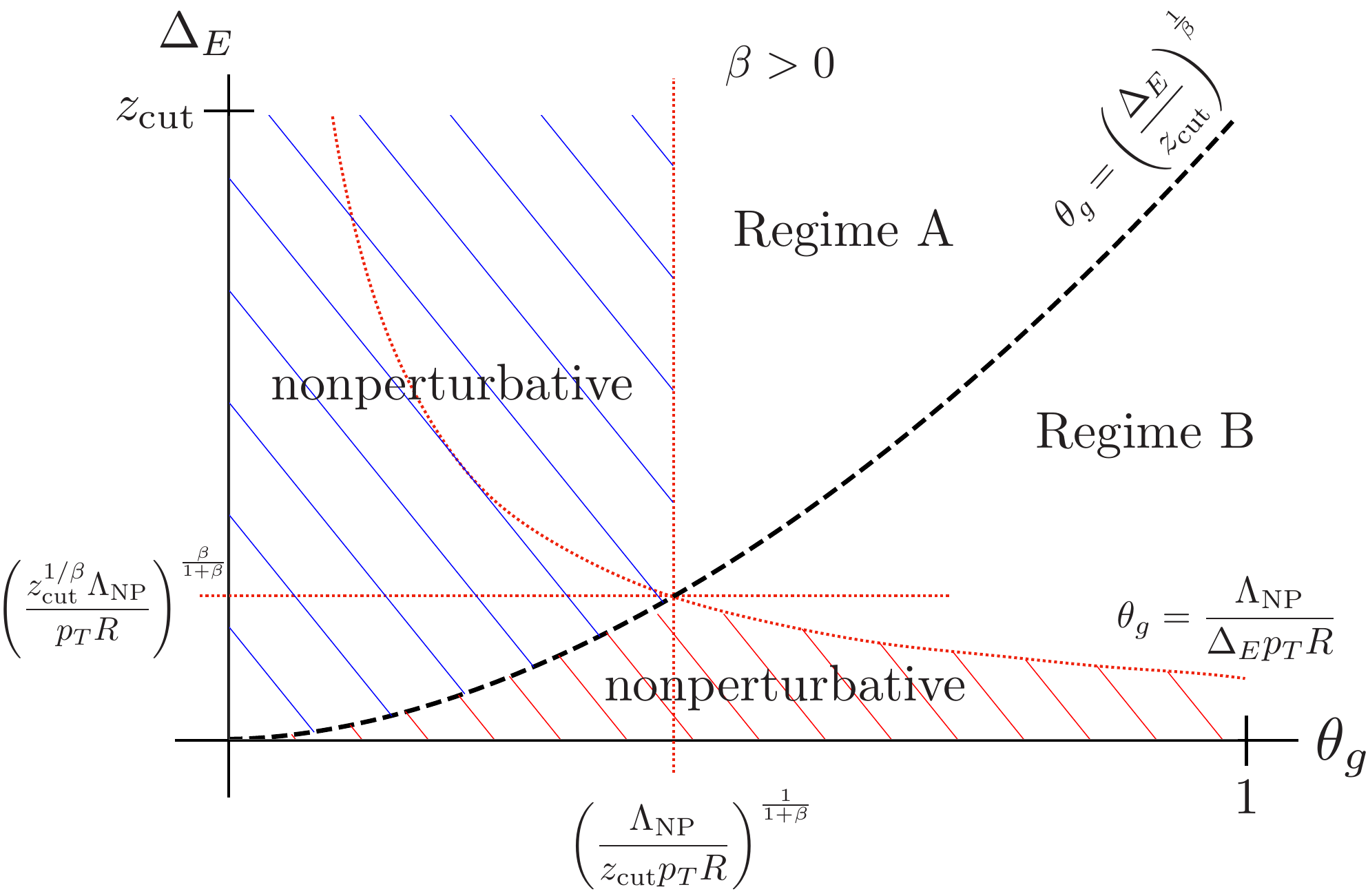} \hfill 
     \includegraphics[width=0.495\textwidth]{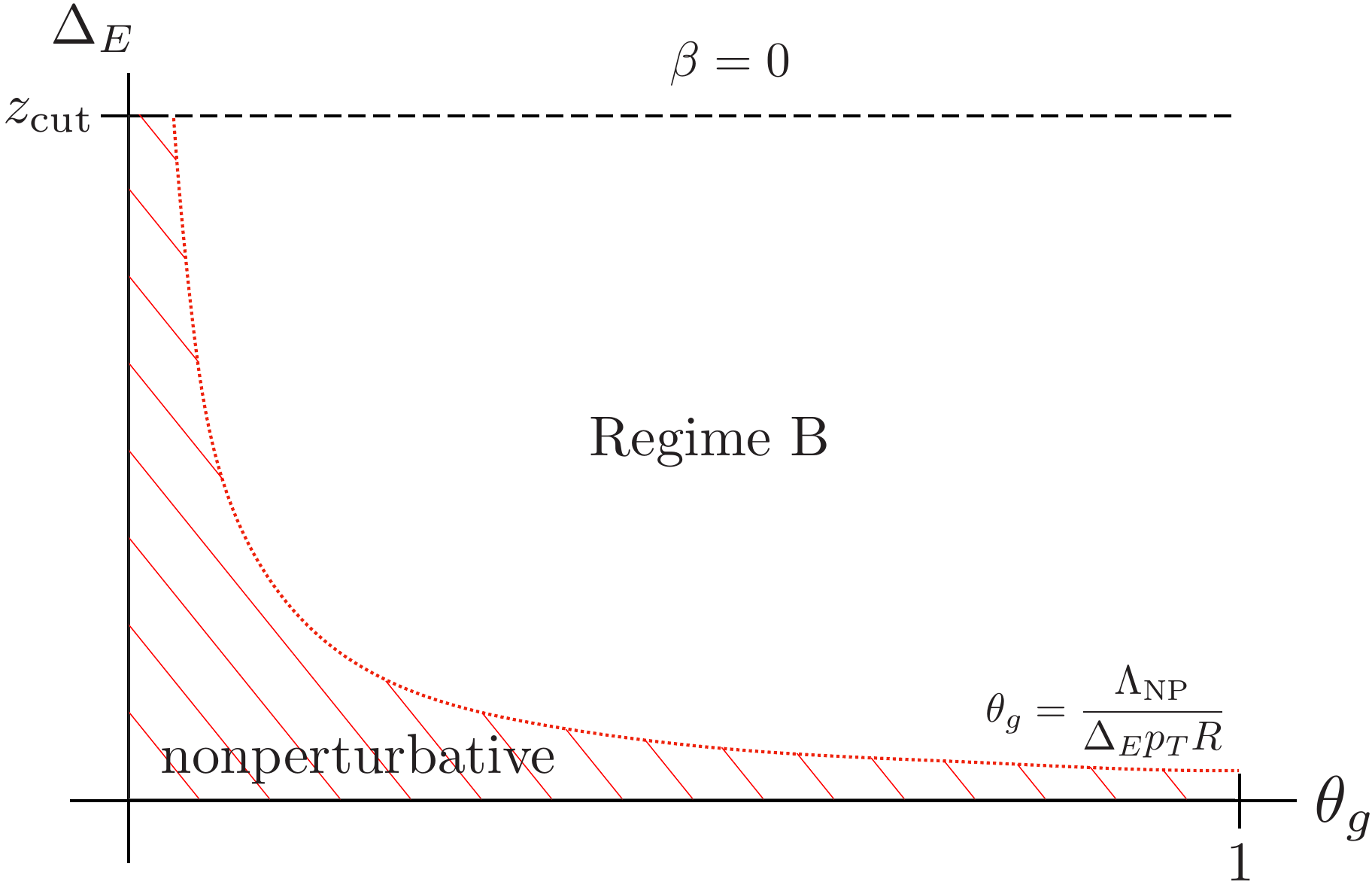} \hfill \phantom{.} 
    \caption{The nonperturbative regions for soft drop with $\beta >0$ (left) and $\beta = 0$ (right). Plots are not to scale to highlight the nonpertubative regions. }~\label{fig:nonpert2D}
\end{figure}

As usual, we determine the onset of the nonperturbative region by considering the softest scales involved in the factorization formulas in \eqs{refact_A}{refact_B}. The softest scales in regime A and regime B are $\mu_{\tilde{\mathcal{S}}_G}\sim z_{\rm cut} \theta_g^{1+\beta}p_T R$ and $\mu_{\tilde{\mathcal{S}}_Z}\sim\Delta_E \theta_g p_T R$, respectively. Therefore, within regime $A$ we use the relation
\bea
\theta_g < \Bigl(\frac{\Lambda_{\rm NP}}{z_{\rm cut} p_T R}\Bigr)^{\frac{1}{1+\beta}}\quad \text{and} \quad \theta_g < \Bigl(\frac{\Delta_E}{z_{\rm cut}}\Bigr)^{\frac{1}{\beta}} \,
\eea 
to determine the nonperturbative region, while for regime $B$
\bea
\label{eq:npregimeB}
\Bigl(\frac{\Delta_E}{z_{\rm cut}}\Bigr)^{\frac{1}{\beta}} < \theta_g < \frac{\Lambda_{\rm NP}}{\Delta_E p_T R} \,.
\eea
As discussed in \sec{SDbeta0}, for $\beta = 0$ we only have regime $B$, and \eq{npregimeB} simplifies to
\bea
\theta_g < \frac{\Lambda_{\rm NP}}{\Delta_E p_T R}\,.
\eea 
These nonperturbative regions are illustrated in \fig{nonpert2D}.

We now highlight some aspects of nonperturbative contributions to the $\Delta_E$ distribution resulting from integrating over (a range of) $\theta_g$.\footnote{Usually one does not associate nonperturbative corrections with variables that are integrated over, but in this case $\Delta_E$ and $\theta_g$ are intertwined by the double-differential factorization. Therefore, the integration of $\theta_g$ does not remove the nonperturbative effects coming from factorization scales associated with the $\theta_g$ measurement.} First, we note that the entire $\Delta_E$ distribution receives a nonperturbative contribution from the $\theta_g$ integration when $\theta_g^{\rm cut} < (\Lambda_{\rm NP}/(z_{\rm cut} p_T R))^{1/(1+\beta)}$, due to region A. If $\theta_g^{\rm cut}$ is above this threshold, the onset of the nonperturbative regions is instead determined by region B (corresponding to the red regions in \fig{nonpert2D}). In this case the nonperturbative contributions become small for much of the $\Delta_E$ distribution, for both $\beta>0$ and $\beta=0$, allowing for a purely perturbative calculation. In our numerical studies presented in \sec{numerics_softdrop}, we always indicate the corresponding onset of the nonperturbative region by a vertical dotted line. 
If the cut is chosen such that $\theta_g^{\rm cut} < (\Lambda_{\rm NP}/(z_{\rm cut} p_T R))^{1/(1+\beta)}$, then we have some nonperturbative contributions for $\Delta_E$ values even above the indicated vertical line.

\subsection{Profile functions and scale variations~\label{sec:SDprofiles}}

In this section we describe our choice of central scales, as well as the variations used to assess the perturbative uncertainty. We start with regime A, which is particularly simple because no function in the factorization formula depends on both $\ed$ and $\tg$, whereas for regime B we need to design two-dimensional profile scales.

In regime A there are no scales that simultaneously depend on both $\Delta_E$ and $\theta_g$. Consequently, we can take the same central scales and scale variations for $S_Z$ and $\CS_X$ as for iterated soft drop, see \sec{ISDprof}. 
For the additional scales associated with the $\tg$ measurement (corresponding to the red region in the Lund plane in the left panel of \fig{LundSD}),  we take
\bea
\mu_{\CS_G}^{\text{cent}} & = f_{\text{pro}}(z_{\rm cut}\theta_g^{1+\beta}p_T R; \Lambda_{\rm freeze})\,,\notag\\
\mu_{C^{\in {\rm gr}}}^{\text{cent}} & = \biggl[\Bigl(\frac{\mu_{\CS_G}^{\text{cent}}}{z_{\rm cut}}\Bigr)^{1/\beta} p_T R\biggr]^{\beta/(1+\beta)}\,,\notag\\
\mu_H^{\text{cent}} &= p_T R\,.
\label{eq:SDAcan}
\eea
Here $f_{\text{pro}}(x;x_0)$ ensures that we avoid the Landau pole. Its expression is given in \eq{ISDfreeze}, and we take the same value $\Lambda_{\rm freeze}=0.2$ GeV as for iterated soft drop. By expressing $\mu_{C^{\in {\rm gr}}}^{\text{cent}}$ in terms of $\mu_{\CS_G}^{\text{cent}}$, we ensure that they stop running simultaneously. 
QCD scale uncertainties are obtained by varying the scales that also appear for ISD in the same way as described in sec.\ \ref{sec:ISDprof}, while identifying the variation of $\mu_H^{\text{cent}}$ with the variation of $\mu_{\mathcal{G}}^{\text{cent}}$ as their canonical forms are identical. We also vary the new $\theta_g$ dependent scales of $\mu_{\CS_G}^{\text{cent}}$ and $\mu_{C^{\in {\rm gr}}}^{\text{cent}}$ individually up and down by a factor of 2 around their central value. The perturbative uncertainty band is obtained by taking the envelope of all the variations. 

\begin{figure}[t]
\centering
\includegraphics[width=0.5\textwidth]{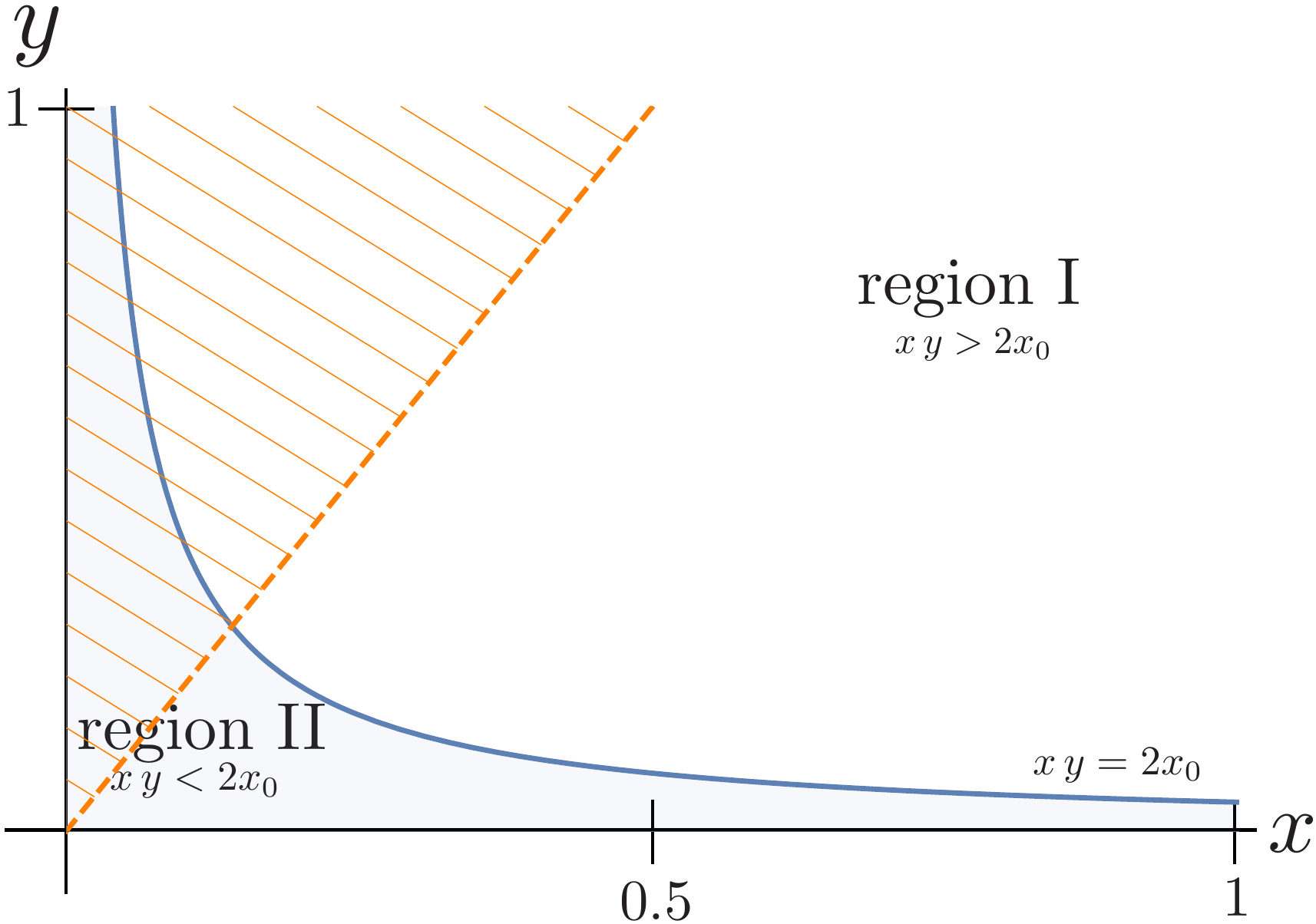} 
\caption{The two regions defined in the two-dimensional profile function given in \eq{SDBfreeze}. All $\theta_g$ and $\Delta_E$-dependent scales take on their canonical values in region I and begin to freeze once they enter region II. The involved scales are properly defined only within the factorization of regime B, satisfying the condition $\Delta_E < z_{\rm cut}\theta_g^\beta$. This condition is indicated by the orange hatched region for $z_{\rm cut} = 0.5$ and $\beta=1$, identifying $x$ and $y$ with $\Delta_E$ and $\theta_g$, respectively. ~\label{fig:freeze2D} }
\end{figure}

Regime B presents a further complication, since the $\Delta_E$ and $\theta_g$ measurements can no longer be treated independently. In this case the softest scale is $\mu_{\CS_Z} \sim \Delta_E\,\theta_g\,p_T R$, which thus becomes nonperturbative before any other scale. As this scale depends on both $\Delta_E$ and $\theta_g$, it can run into the nonperturbative region by either $\Delta_E$ or $\theta_g$ becoming small, corresponding to region II in  \fig{freeze2D}. Therefore, we now need a two-dimensional profile to implement the freezing of the scale in the nonperturbative region 
\bea
\label{eq:SDBfreeze}
f^{2d}_{\text{pro}}(x,y;x_0)=&\left\{
    \begin{array}{ll}
      x\,y\hspace{3.7cm} x\,y>2x_0 \hspace{1cm} \text{region I}\,,\\ 
      x_0[1+(x\,y/x_0)^2/4] \hspace{1cm}x\,y\leq 2x_0 \hspace{1cm}\text{region II}\,.
    \end{array}
  \right.
\eea
With this new profile function, we define the central  scale for $\mu_{\CS_Z}$ as
\bea
\mu_{\CS_Z}^{\rm cent} = f^{2d}_{\rm pro}\Bigl(\Delta_E, \theta_g; \frac{\Lambda_{\rm freeze}}{p_T R}\Bigr)\,p_T R\,,
\eea
to smoothly freeze $\mu_{\CS_Z}$ at $\Lambda_{\rm freeze}$. The condition $\Delta_E < z_{\rm cut} \theta_g^\beta$ is necessary to ensure that we stay within regime B, and is indicated in \fig{freeze2D} by the orange hatched region for $z_{\rm cut} = 0.5$ and $\beta = 1$, when $x$ and $y$ of \eq{SDBfreeze} are identified with $\Delta_E$ and $\theta_g$, respectively.

When $\mu_{\CS_Z}^{\rm can}$ starts freezing in region II, we want to ensure that all other scales also stop running. To accomplish this, we define the following profile function
\bea
&g^{2d}_{\text{pro}}(x,y;A,B,c_0,x_0)
\nn\\& \quad =
\left\{
    \begin{array}{ll}
      x^{1+B}\hspace{7.0cm} &x\,y>2x_0 \hspace{0.7cm}\text{region I}\,,\\ 
      c_0\biggl[1+\Bigl(\frac{x\,y}{2x_0}\Bigr)^{1+A}\biggr(\frac{1}{c_0}\Bigl(\frac{2x\,x_0}{y}\Bigr)^{\frac{1+B}{2}} - 1\biggr)\biggr] &x\,y\leq 2x_0 \hspace{0.7cm} \text{region II}\,,
    \end{array}
  \right.
\eea
where $A> -1$ controls the rate at which scale freezes to $c_0$ in region II. The parameter $B$ is chosen according to the canonical behaviors of different scales in region I. The profile function is continuous everywhere and smoothly approaches $c_0$ as $x$ or $y$ become small. We then take the remaining central scales to be
\bea
\mu_H^{\text{cent}} & = p_T R\,,\notag\\
\mu_{C^{\in {\rm gr}}}^{\text{cent}} & = g^{2d}_{\text{pro}}\Bigl(\theta_g,\Delta_E; 0,0,\lambda_{C^{\in {\rm gr}}},\frac{\Lambda_{\rm freeze}}{p_T R}\Bigr)p_T R\,,\notag\\
\mu_{\CS_G'}^{\text{cent}} &= g^{2d}_{\text{pro}}\Bigl(\theta_g,\Delta_E; 0,\beta,\lambda_{\CS_G'},\frac{\Lambda_{\rm freeze}}{p_T R}\Bigr) z_{\rm cut} p_T R\,,\notag\\
 \mu_{S_Z}^{\text{cent}} &= g^{2d}_{\text{pro}}\Bigl(\Delta_E,\theta_g; 0,0,\lambda_{S_Z},\frac{\Lambda_{\rm freeze}}{p_T R}\Bigr) p_T R\,.
\label{eq:SDBcan}
\eea
This ensures that all $\theta_g$ and $\Delta_E$-dependent scales take on their canonical values in region I, given in \eq{scales_SD-B}, while freezing them in the same region II. Because different scales enter region II with different values, it is natural to freeze them to different $\lambda_i$ to maintain their relative hierarchy. We take $\lambda_i$ to be the average value of the scale $x^{1+B}$, with $x$ and $B$ defined in \eq{SDBcan}, along $\Delta_E\,\theta_g = 2\frac{\Lambda_{\rm freeze}}{p_T R}$ within regime B, divided by $2$. 
We then vary all scales individually by factors of 2 around their central choices and all simultaneously, taking the envelope to obtain the uncertainty band.

\begin{figure}[b]
     \centering
     \includegraphics[width=0.48\textwidth]{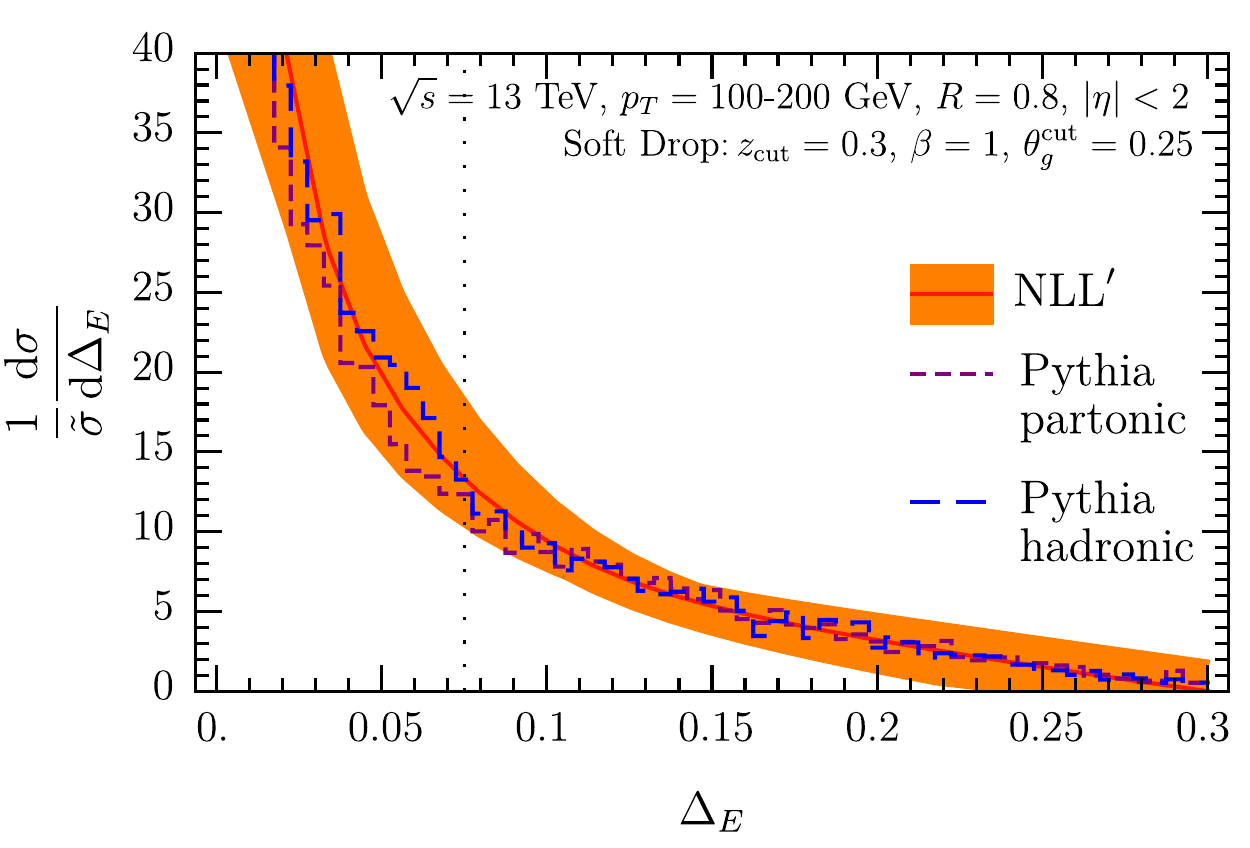} 
     \includegraphics[width=0.48\textwidth]{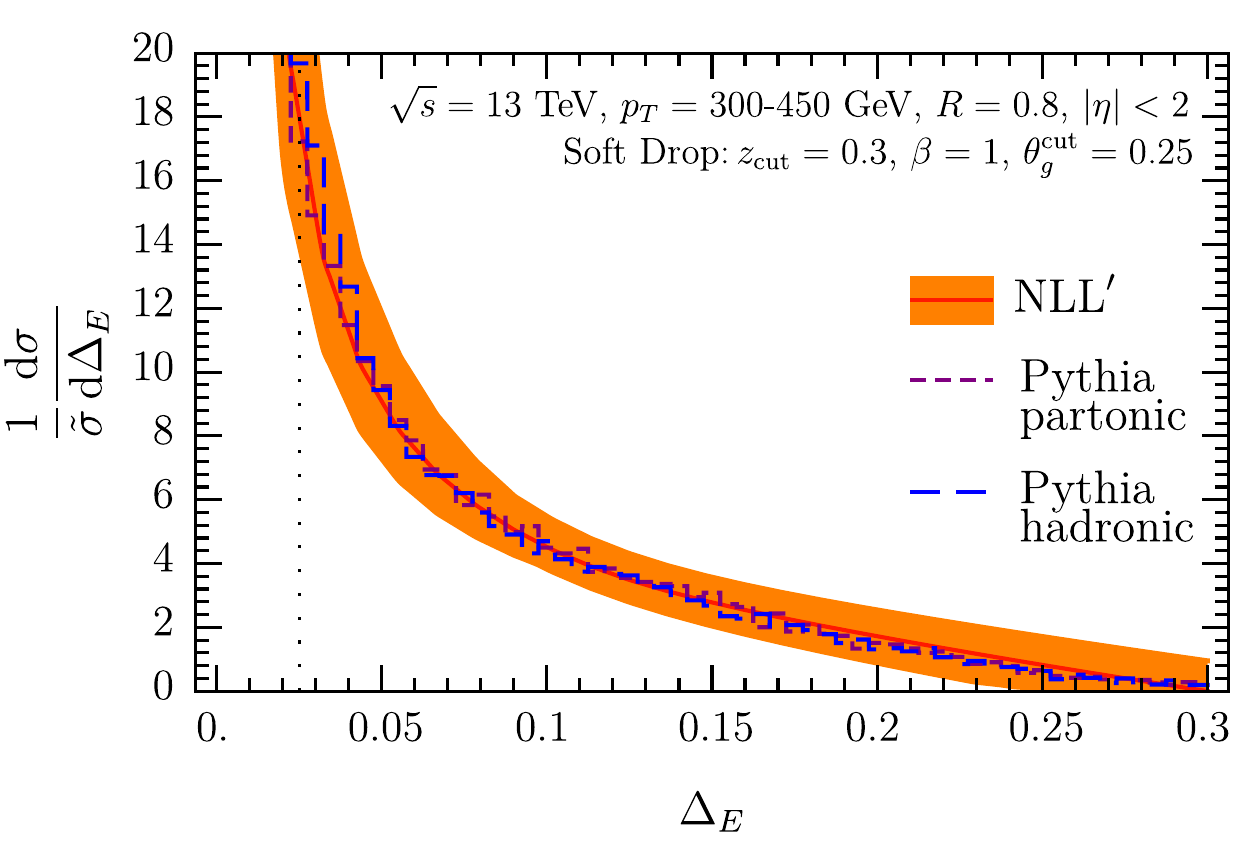} \hfill \phantom{.} \\
     \hfill \includegraphics[width=0.48\textwidth]{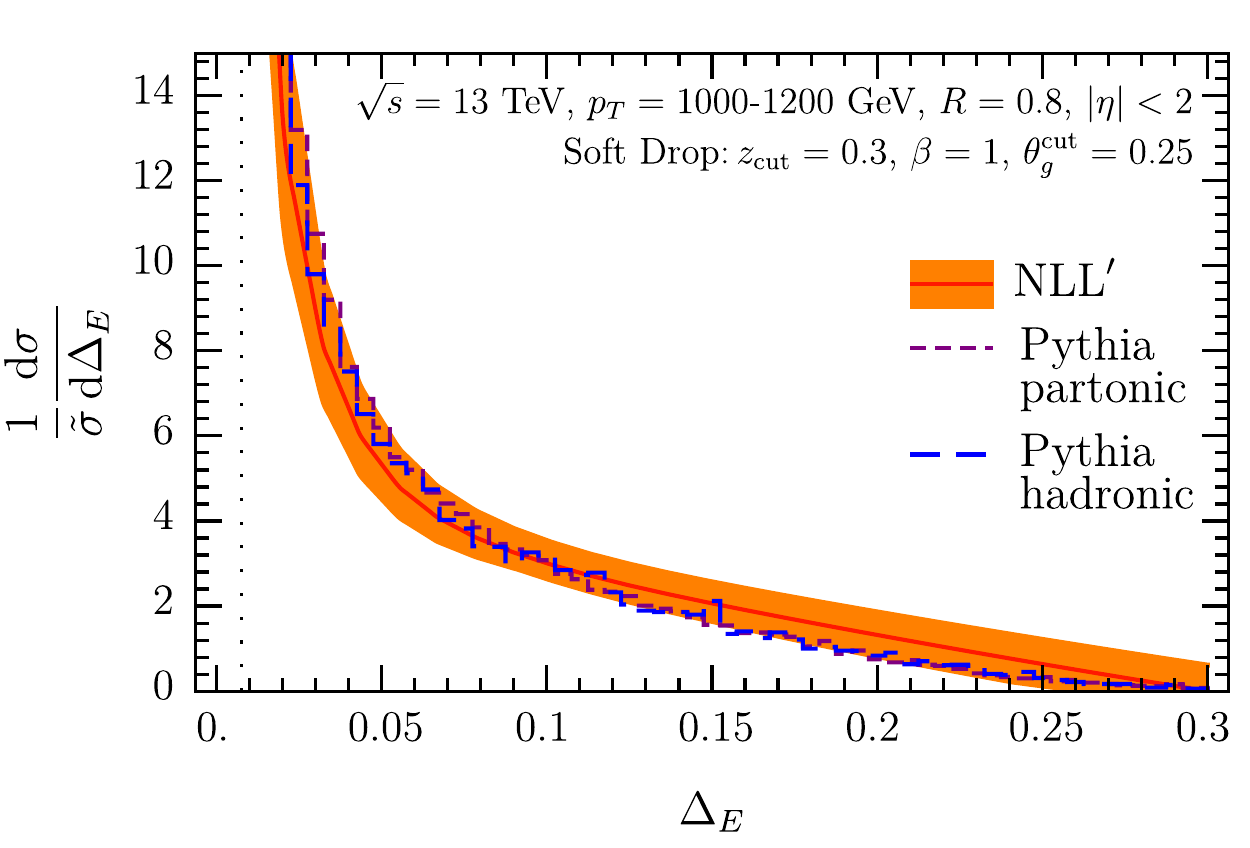} \hfill \phantom{.} 
    \caption{Numerical results at NLL$'$ (orange) for the jet energy drop for soft drop with $z_{\rm cut}=0.3$, $\beta=1$ and $\theta_g^{\rm cut}=0.25$. In addition, we show \Pythia results at parton (purple dashed) and hadron level (blue dashed) for comparison. The different panels correspond to different jet transverse momenta.  The central curves are normalized to unity between the dotted vertical line and the endpoint $\Delta_E=z_{\rm cut}$.~\label{fig:numerics_SD1}} 
\end{figure}

\subsection{Numerical results~\label{sec:numerics_softdrop}}

\begin{figure}[t]
\centering
     \includegraphics[width=0.48\textwidth]{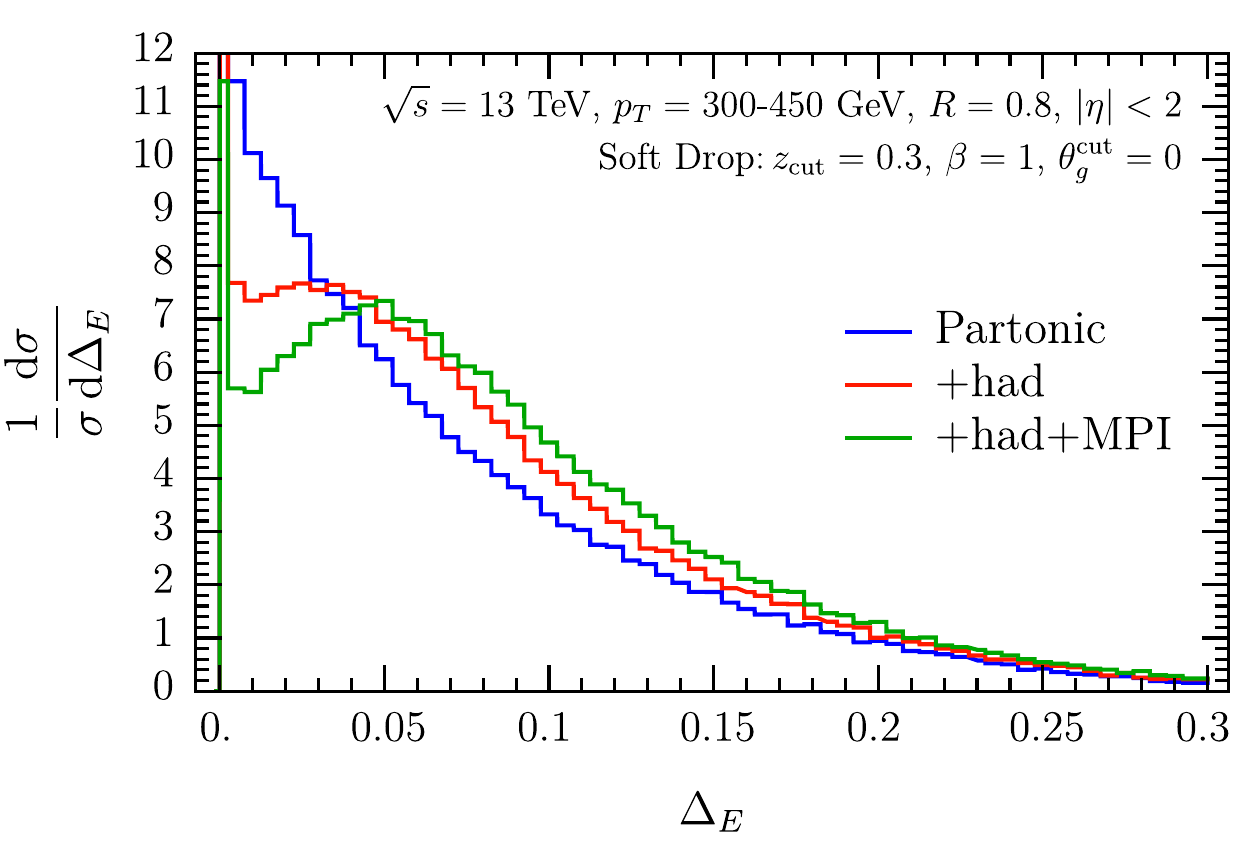}  
     \includegraphics[width=0.48\textwidth]{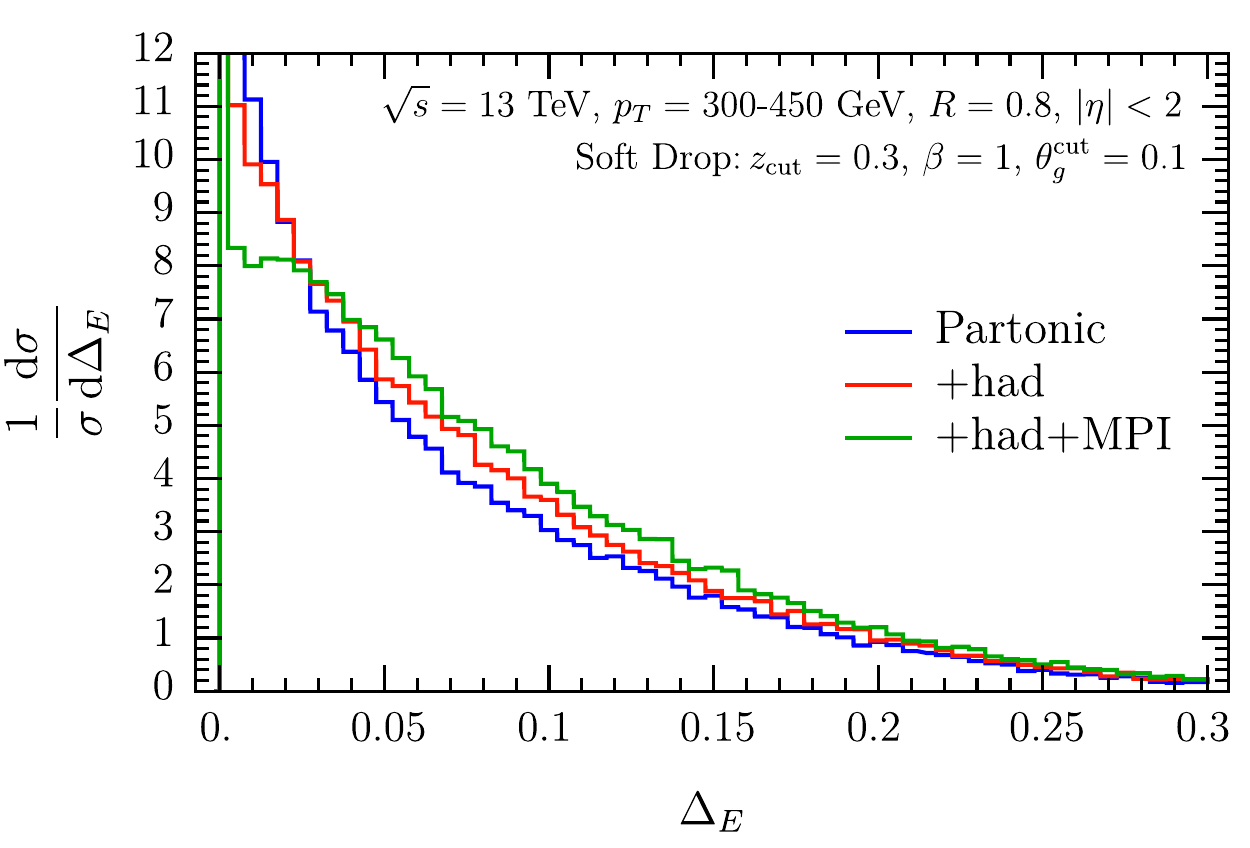} \hfill \phantom{.} \\
     \hfill \includegraphics[width=0.48\textwidth]{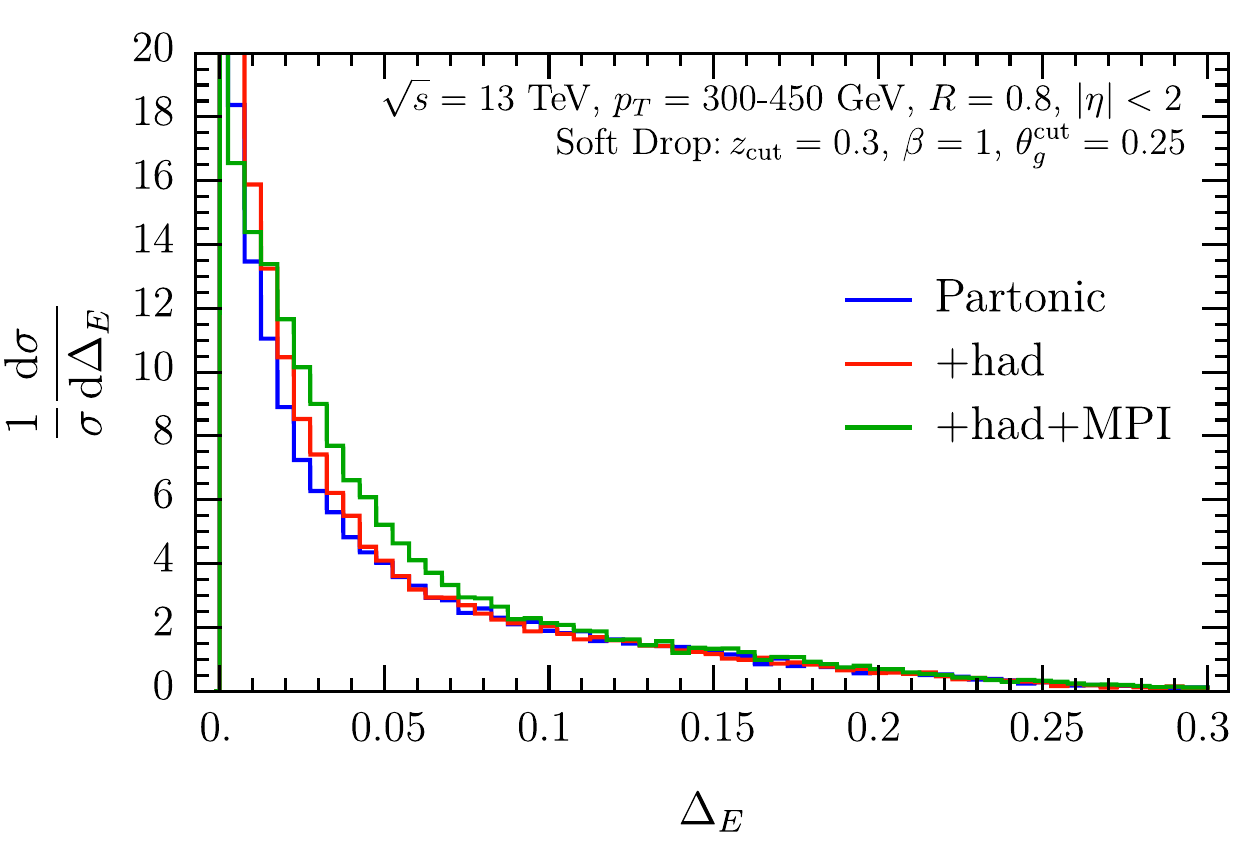}  \hfill \phantom{.} 
    \caption{\Pythia results for the jet energy drop with soft drop at parton level  (blue) and including hadronization (red) and MPI (green). The three panels correspond to $\theta_g^{\rm cut} = 0$ (top left), 0.1 (top right), 0.25 (bottom). Note that these curves are normalized on the full $\ed$ interval.~\label{fig:Pythia_SD}}
\end{figure}

We start by presenting results for the jet energy drop with soft drop for $z_{\rm cut}=0.3,\;\beta=1$, and compare to \Pythia at parton and hadron level (without MPI). The NLL$'$ results for $\theta_g^{\rm cut}=0.25$ and three jet $p_T$ intervals are presented in \fig{numerics_SD1}. Here we leave out the interval $p_T=30-50$~GeV, which is shown for the other grooming procedures, as the energy drop distribution is nonperturbative over most of the $\ed$ range in this case. We indicate the nonperturbative region by the dotted vertical line, and we normalize our results over the perturbative range. We find very good agreement with the \Pythia results at parton and hadron level, which are also normalized on the same range. For the chosen kinematics, and in particular the $\theta_g^{\rm cut}$ value, the cross section is dominated by perturbative dynamics. 

\begin{figure}[t]
     \centering 
     \includegraphics[width=0.48\textwidth]{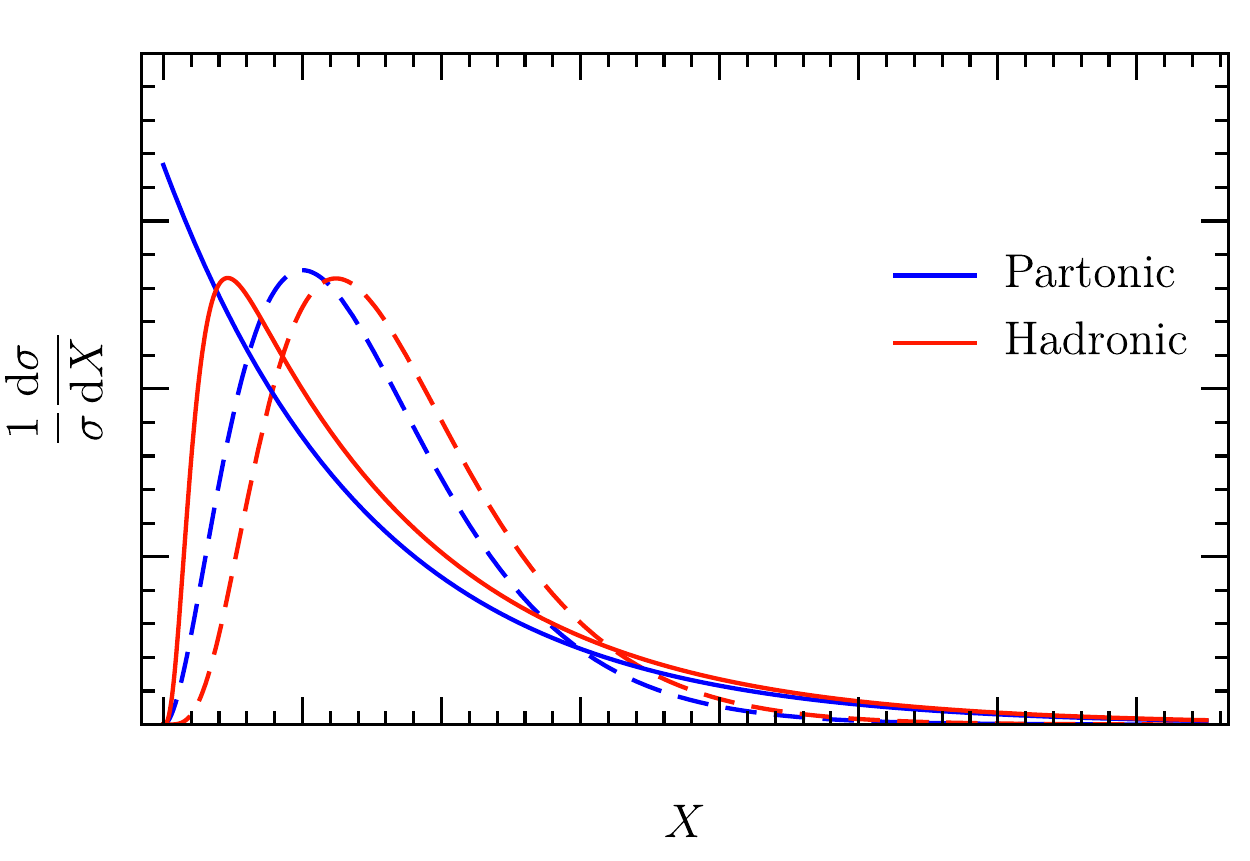}
     \includegraphics[width=0.48\textwidth]{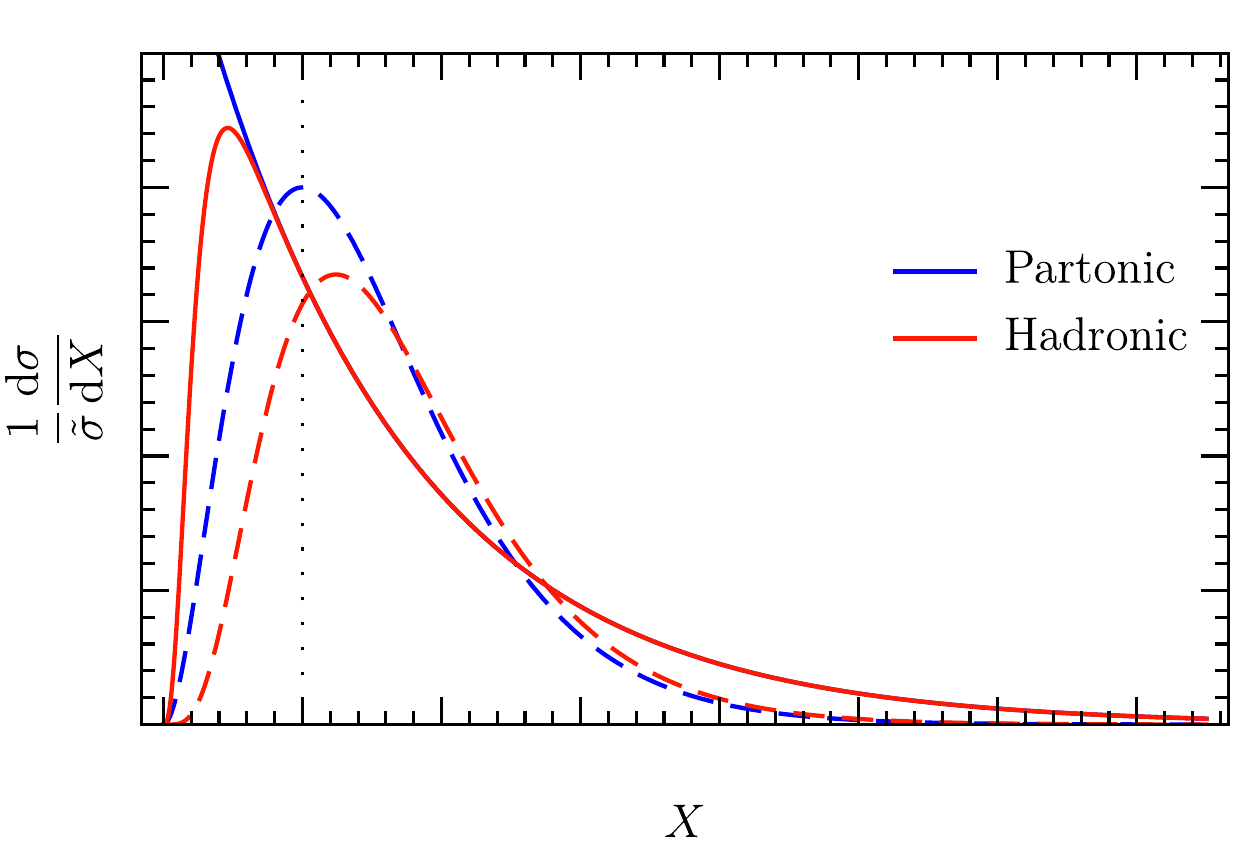} \hfill \phantom{.} \\
    \caption{We explore the effect of hadronization by taking a falling distribution (solid) or a peaked distribution (dashed), as proxy for the perturbative result (blue) of energy drop (for soft drop) and jet mass, and convolving with a shape function to obtain the red curve. While the effects are similar when normalizing on the full range ($1/\sigma$, left panel), this is no longer the case when normalizing on the region to the right of the vertical dotted line ($1/\tilde \sigma$, right panel).~\label{fig:nonp_shape}} 
\end{figure}

\begin{figure}[t]
     \centering 
     \includegraphics[width=0.48\textwidth]{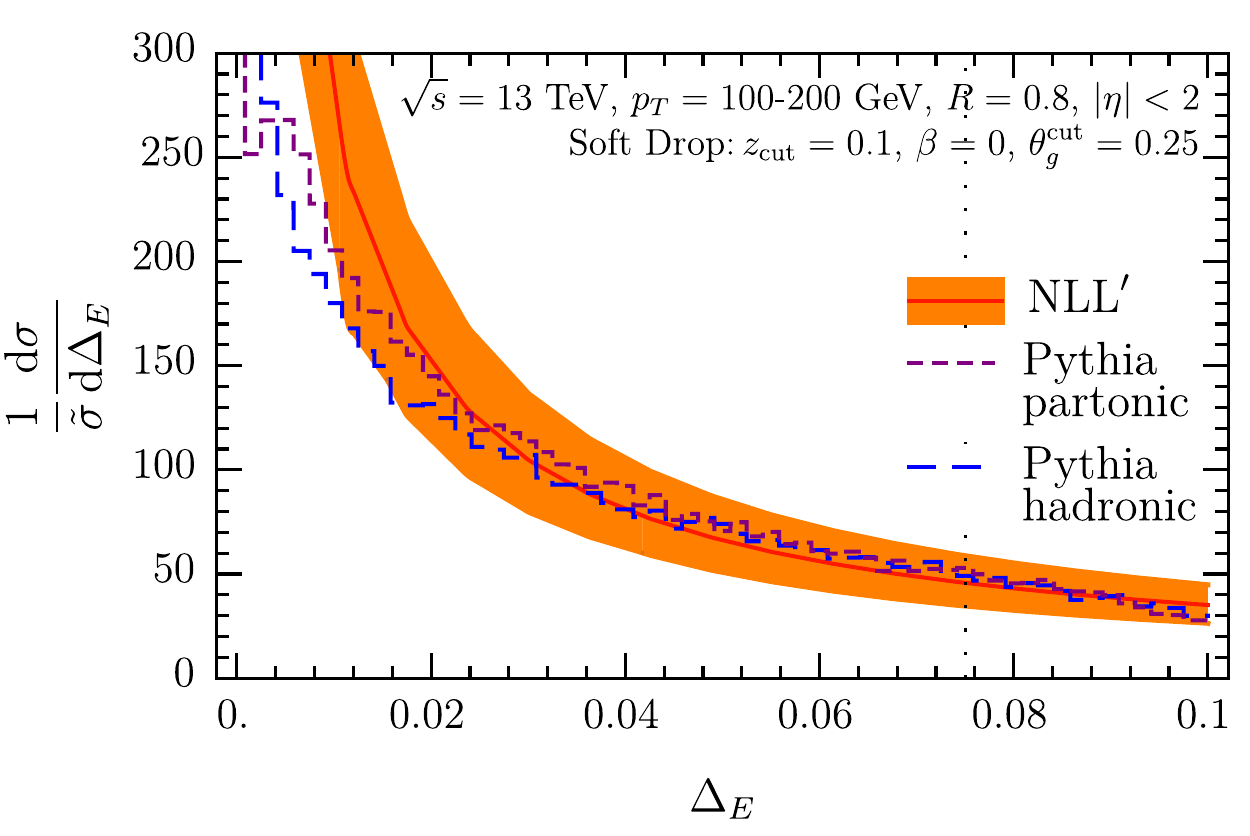}
     \includegraphics[width=0.48\textwidth]{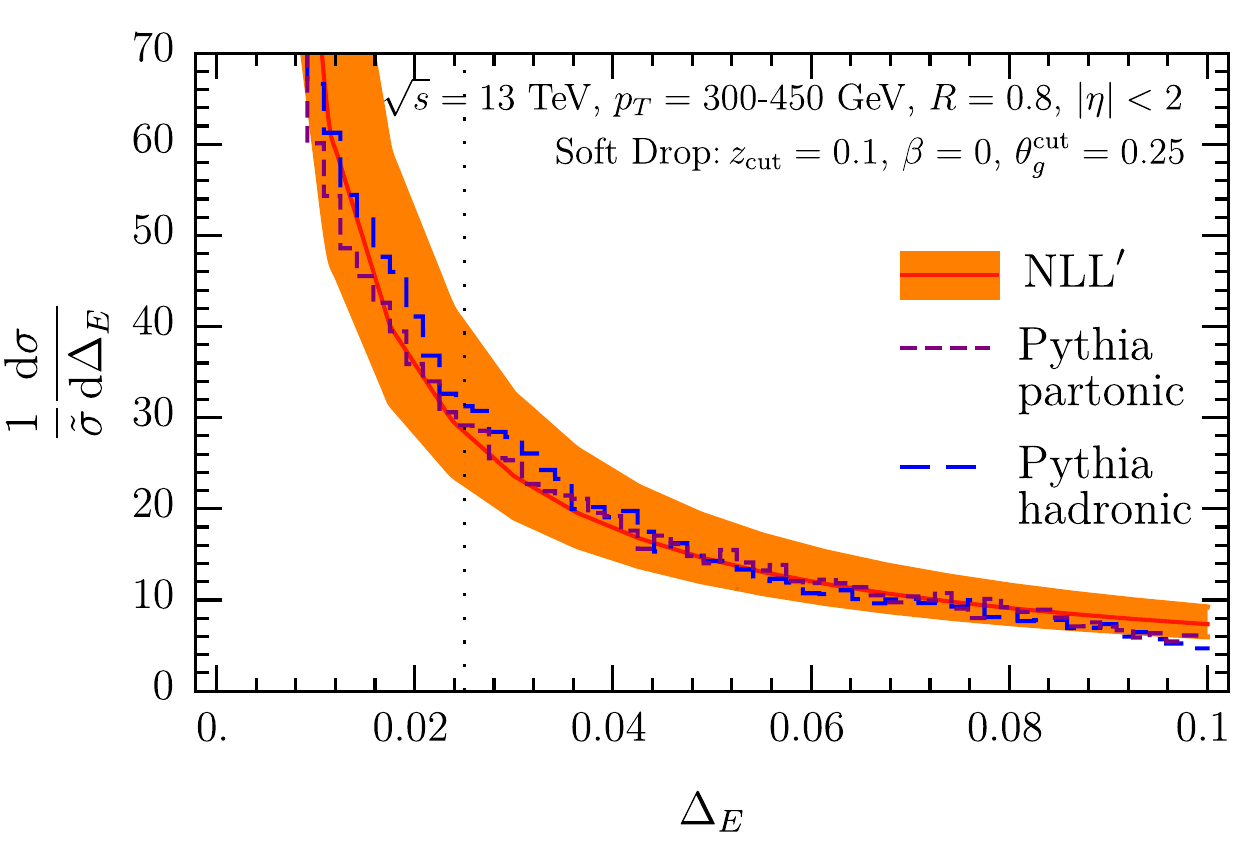} \hfill \phantom{.} \\
     \hfill \includegraphics[width=0.48\textwidth]{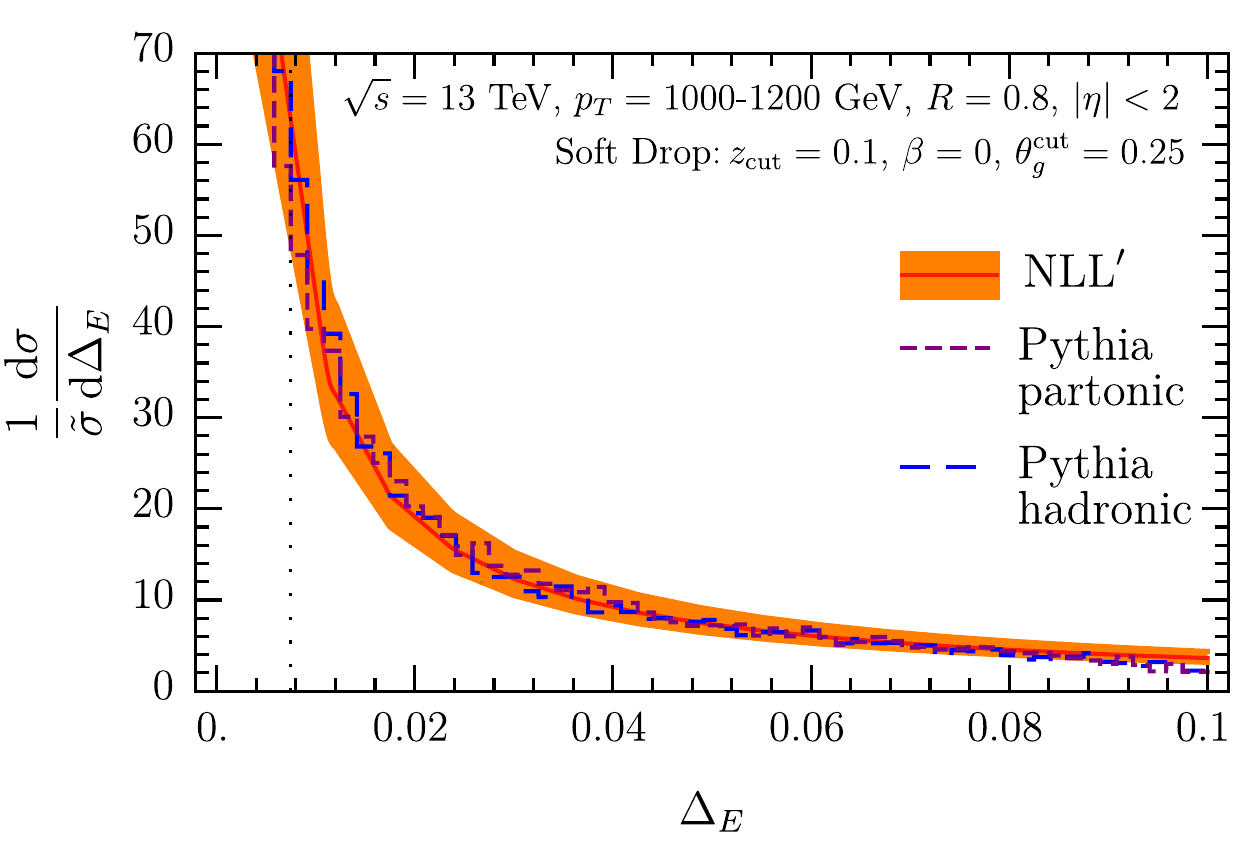} \hfill \phantom{.} 
    \caption{Numerical results at NLL$'$ (orange) for the jet energy drop with soft drop and $z_{\rm cut}=0.1$, $\beta=0$ and $\theta_g^{\rm cut}=0.25$. In addition, we show \Pythia results at parton (purple dashed) and hadron level (blue dashed) for comparison. The different panels correspond to different jet transverse momenta.~\label{fig:numerics_SD2}} 
\end{figure}

In \fig{Pythia_SD}, we investigate in more detail the impact of nonperturbative effects using \Pythia. We show the parton level results and including corrections due to hadronization and MPI for the jet transverse momentum interval of $p_T=300-450$~GeV. The three panels correspond to three different values of $\theta_g^{\rm cut}=0,\,0.1,\, 0.25$. Here we normalize the result over the entire $\Delta_E$ range. We find that nonperturbative effects are small when the relatively large value of $\theta_g^{\rm cut}=0.25$ is chosen, which corresponds to the results in \fig{numerics_SD1}. However, if we lower $\theta_g^{\rm cut}$ nonperturbative effects become more important. We observe that hadronization corrections dominate but also MPI leads to a shift of the distribution to larger values of $\Delta_E$. Interestingly, these differences are substantially reduced when normalizing to the perturbative region, indicating that the \emph{shape} in the perturbative region is not much affected by hadronization. This is due to the shape of the perturbative distribution, which we illustrate in \fig{nonp_shape}. We consider the case where the perturbative distribution is falling (like here) or peaked (as for jet mass), and convolve with a nonperturbative shape function to model the effect of hadronization. Before normalizing, the effect of this convolution is similar, but this is no longer true after normalizing on a restricted range that does not include the nonperturbative region. When no cut on $\theta_g$ is imposed (upper left panel of \fig{Pythia_SD}), the nonperturbative corrections are very large. We thus conclude that imposing a cut on $\theta_g$ allows us to control the soft sensitivity of the jet energy drop. 

 Next, we consider the jet energy drop for $\beta=0$, which is a Sudakov safe observable, as discussed in \sec{SDbeta0}. In \fig{numerics_SD2}, we show the NLL$'$ results for $z_{\rm cut}=0.1$ with $\theta_g^{\rm cut}=0.25$, choosing the same jet kinematics as in \fig{numerics_SD1}. In addition, we show \Pythia results at parton and hadron level, finding again good agreement.

\begin{figure}[t]
    \centering
    \includegraphics[width=0.6\textwidth]{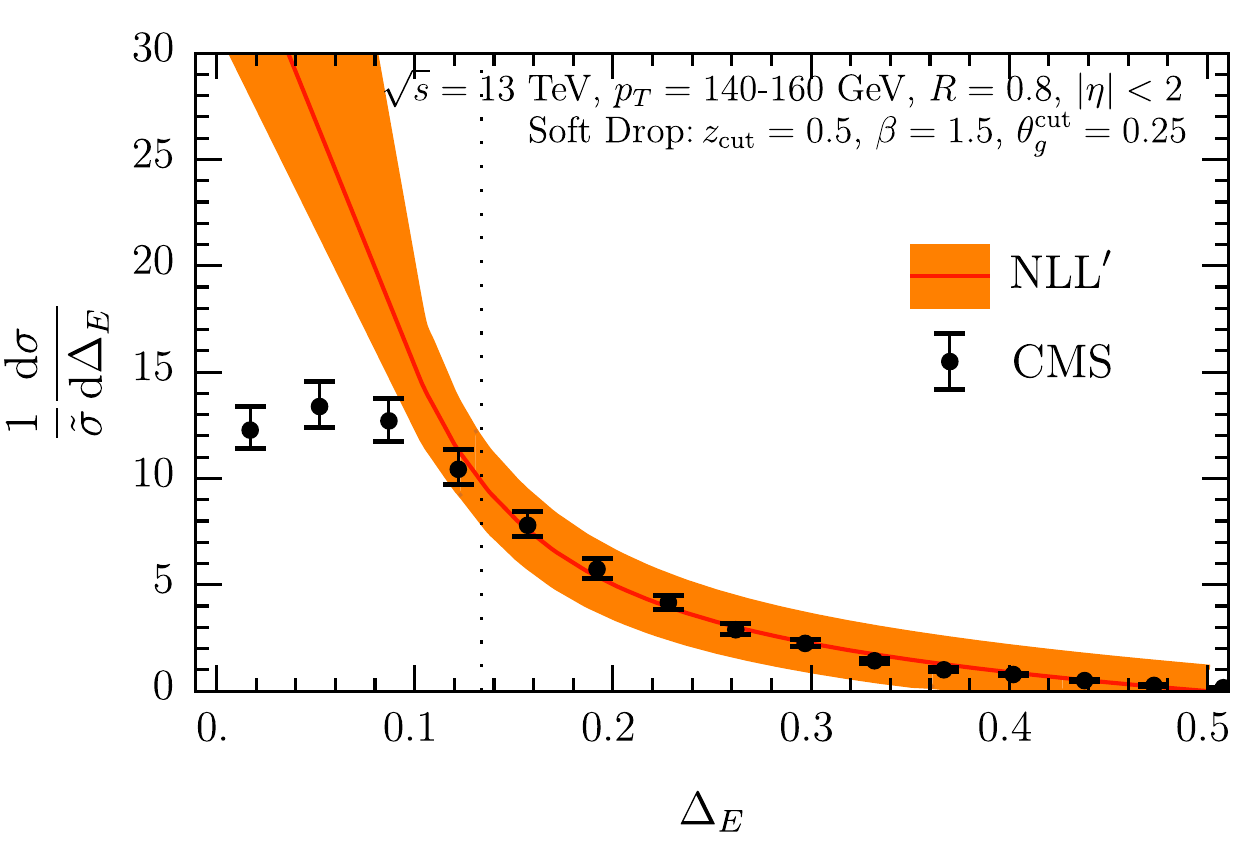}
    \caption{Comparison of our theoretical calculation for soft drop and the preliminary CMS data of ref.~\cite{CMS:2017xdn}.~\label{fig:SDCMS}} 
\end{figure}

We end this section by comparing in \fig{SDCMS} our numerical results to the preliminary CMS data of ref.~\cite{CMS:2017xdn}. The grooming parameters chosen by CMS are $z_{\rm cut}=0.5$ and $\beta=1.5$, and a cut on the groomed jet radius of $\theta_g^{\rm cut}=0.25$ was imposed. We observe very good agreement in the perturbative region which is indicated by the dotted vertical line. Note that both our theoretical results and the data are normalized in the perturbative region. We note that the data of ref.~\cite{Sirunyan:2017bsd} with $\beta=0$ and $z_{\rm cut}=0.1$ are in the nonperturbative region or $\ed> \zc$, where our factorization theorem does not apply. 

\section{Trimming~\label{sec:trimming}}

In this section we consider the jet energy drop of a trimmed jet. We start by introducing the trimming algorithm in~\sec{trim}. We then present fixed-order results of the corresponding jet function in~\sec{cG_trimming}, and introduce the factorization and resummation in~\sec{refact_trimming}, including a discussion of non-global logarithms. In~\sec{numerics_trimming}, we present numerical results and compare to \Pythia.

\subsection{The trimming algorithm~\label{sec:trim}}

Trimming (TR)~\cite{Krohn:2009th} is one of the first jet-grooming algorithms. It improves the event reconstruction at high luminosity colliders and is frequently used for experimental analyses (particularly ATLAS), see e.g.~refs.~\cite{Aad:2019wdr,Sirunyan:2019vxa}. The grooming proceeds as follows: First jets are reconstructed with the anti-k$_T$ algorithm and jet radius parameter $R$. The constituents of the identified jets are then reclustered with a smaller jet radius $R_{\rm sub}<R$. Subjets are removed from the jet if their transverse momentum $p_{T i}$ (or energy) is below a threshold given by $p_{T i}<f_{\mathrm{cut}}  \Lambda_{\mathrm{hard}}$. Here $f_{\rm cut}$ is a dimensionless quantity and $\Lambda_{\rm hard}$ is a hard scale, which we choose as $p_T$ of the initial ungroomed jet. The trimmed jet is then given by all the particles in the remaining subjets. 

In ref.~\cite{Dasgupta:2013ihk} the jet mass of a trimmed jet was calculated, and in refs.~\cite{Dai:2016hzf,Kang:2017mda} the longitudinal momentum fraction $z=p_{Ti}/p_T$ of the reclustered inclusive subjets was considered. Here we consider the jet energy drop induced by the trimming procedure. Similar to soft drop, this observable is particularly sensitive to the soft aspects of jets. In ref.~\cite{Krohn:2009th} the k$_T$ algorithm was used for reclustering the jets into smaller subjets, as k$_T$ subjets better share the total available jet energy amongst themselves. We use instead the C/A algorithm, such that the clustering effects are the same as for soft drop. (We have checked in \Pythia that this has a minimal effect on the jet energy drop distribution.) Typical values of the trimming parameters used in experimental analyses are  $R_{\rm sub}=0.2$ and $\fc=0.05$, though they depend on the observable under consideration. For our numerical results in \sec{numerics_trimming}, we choose a relatively large value of $\fc=0.3$ to ensure that a large part of the distribution can be described perturbatively.

\subsection{Fixed-order results~\label{sec:cG_trimming}}

We denote the jet function that measures the jet energy drop for trimming by $\Delta\cG^{\rm TR}_i$. It depends on the grooming parameters $\theta_t\equiv R_{\rm sub}/R,\, \fc < \tfrac12$ and the jet energy drop $\Delta_E$, which is the total energy fraction of the subjets removed by trimming. At NLO, the jet function can be calculated as 
\begin{align}
&\Delta \cG_i^{\rm TR}(\Delta_E, p_T R, \theta_t,\fc,\mu)
 \\ & \quad
=
\int \df \Phi_2\, \si_{2,i}^c\; \Theta \left(\theta < R \right) \bigl\{  \Theta \left(\theta > \theta_t R \right) 
 \bigl[ \Theta \left(x> \fc \right)\Theta \left(1-x >\fc \right)  \delta (\ed) 
 \nn \\ & \qquad
+ \Theta \left(x> \fc  \right)\Theta \left(1-x <\fc  \right) \delta (\Delta_E-x)
\nn  \\ & \qquad 
+ \Theta \left(x< \fc  \right)\Theta \left(1-x >\fc  \right) \delta (\Delta_E-(1-x))\bigr] 
+  \Theta \left(\theta < \theta_t R \right)\delta(\ed)  - \delta(\ed) \bigr\} \,.
\nn \end{align}
If the two partons are clustered into different subjets $\theta>\theta_t R$, they are individually tested against the trimming condition. As before, the very last term subtracts the contribution already contained in the semi-inclusive jet function.

\begin{figure}[t]
    \centering
    \includegraphics[width=0.48\textwidth]{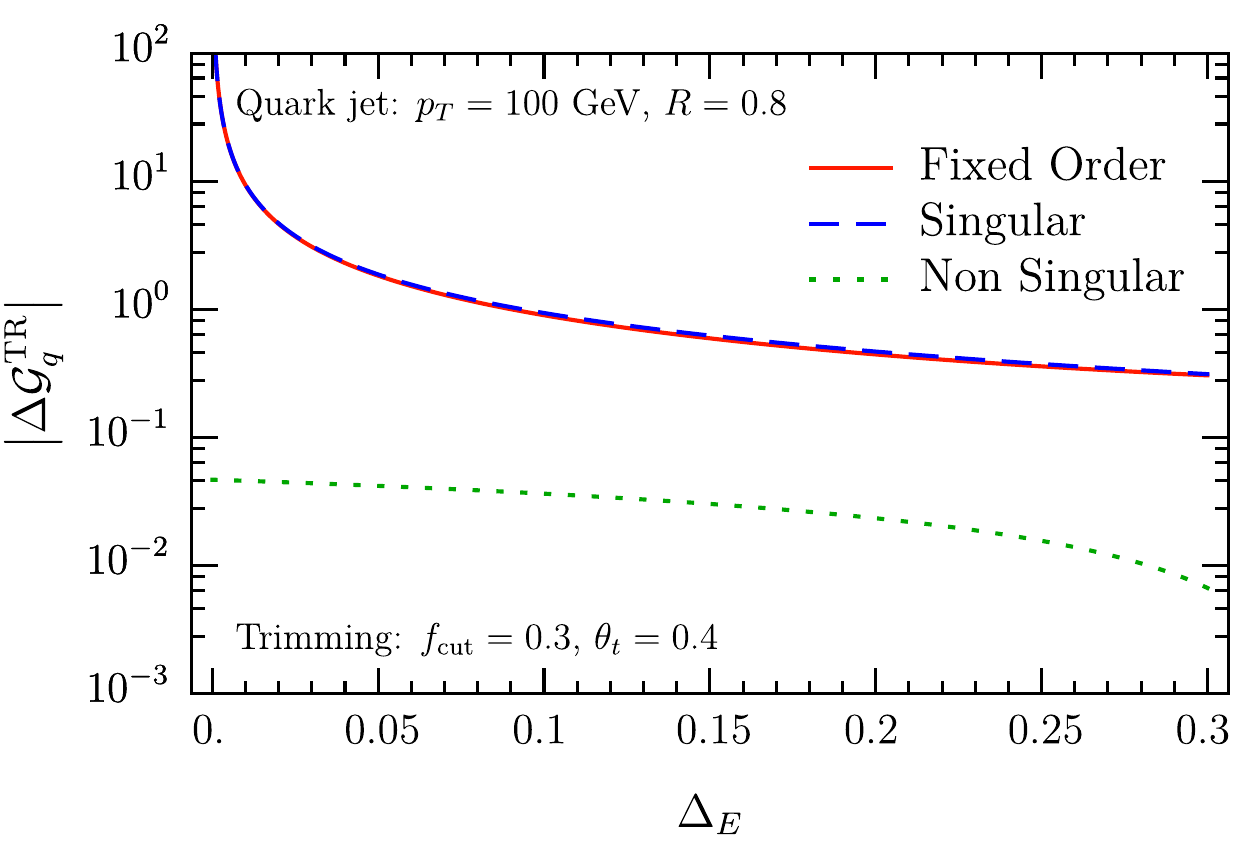} \hfill 
       \includegraphics[width=0.48\textwidth]{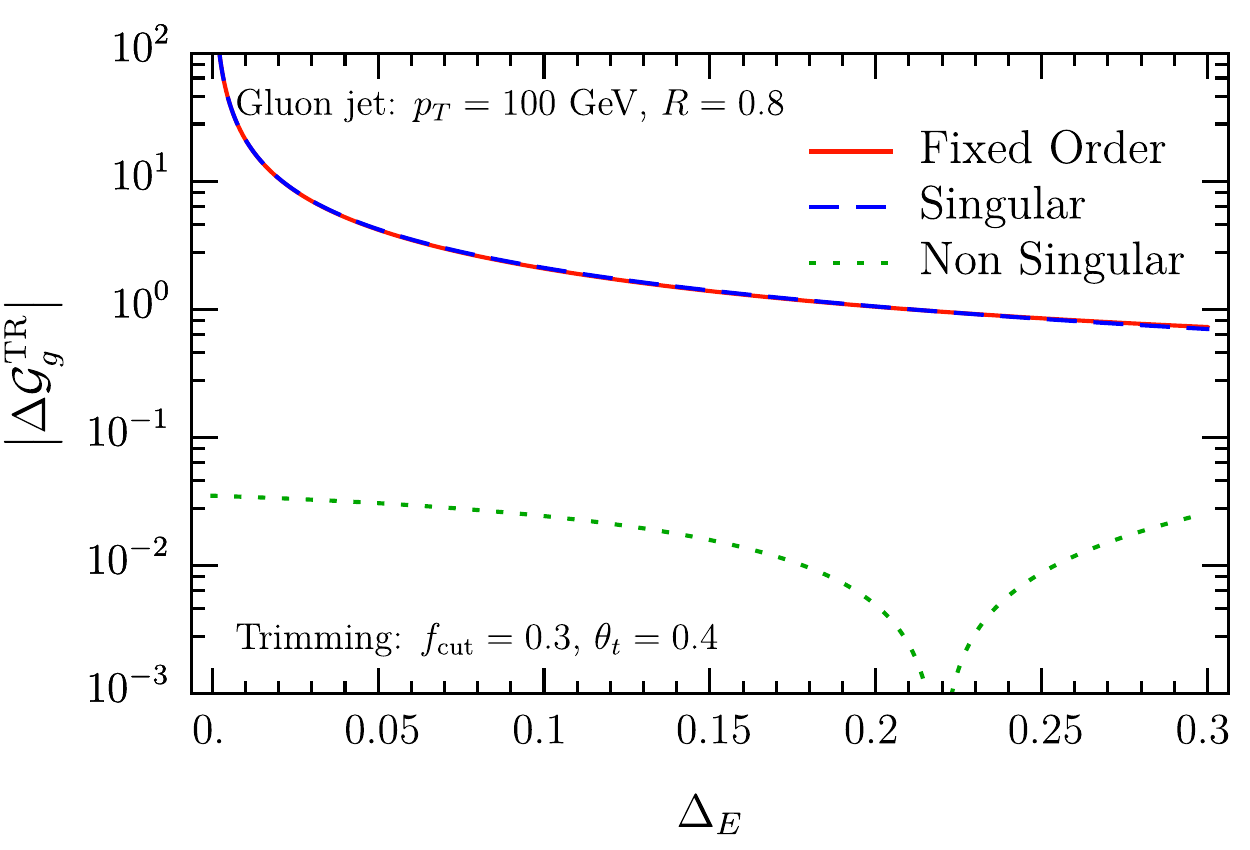} \hfill \phantom{.} \\
    \caption{Numerical comparison of the size of singular (dashed blue) and non-singular (dotted green) terms of the fixed-order (red) jet function for trimming. We show the quark and gluon result in the left and right panel, respectively.~\label{fig:refactorization_trimming}}
\end{figure}

For quark and gluon jets, we find
\begin{align}
&\Delta \cG_q^{\rm TR} (\Delta_E, p_T R, \theta_t,\fc,\alpha_s(\mu))
\nn\\&\quad =
\, \f{\as C_F}{\pi} \, \ln \tet\, \bigg\{ \Theta(\ed < \fc) \bigg[-\f{2}{1-\ed}-\f{2}{[\ed]}_+ +3 \bigg] + \delta (\ed) \bigl(-2 \ln (1-\fc)
\nn \\&\quad\quad
  +2 \ln \fc - 3 \fc \bigr)  \bigg\}
\,,\\[.5em] 
&\Delta \cG_g^{\rm TR}(\Delta_E, p_T R, \theta_t,\fc,\alpha_s(\mu))
\nn\\&\quad=
\, \f{\as}{\pi} \, \ln\tet\, \bigg\{ \Theta(\ed < \fc) \bigg[ -2n_f T_F ( \ed^2+(1-\ed)^2) + C_A\bigg(4 - \f{2}{[\ed]}_+ -\f{2}{1-\ed}
\nn \\&\quad\quad
-2(1-\ed) \ed \bigg) \bigg] +  \delta (\ed)\bigg[  n_f T_F \bigg(\f43 \fc^3 -2 \fc^2 +2\fc \bigg) 
\nn \\ &\quad\quad
+ C_A \bigg( -\f{2}{3}\fc^3 +\fc^2 -4\fc-2 \ln (1-\fc)+2 \ln \fc \bigg) \bigg] \bigg\} 
\,.
\end{align}
We observe that at NLO the jet energy drop $\Delta_E$ is always less than $\fc$, similar to (iterated) soft drop where $\Delta_E<\zc$. The plus distribution here is defined on the interval  $0<\Delta_E<1$. If we rewrite it to be defined on the interval of the theta function $\Theta(\Delta_E<f_{\rm cut})$, the $\ln f_{\rm cut}$ term in the last lines for both the quark and gluon jet function cancels, and we can safely take the limit $f_{\rm cut}\to 0$ (similar to $\zc \to 0$ for iterated soft drop in \sec{fixedorder_iteratedsoftdrop}). In this limit the trimming is removed and the jet function $\Delta \cG^{\rm TR}$ thus vanishes.
\begin{figure}[t]
    \centering
    \includegraphics[width=0.48\textwidth]{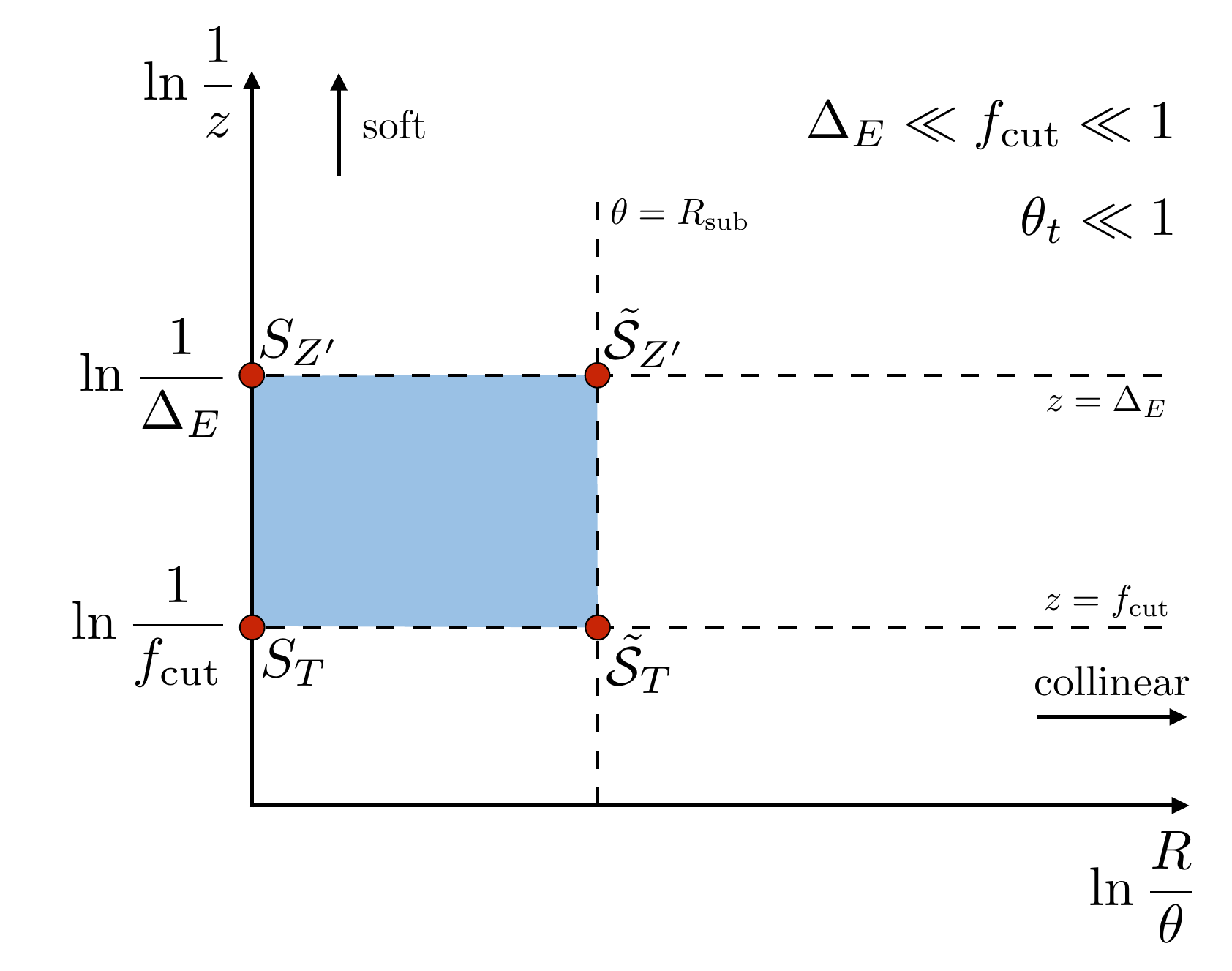} \\
    \caption{Lund diagram for the energy drop of a trimmed jet. The relevant SCET modes are indicated by red dots.~\label{fig:Lund_trimming}}
\end{figure}
Next we consider the relative size of the singular and non-singular terms for trimming at fixed order in \fig{refactorization_trimming}. We observe that the singular terms dominate over the entire range of $\Delta_E$, suggesting that the resummation is likely important, and that the matching to NLO does not need to be included in our numerical results. We note that different from iterated soft drop (see \fig{refactorization_ISD}) the NLO distribution does not smoothly go to zero at the endpoint $\Delta_E=f_{\rm cut}$.

\subsection{Factorization and resummation~\label{sec:refact_trimming}}

For the trimmed jet energy drop there are three parameters that enter in the large logarithms requiring resummation, namely the energy drop $\Delta_E$ and the grooming parameters $f_{\rm cut}\,,\theta_t$. We will assume $\ed \ll f_{\rm cut} \ll 1$ and $\theta_t \ll 1$, for which the corresponding Lund diagram is shown in \fig{Lund_trimming}. The two horizontal dashed lines correspond to the measurement of the jet energy drop $z=\Delta_E$ and the cutoff imposed on the reclustered subjets $z=f_{\rm cut}$, respectively. The vertical line corresponds to the size of the subjets $R/\theta=R/R_{\rm sub}=1/\theta_t$. We note that the Lund diagram here looks quite similar to that for the jet energy drop for soft drop with $\beta=0$ in \fig{LUNDb0}. However, in contrast to the groomed radius $\theta_g$, the subjet radius $\theta_t$ has a fixed value. In particular, it is not integrated over after the resummation is performed, as in the case of soft drop. Emissions in the shaded region of the Lund diagram are vetoed and we can read of the resummed LL expression,
\begin{align}
 \tilde \cG_{i}^{\rm TR}\bigl(\ed , p_T R, \fc, \theta_t,\al_s(\mu)\bigr) &\stackrel{{\rm LL}}{=}\,\frac{\df}{\df \ed} \,\exp\biggl[- \frac{2\al_s C_i}{\pi}  \ln \biggl(\f{\ed}{\fc} \biggr)\ln \tet \biggr]
\,.\end{align}

The relevant modes in SCET, needed to achieve NLL$'$ resummation, again correspond to  the corners of the shaded region in \fig{Lund_trimming}. Their power counting is summarized in table~\ref{tab:modes_tr}. The refactorization of the jet function for trimming is at NLL$'$ accuracy given by 
\begin{align}\label{eq:trimming_NLL}
 &\tilde \cG_{i}^{\rm TR}\bigl(\ed , p_T R, f_{\rm cut},\theta_t, \al_s(\mu)\bigr) 
\\
 & \quad \stackrel{{\rm NLL}
'}{=} 
 S_{i, T}(\fc p_T R,\mu)\,\CS_{i,T}(\fc\,  \tet \, p_T R,\mu ) \,
\int\! \df \ed'\, S_{i, Z'}(\ed', p_T R,\mu) \, 
\nn \\ & \qquad \times 
\CS_{i, Z'}(\ed - \ed', \tet\, p_T R, \mu)
 S_{i}^{\rm NG}(\ed/\fc)\,  
\CS_{i}^{\rm NG+AC}(\ed/\fc) \,.
\nn \end{align}
We now provide the NLO expressions for the various ingredients in this factorization. The one-loop soft function $S_{i,\, Z'}$ is the same as $S_{i,\, Z}$, which appeared in the factorization formula of (iterated) soft drop, and is given in \eq{factorization_ISD}. The other soft and collinear-soft functions are given by
\begin{align}
S_{i,T}(f_{\rm cut} p_T R,\mu)&= 
1 + \f{\alpha_s C_i }{ \pi }\left\{- \ln^2 \left( \f{\mu}{\fc p_T R } \right) +\f{\pi^2}{24} \right\}
\,, \\[.5em]
\CS_{i,T}(\fc \tet \, p_T R,\mu )&= 
1+  \f{\alpha_s C_i }{ \pi}\left\{\ln ^2 \left( \f{\mu}{\fc \tet \, p_T R } \right) -\f{\pi^2}{24} \right\}
\,, \\[.5em]
\CS_{i, Z}(\Delta_E,\tet\, p_T R,\mu)&=
\delta(\ed)+\f{\as C_i}{\pi} \bigg\{ -2\left[ \f{\ln \ed}{\ed} \right]_+ + \f{2}{[\ed]}_+ \ln \left( \f{\mu}{\tet \, p_T R}\right)  
\nn \\& \quad
+\delta(\ed) \bigg[- \ln^2 \left( \f{\mu}{\tet \, p_T R}\right) +\f{\pi^2}{24}\bigg] \bigg\}
\,.
\end{align}
These functions are similar to the ones for soft drop, as they can be obtained by appropriately replacing trimming parameters by soft drop parameters.  For example, $\CS_{i, Z}$ can be obtained from \eq{B_CSZ} by replacing $\theta_g$ by $\theta_t$. 
\begin{table}
   \centering
   \begin{tabular}{l|l|c}
     \hline \hline
     Mode: & Function: & Scaling\\ \hline
     soft & $S_T$ & $\fc\,  p_T(R^2,1,R)$ \\
     soft & $S_{Z'}$ & $\ed\, p_T (R^2, 1, R) $ \\ 
     collinear-soft & $\tilde{\cal S}_T$ & $\fc\,  p_T(\rs^2,1,\rs) $  \\
     collinear-soft & $\tilde {\cal S}_{Z'}$ &  $\ed\,  p_T(\rs^2,1,\rs) $  \\
     \hline \hline
   \end{tabular}
   \caption{The parametric scaling of the momenta of the various modes in SCET, needed to describe the jet energy drop cross section for trimming with $\ed \ll f_{\rm cut} \ll 1 $ and $\theta_t=R_{\rm sub}/R \ll 1$. ~\label{tab:modes_tr}}
\end{table}
Eq.~\eqref{eq:trimming_NLL} contains two contributions from non-global logarithms: First of all, $S_{i}^{\rm NG}$  arises at the boundary $\theta=R$ of the initial jet, because emissions inside the jet must have $z<\ed$ or $z>\fc$ (for details, see the discussion of the corresponding NGL for iterated soft drop in \sec{NGL_iteratedsoftdrop}). Since the jet is obtained by anti-k$_T$ reclustering, it has a hard boundary that is not perturbed by the clustering of soft radiation. 
The second contribution from NGLs arises at the boundary $\theta = R_g$ of the trimmed subjets. It has a very similar origin: emissions outside of a trimmed subjet must have $z<\ed$ or $z>\fc$. However, because the subjets are obtained with the C/A algorithm, they are sensitive to clustering effects, and there are also Abelian clustering effects. It is described by the same $\CS_{i}^{\rm NG+AC}$ given in \eq{NGAC} for soft drop, but with a different argument.

The large logarithms are resummed by evaluating each of the four (global) functions in \eq{trimming_NLL} at their characteristic scale
\begin{equation}\label{eq:scales_trimming}
\mu_{S_T}\sim f_{\rm cut}p_T R\,,\quad \mu_{\CS_T}\sim f_{\rm cut}\theta_t\, p_T R\,,\quad \mu_{S_{Z'}}\sim \Delta_E\, p_T R\,,\quad \mu_{\CS_{Z'}}\sim \Delta_E\theta_t \,p_T R \,,
\end{equation}
and using the RG equations 
\begin{align}
\mu\f{\df}{\df\mu}\,S_{i,T}(\fc p_T R,\mu) =&\,\ga_i^{S_T}(\fc p_T R,\mu) \,S_{i,T}(\fc p_T R,\mu)
\,,\\[.5em]
\mu\f{\df}{\df\mu}\,\CS_{i,T}(\fc \theta_t \,p_T R,\mu) =&\,\ga_i^{\CS_T}(\fc\theta_t\, p_T R,\mu) \,\CS_{i,T}(\fc\theta_t\, p_T R,\mu)
\,,\\[.5em]
\mu\f{\df}{\df\mu}\,\CS_{i,Z'}(\ed, \theta_t\,p_T R,\mu) =&\,\int \df \ed' \gamma^{S_{Z'}}_i(\ed-\ed',\theta_t\, p_TR,\mu)\,\CS_{i,Z'}(\ed', \theta_t\,p_T R,\mu)
\,,
\end{align}
to evolve them to a common scale. 
The anomalous dimensions are summarized in the appendix~\ref{app:anom}. 

\subsection{Numerical results~\label{sec:numerics_trimming}}

\begin{figure}[t]
\centering
      \includegraphics[width=0.48\textwidth]{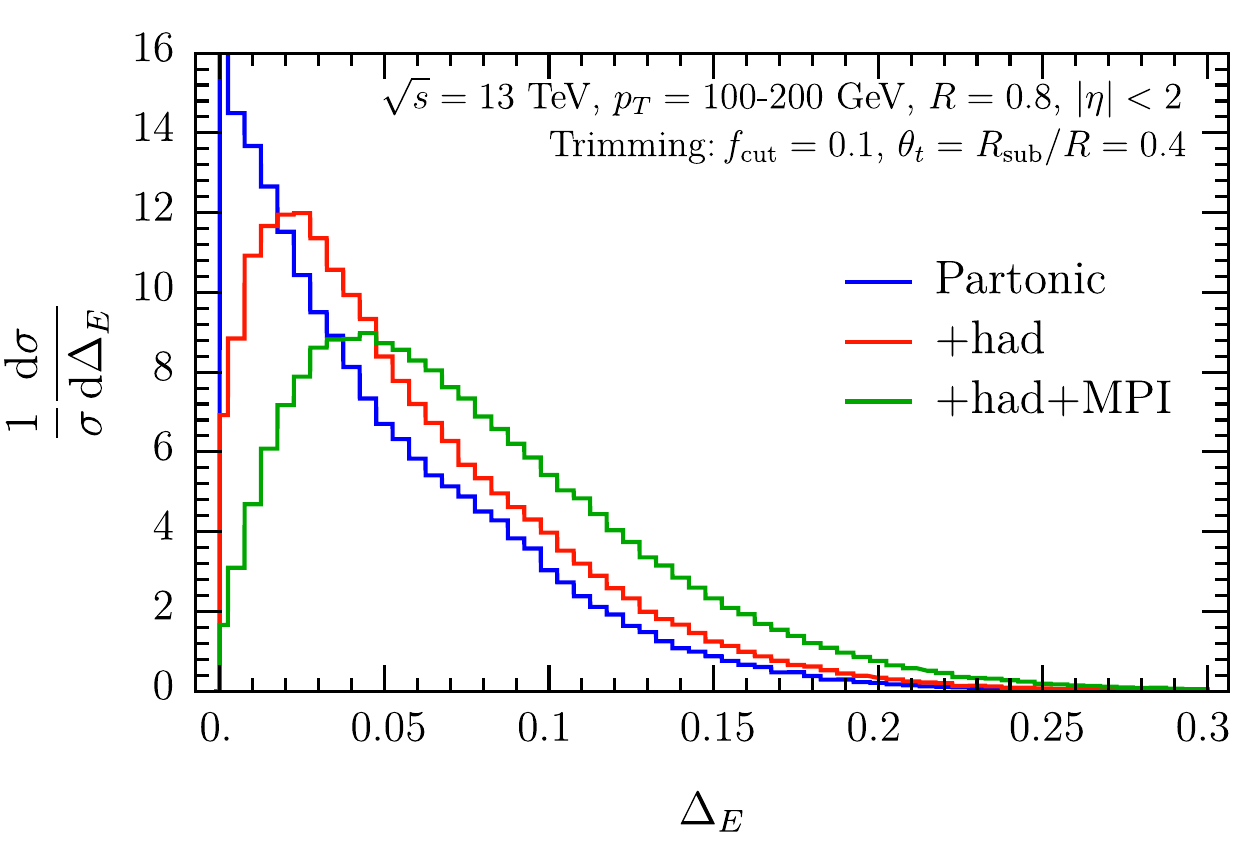}  
     \includegraphics[width=0.48\textwidth]{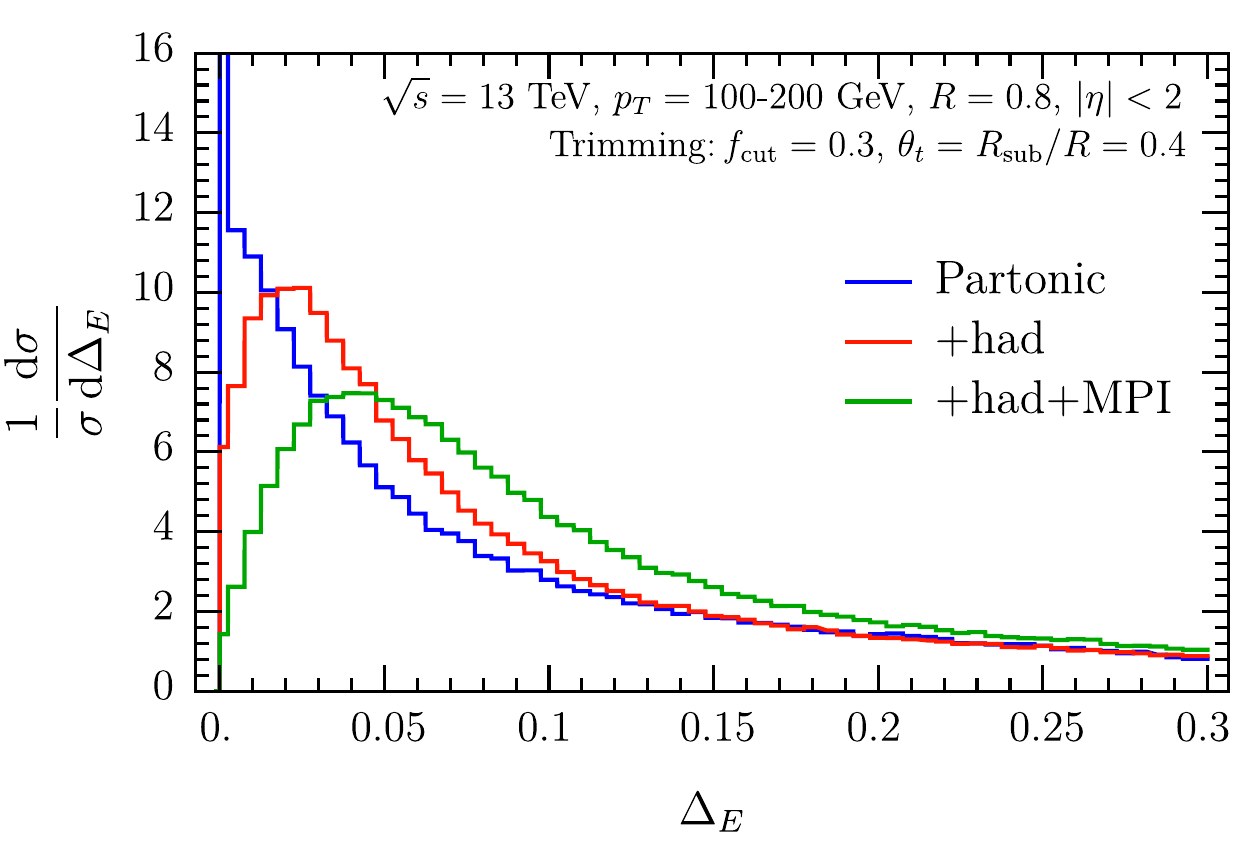} \hfill \phantom{.} \\
    \caption{\Pythia results for the jet energy drop with trimming for $f_{\rm cut} = 0.1$ (left panel) and $f_{\rm cut} = 0.3$, at parton level (blue), hadron level (red), and including MPI (green). Note that these curves are normalized on the full $\ed$ interval.~\label{fig:pythia_trim}}
\end{figure}

\begin{figure}[t]
     \hfill \includegraphics[width=0.48\textwidth]{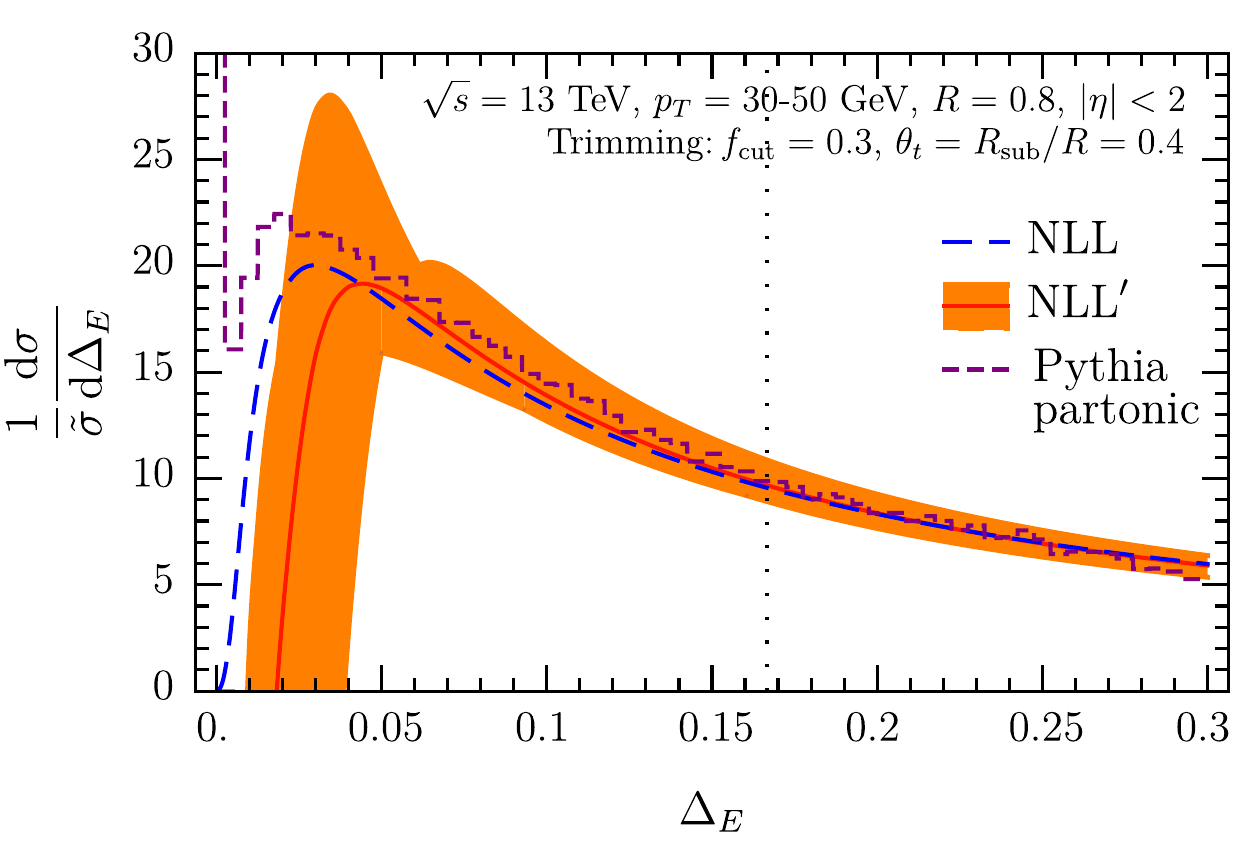} \hfill 
     \includegraphics[width=0.48\textwidth]{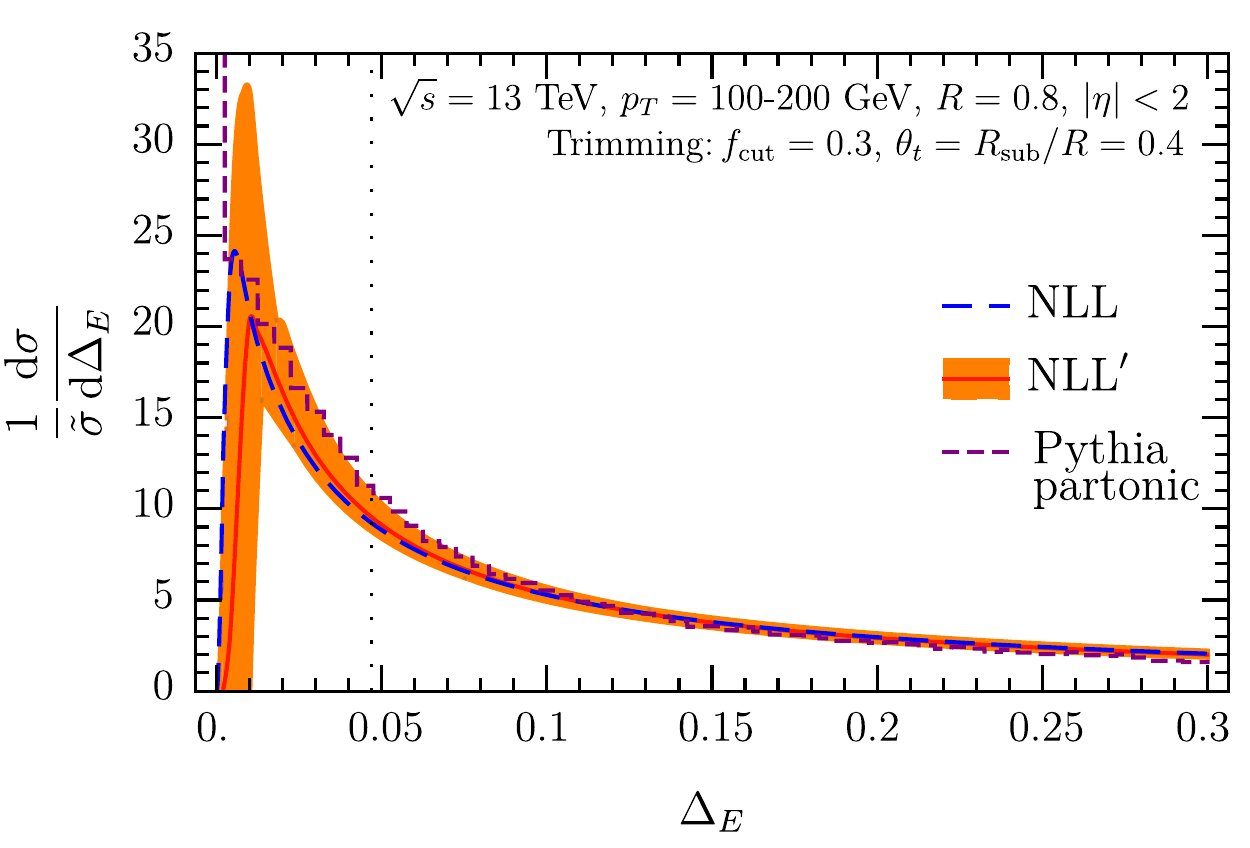} \hfill \phantom{.} \\
     \hfill \includegraphics[width=0.48\textwidth]{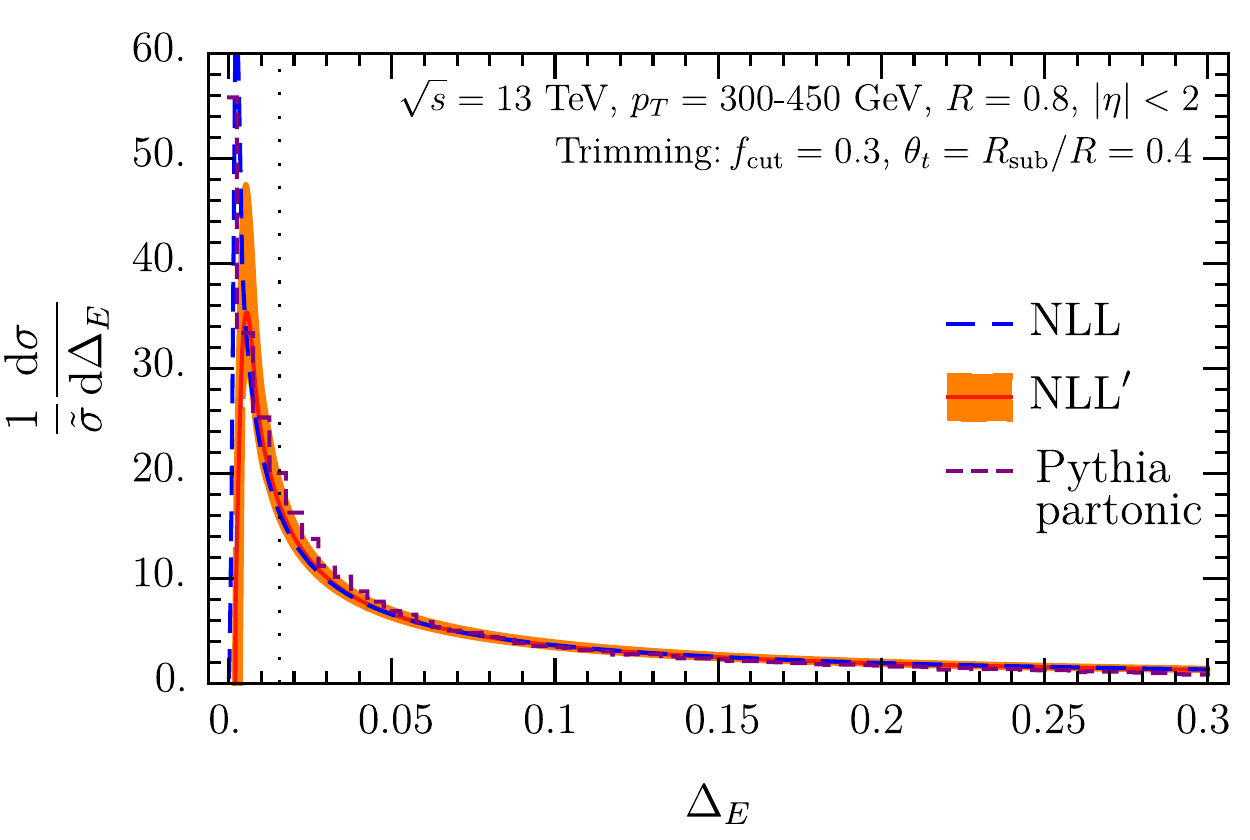} \hfill 
     \includegraphics[width=0.48\textwidth]{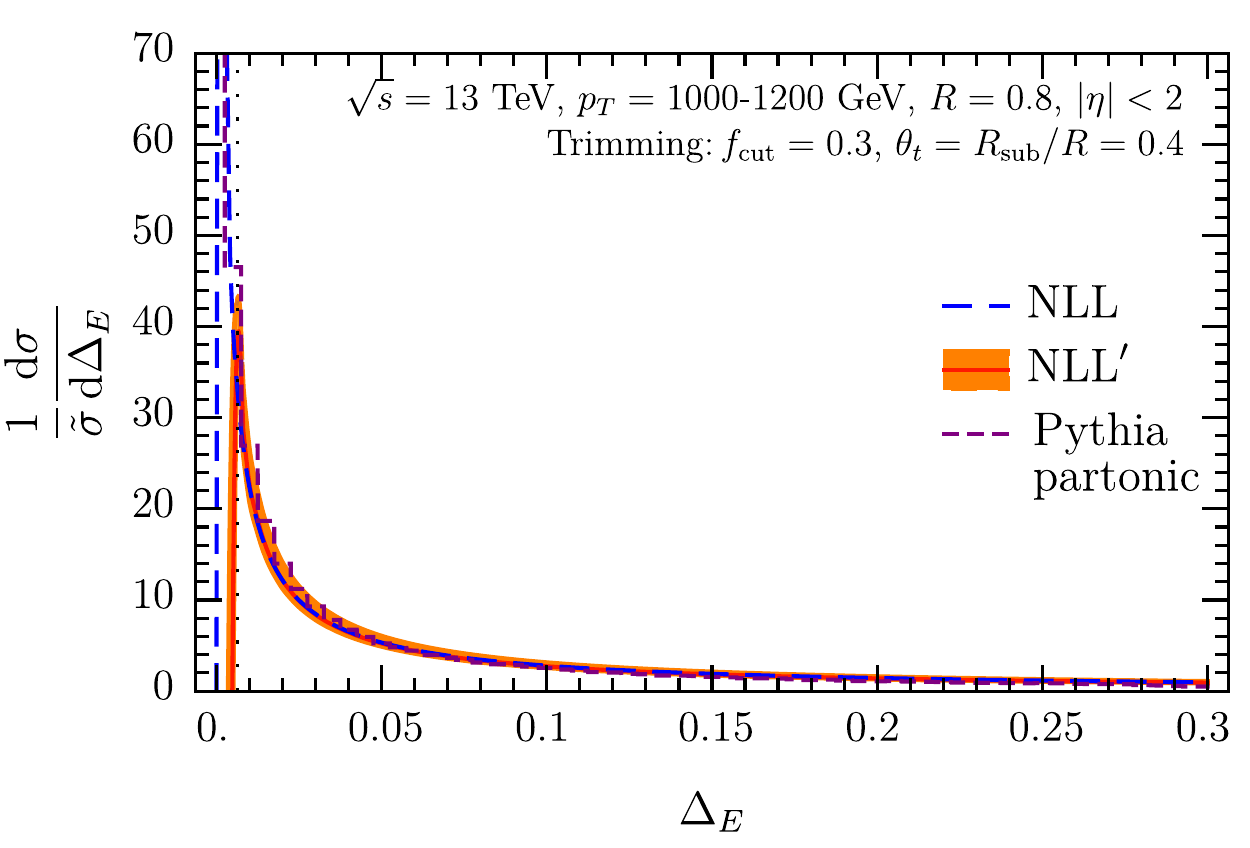} \hfill \phantom{.} 
    \caption{Numerical results for the jet energy drop obtained with the trimming algorithm at NLL (blue dashed) and NLL$'$ (orange) order, compared to \Pythia partonic (purple dashed). We choose trimming parameters $f_{\rm cut}=0.3$ and $\theta_t=0.4$. The different panels correspond to different jet transverse momenta.  The central curves are normalized to unity between the dotted vertical line and $\Delta_E=f_{\rm cut}$.~\label{fig:numerics_trimming}} 
\end{figure}

We start by presenting numerical results for the jet energy drop with trimming. For the four panels in \fig{numerics_trimming}, we choose the same LHC jet kinematics as in the previous sections. The grooming parameters are taken to be $f_{\rm cut}=0.3$ and $\theta_t =0.4$.  As mentioned before, the $f_{\rm cut}$ value here is  larger than what is typically used in experimental analyses. Our choice is motivated by the fact that for relatively large values of $f_{\rm cut}$ a significant fraction of the jet energy drop cross section is in the perturbative range, where the resummation techniques studied in this work are applicable. This point is illustrated in \fig{pythia_trim}, where \Pythia predictions at parton and hadron level are compared. Explicitly, in the right panel the red and green curves overlap for $\ed\gtrsim 0.15$, while they never completely overlap in the left panel. Similar to soft drop, MPI affects the whole distribution.

In \fig{numerics_trimming} we show both the NLL$'$ and NLL results, finding again that the NLL curve is within the uncertainty band of the NLL$'$ result in the perturbative region, indicating a good convergence of the resummed cross section. We note that the cross section does not vanish at $\Delta_E=f_{\rm cut}$, consistent with the NLO result shown in \fig{refactorization_trimming}. Indeed, the cross section can extend to $\Delta_E$ values well above $f_{\rm cut}$. However, this requires a different factorization formula than \eq{trimming_NLL}, since we formally assumed $\Delta_E \ll f_{\rm cut}$, and we leave an analysis of the region $\ed \gtrsim f_{\rm cut}$ for future work.

\begin{figure}[t]
     \hfill \includegraphics[width=0.48\textwidth]{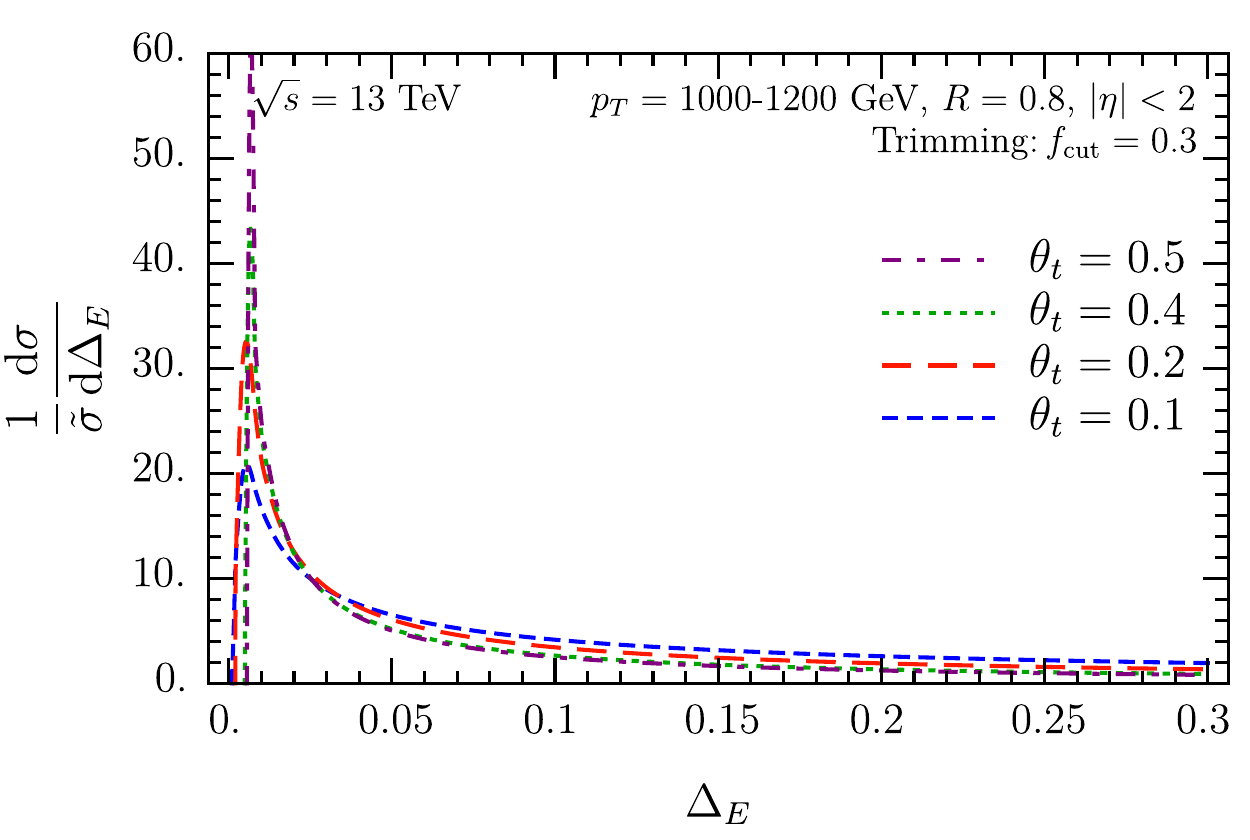} \hfill 
     \includegraphics[width=0.48\textwidth]{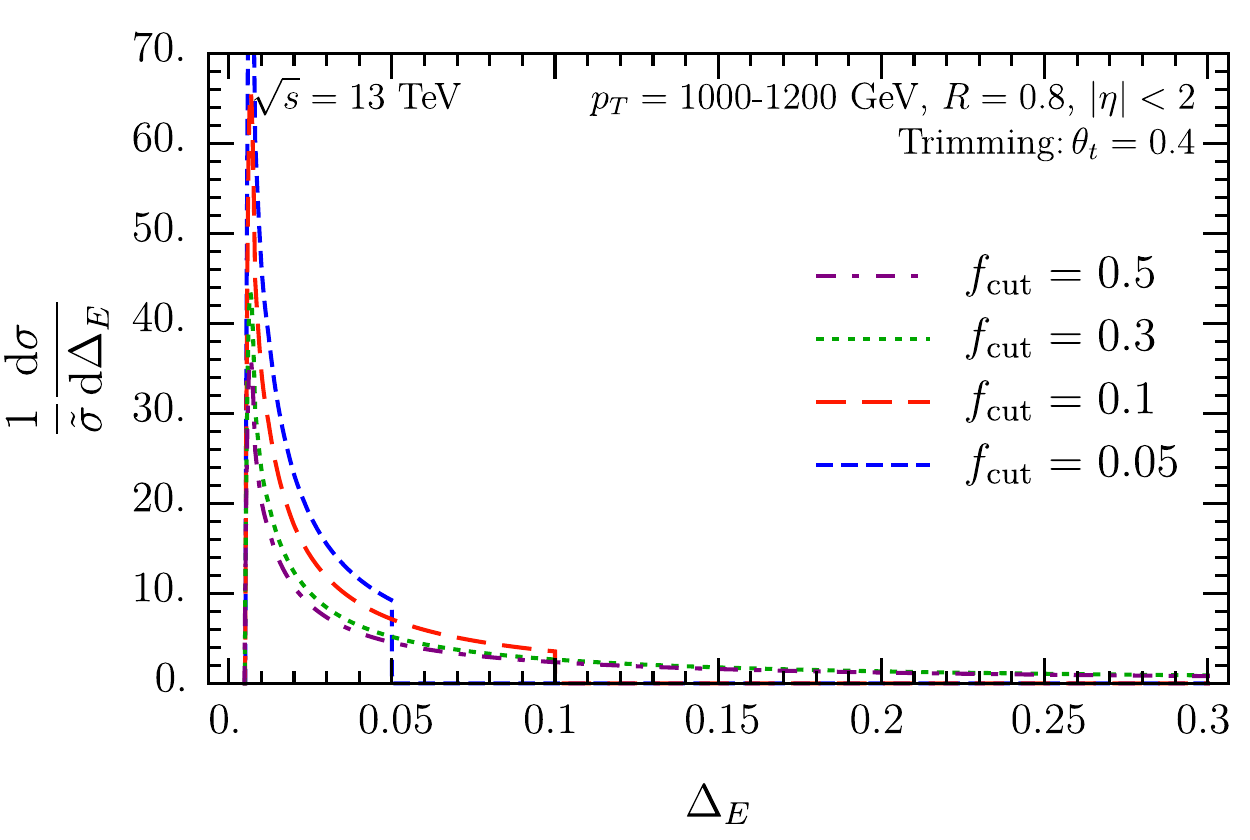} \hfill \phantom{.}
    \caption{Dependence of the jet energy drop for trimming on the grooming parameters $\theta_t$ (left) and $f_{\rm cut}$ (right).~\label{fig:trimming_parameterdep}}
\end{figure}

In \fig{trimming_parameterdep} we consider the dependence of the jet energy drop on the grooming parameters $\theta_t$ and $f_{\rm cut}$, for a jet $p_T=1000-1200$~GeV (as in the lower-right panel of \fig{numerics_trimming}). A larger value of $\theta_t$ leads to larger subjet energies, which are more likely to cross the threshold set by $f_{\rm cut}$. This leads to the larger spike near $\Delta_E\approx 0$ in the left panel of \fig{trimming_parameterdep}. Similarly, a smaller value of $f_{\rm cut}$ allows more subjets to pass the grooming condition, reducing the jet energy drop. Note that in this case we only plot  distributions for $\ed < f_{\rm cut}$, because our factorization formula does not lead to reliable predictions beyond that. 

\section{Conclusions~\label{sec:conclusions}}

We have studied the jet energy drop, which is the relative difference in jet energy (or transverse momentum) of a groomed and ungroomed jet, and is a key observable for characterizing the impact of grooming on jets. We considered three different grooming algorithms, frequently used in experimental analyses: i) soft drop, ii) iterated soft drop, and iii) trimming.  The jet energy drop is particularly sensitive to soft radiation, making it ideally suited for tuning parton shower event generators to data, particularly to constrain the hadronization model. Since it maps out the soft substructure of jets, it also has significant potential for studying the modification induced by medium effects in proton-nucleus and nucleus-nucleus collisions.

We have developed factorization formulae which allow for an evaluation of the cross sections at next-to-leading logarithmic (NLL$'$) accuracy, resumming logarithms of the jet radius, jet energy drop, and grooming parameters.
We also include the non-global logarithms with C/A clustering effects and Abelian clustering effects at order $\al_s^2$. Formally, one should also resum the NGLs, but from earlier work we note that their effect beyond the leading term is negligible for our phenomenological results. The factorization for soft drop requires a joint resummation of the jet energy drop and  groomed jet radius $\theta_g$, and is very sensitive to nonperturbative effects when integrating the resummed cross section over $\theta_g$. This sensitivity can be reduced in a controlled manner by imposing a minimum cut on $\theta_g$. The energy drop for soft drop with $\beta=0$ is a Sudakov safe observable, which we calculate also to NLL$'$. We have presented numerical results for all three algorithms and compared to \Pythia simulations, finding very good agreement in the perturbative region. For soft drop we also compared to CMS data.

\acknowledgments
We thank Y.~Chen, Y.~J.~Lee, J.~Mulligan, B.~Nachman and D.~Neill for helpful discussions, and A.~Larkoski for feedback on the manuscript. In addition, we thank Z.~B.~Kang for collaborating during the early stages of this project and A.~Papaefstathiou for assistance with Monte Carlo simulations. This work is supported by the U.S.~Department of Energy under Contract No.~DE-AC02-05CH11231, the LDRD Program of Lawrence Berkeley National Laboratory, the National Science Foundation under Grants PHY-1316617, 1620628, 1915093, and No.~ACI-1550228, by the ERC grant ERC-STG-2015-677323, the NWO projectruimte 680-91-122 and the D-ITP consortium, a program of NWO that is funded by the Dutch Ministry of Education, Culture and Science (OCW).

\appendix

\section{Anomalous dimensions~\label{app:anom}}

The coefficients in the QCD beta function are given by 
\begin{align} \label{eq:bt}
\bt_0& = \f{11}{3}C_A -\f{4}{3}T_F n_f \, , \nn \\
\bt_1& =\f{34}{3} C_A^2-\bigg(\f{20}{3}C_A +4 C_F \bigg)T_f n_f  \,
\,,\end{align}
and the cusp anomalous dimension is expanded in terms of 
\begin{align} \label{eq:ga}
\Gamma_0^i&= 4 C_i  \, , \nn \\
\Gamma_1^i&= 4 C_i \bigg[ \bigg(\f{67}{9}-\f{\pi^2}{3}\bigg) C_A-\f{20}{9} T_F n_f \bigg]
\end{align}
The one-loop Altarelli-Parisi splitting functions are given by
\begin{align} \label{eq:split}
P_{qq}(z) &= C_F\,\bigg[\frac{1+z^2}{(1-z)}_+ + \frac32 \de(1-z)\bigg] \,,
&
P_{gq}(z) &= C_F\,\f{1+(1-z)^2}{z}
\,, \nn \\
   P_{gg}(z) &= 2C_A \bigg[ \frac{z}{(1-z)}_+  + \frac{1-z}{z}+z(1-z)\bigg] + \frac{\beta_0}{2}\,\de(1-z)\,,
   &
   P_{qg}(z) &= T_F \big[z^2+(1-z)^2\big]
\,.\end{align}
Here we list all relevant anomalous dimensions
\begin{align}\label{eq:anomdim}
\ga_i^{S_G}(\zc p_T R,\beta,\mu) 
&= -\frac{2\as C_i}{\pi}  \f{1}{1+\beta} \, \ln\bigg(\frac{\mu}{\zc p_T R}\bigg)
\,, \nn \\[.5em] 
\gamma^{S_Z}_i(\ed,p_TR,\mu)
& = \gamma^{S_{Z'}}_i(\ed,p_TR,\mu) = -\frac{2\as C_i}{\pi}\bigg[\f{1}{[\ed]}_+ -  \ln\bigg(\frac{\mu}{ p_T R}\bigg) \delta(\ed) \bigg]
\,,\nn\\[.5em]
\ga^{\CS_X}_i(\Delta_E,z_{\rm cut}^{-1/\beta} p_TR,\beta,\mu)
&= \frac{2\as C_i}{\pi } \bigg[\f{1}{[\ed]}_+ - \frac{\beta}{1+\beta}\ln\bigg(\frac{\mu}{\zc^{-1/\bt }p_T R}\bigg) \delta (\ed) \bigg]
\,, \nn \\[.5em]
 \ga^{\tilde H}_{q}(p_{T} R,\mu)
  &= \f{\as C_F}{\pi}  \bigg[2 \ln\Big(\f{p_T R}{\mu} \Big)  - \frac32\bigg]
, \nn \\[.5em]
\ga^{\tilde H}_{g}(p_{T} R,\mu)
  &= \f{\as}{\pi} \bigg[2 C_A \ln\Big(\f{p_T R}{\mu} \Big) - \frac12 \beta_0 \bigg]
, \nn \\[.5em]
 \ga^{C^{\in {\rm gr}}}_{q}(\theta_g^c p_{T} R,\mu)
  &= \f{\as C_F}{\pi} \Big[2 \ln\bigg(\frac{\mu}{\theta_g^c p_T R}\Big) + \frac32\bigg]
\,, \nn \\[.5em]
\ga^{C^{\in {\rm gr}}}_{g}(\theta_g^c p_{T} R,\mu)
  &= \f{\as}{\pi} \Big[2C_A \ln \bigg(\frac{\mu}{\theta_g^c p_T R}\Big) + \frac12 \beta_0 \bigg]
\,, \nn \\[.5em]
\ga^{\CS_Z}_i(\ed,\theta_g\, p_T R,\mu)
&=\f{2\as C_i}{\pi} \bigg[\f{1}{[\ed]}_+ - \ln \bigg(\f{\mu }{\tg \, p_T R} \bigg)\delta (\ed) \bigg]
\,, \nn \\[.5em]
\ga^{\CS_{Z'}}_i(\ed,\theta_t\, p_T R,\mu)
&=\f{2\as C_i}{\pi} \bigg[\f{1}{[\ed]}_+ - \ln \bigg(\f{\mu }{\theta_t \,p_T R} \bigg)\delta (\ed) \bigg]\,, 
  \nn \\[.5em]
\ga_i^{S_T }(\fc\, p_T R,\mu) 
&= -\frac{2\as C_i}{\pi}   \, \ln\bigg(\frac{\mu}{\fc\,  p_T R}\bigg)
\,, \nn \\[.5em]
\ga^{\CS_T}_i(\fc\theta_t\, p_T R,\mu)
&=\f{2\as C_i}{\pi}  \ln \bigg(\f{\mu }{ \fc  \theta_t\, p_TR} \bigg) 
\,. 
\end{align}
where $C_i = C_F$ ($C_A$) for $i=q$ ($i=g$). 
We achieve full NLL$'$ accuracy by including the two-loop cusp anomalous dimension, which multiplies the $\ln \mu$ terms in~\eq{anomdim}.

\bibliographystyle{JHEP}
\bibliography{bibliography}

\end{document}